\documentclass[unsortedaddress,floatfix,nofootinbib,aps,prd,preprintnumbers,showpacs,showkeys]{revtex4-1}

\usepackage{amssymb}
\usepackage{amsmath}
\usepackage{graphicx}
\usepackage{dcolumn}
\usepackage{bm}
\usepackage{epsfig}

\def\figdir{./}

\newcommand\figWidthHalf{.48\textwidth}

\newcommand\figWidthQuarters{.24\textwidth}
\newcommand\figWidthFifths{.19\textwidth}
\newcommand\Eq[1]{Eq.~\ref{eq:#1}}
\newcommand\Fig[1]{Fig.~\ref{fig:#1}}
\newcommand\Sec[1]{Sec.~\ref{sec:#1}}
\newcommand\Appendix[1]{Appendix~\ref{appendix:#1}}
\newcommand\Tab[1]{Table~\ref{tab:#1}}

\newcommand\bfn{\mathbf n}

\newcommand\bfe{\mathbf e}

\newcommand\calO{\mathcal O}

\newcommand\calP{\mathcal P}
\newcommand\calQ{\mathcal Q}
\newcommand\calR{\mathcal R}

\newcommand\calM{\mathcal M}
\newcommand\Tr{\textrm{Tr}}
\newcommand\Transpose{\intercal}

\begin{document}

\preprint{MIT-CTP/4726}

\title{Multiscale Monte Carlo equilibration: Pure Yang-Mills theory}

\author{Michael G. Endres}
\email{endres@mit.edu}
\affiliation{Center for Theoretical Physics, Massachusetts Institute of Technology, Cambridge, Massachusetts 02139, USA}

\author{Richard C. Brower}
\email{brower@bu.edu}
\affiliation{Department of Physics, Boston University, Boston, Massachusetts 02215, USA}

\author{William Detmold}
\email{wdetmold@mit.edu}
\affiliation{Center for Theoretical Physics, Massachusetts Institute of Technology, Cambridge, Massachusetts 02139, USA}

\author{Kostas Orginos}
\email{kostas@wm.edu}
\affiliation{Department of Physics, College of William and Mary, Williamsburg, Virginia 23187-8795, USA}
\affiliation{Jefferson Laboratory, 12000 Jefferson Avenue, Newport News, Virginia 23606, USA}

\author{Andrew V. Pochinsky}
\email{avp@mit.edu}
\affiliation{Center for Theoretical Physics, Massachusetts Institute of Technology, Cambridge, Massachusetts 02139, USA}


\date{\today}

\begin{abstract}
We present a multiscale thermalization algorithm for lattice gauge theory, which enables efficient parallel generation of uncorrelated gauge field configurations.
The algorithm combines standard Monte Carlo techniques with ideas drawn from real space renormalization group and multigrid methods.
We demonstrate the viability of the algorithm for pure Yang-Mills gauge theory for both heat bath and hybrid Monte Carlo evolution, and show that it ameliorates the problem of topological freezing up to controllable lattice spacing artifacts.
\end{abstract}

\maketitle

\section{Introduction}
\label{sec:introduction}

Numerical simulations of lattice quantum chromodynamics (QCD) and other lattice gauge theories rely on Markov chain Monte Carlo techniques to evaluate the path integral that defines the theory and its correlation functions.
Ensembles generated in a Markov process, however, are often highly correlated due to slow modes in the stochastic evolution.
Such correlations reduce the effective sample size of the generated ensemble, and thus directly influence the efficiency of such simulations.
The problem of slow modes becomes particularly acute in the vicinity of critical points where the continuum limit is defined, resulting in what is commonly known as critical slowing down.
Although a variety of algorithmic developments, such as cluster algorithms~\cite{PhysRevLett.58.86, Wolff:1988uh} and the worm algorithm~\cite{Prokof'ev:2001zz}, have dramatically reduced the problem of critical slowing down for some simple statistical models, they appear to have limited utility for gauge theories such as QCD, where simulations at lattice spacings $a<0.05$ fm remain extremely challenging.

In gauge theories, topological quantities\footnote{Below, we will specify to a particular definition of topology, but the evolution properties and connection to slow modes are insensitive to these details.} (e.g., topological charge and susceptibility) are examples of observables that couple strongly to slow modes of the stochastic evolution.
In the continuum and at infinite volume, topological charge is invariant under continuous local deformations of a field configuration.
By contrast, on a finite lattice, changes in topology are possible through local updates.
Such changes, however, require traversals over large action barriers in configuration space, which in the continuum become infinite and result in the breakup of the configuration space into distinct topological sectors.
The likelihood of such tunneling events rapidly diminishes in the approach to the continuum as the height of such topological barriers diverge, resulting in a problem known as topological freezing.
This phenomenon was first observed in quenched calculations with improved gauge actions~\cite{PhysRevD.73.094507} as well as in more recent dynamical simulations~\cite{Schaefer:2010hu}.
Recently, open boundary conditions (BCs) in time~\cite{Luscher:2011kk} were proposed as a method for enhancing changes in topology by allowing charge to flow in and out through the boundaries.
Although offering an improvement in topological tunneling over periodic BCs, open BC simulations still suffer from critical slowing down~\cite{McGlynn:2014bxa}.

It is important to note that for gauge theories, critical slowing down persists even in the absence of topological freezing, because the evolution of long distance (slow) modes can only arise through the application of many local updates at the scale of the lattice spacing.
As the lattice spacing is reduced, the number of updates required to move modes at a given physical scale increases.
Multiscale evolution algorithms offer the prospect of performing Markov process updates that change modes at different physical scales more efficiently.
A number of such approaches have been explored in the literature, primarily for models that are simpler than QCD, and have met with some success (see e. g.,~\cite{Goodman:1986pv,Edwards:1990hu,Edwards:1991eg,Grabenstein:1993nh,Janke:1993et,Grabenstein:1994ze}).
We are unaware, however, of any successful work in this direction relevant to QCD.

In this study we investigate a less ambitious direction, namely a multiscale {\it thermalization} algorithm, which combines standard heat bath (HB) or hybrid Monte Carlo (HMC) updating methods with the real-space renormalization group (RG) and multigrid concepts of restriction and prolongation between pairs of matched coarse and fine lattices and lattice actions.
The algorithm proceeds in four steps:
\begin{enumerate}
\item A coarse action is determined by a RG transformation from the target (fine) lattice action\footnote{In practice, an approximation to the RG transformation is used.};
\item A set of $N_s$ independent equilibrated coarse configurations are subsequently generated by a conventional Monte Carlo process;
\item Each coarse configuration is then prolongated (or refined), thereby producing a set of $N_s$ configurations defined on the fine lattice;
\item The prolongated (fine) ensemble is then equilibrated (or rethermalized) and evolved in parallel using a conventional algorithm to produce an ensemble of $N_e$ decorrelated configurations for each of the $N_s$ independent streams.
\end{enumerate}
This procedure may be generalized to have several levels of refinement proceeding from the coarsest to the finest target ensemble.
At each level, the coarse action should follow an RG flow of the underlying gauge dynamics. 

Assuming that the computational cost of the coarse evolution and prolongation are negligible compared to the fine evolution, the efficiency of this strategy is determined by the rethermalization time of the prolongated ensemble compared to the decorrelation time for fine evolution.
Under the physically reasonable assumption that the distribution of prolongated configurations only differs from that of the target distribution for fine configurations by cutoff artifacts, one might expect the former time scale to be shorter than the latter. 
Given this is indeed the case, the scheme will provide an efficient method for initializing field configurations at a fine lattice spacing for subsequent parallel evolution, ultimately yielding decorrelated ensembles of size $N_s \times N_e$.

In many cases, thermalization is considerably more challenging than evolution, and therefore we expect the approach to have significant advantages.
Computationally, the parallel nature of the fine evolution of multiple independent streams means that ensembles can be generated more efficiently using fewer computational resources.
The trade-off between the parameters $N_s$ and $N_e$ opens possibilities for optimizing the statistical power of subsequent analysis and the use of hardware resources.
Furthermore, the strategy can be implemented on a hierarchy of different coarse/fine pairs resulting in rapid thermalization at multiple scales, thus enabling simulations at very fine lattice discretizations.

\begin{figure} 
\includegraphics[width=\figWidthHalf]{\figdir 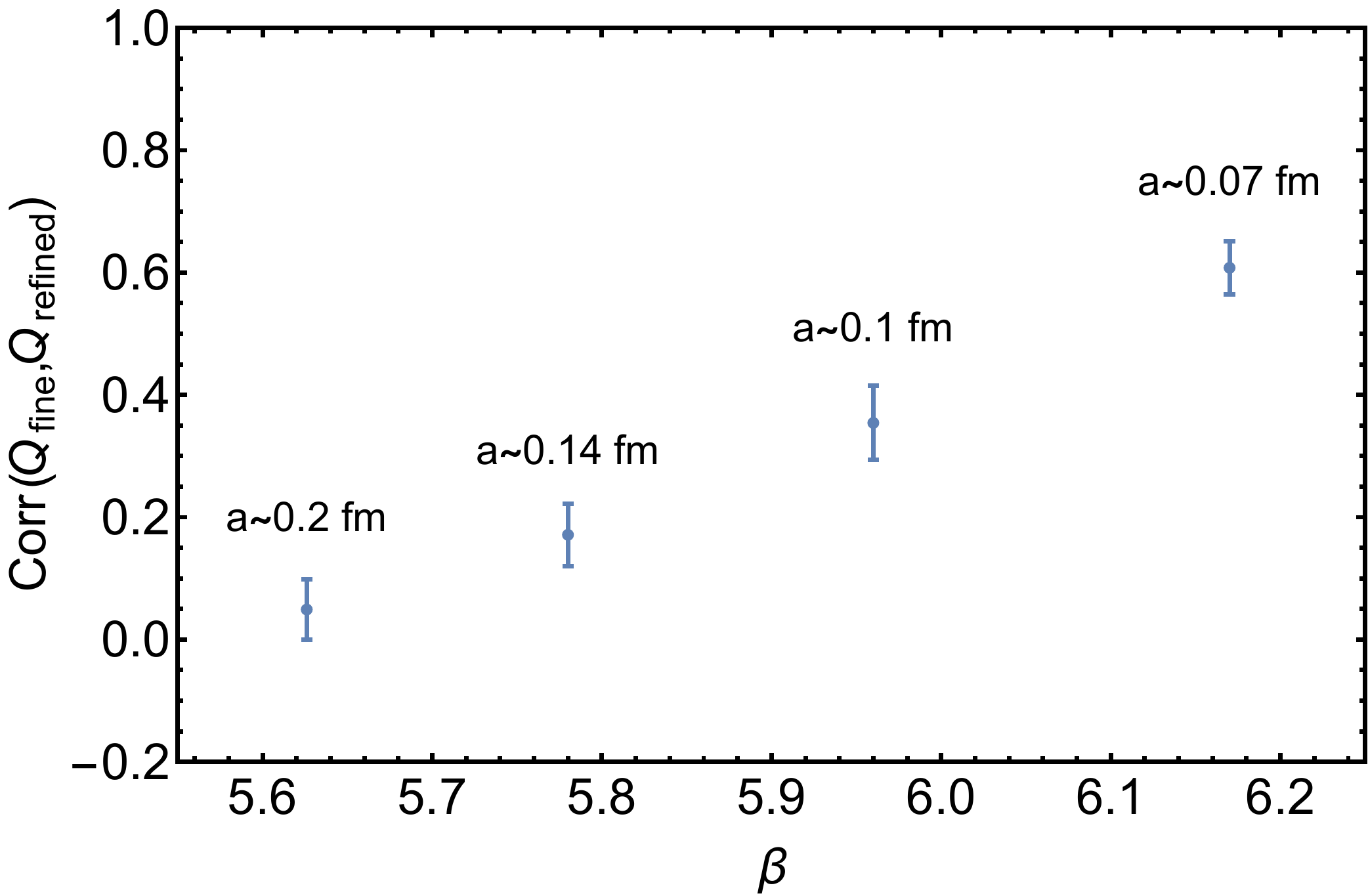}
\hspace{10pt}
\includegraphics[width=\figWidthHalf]{\figdir 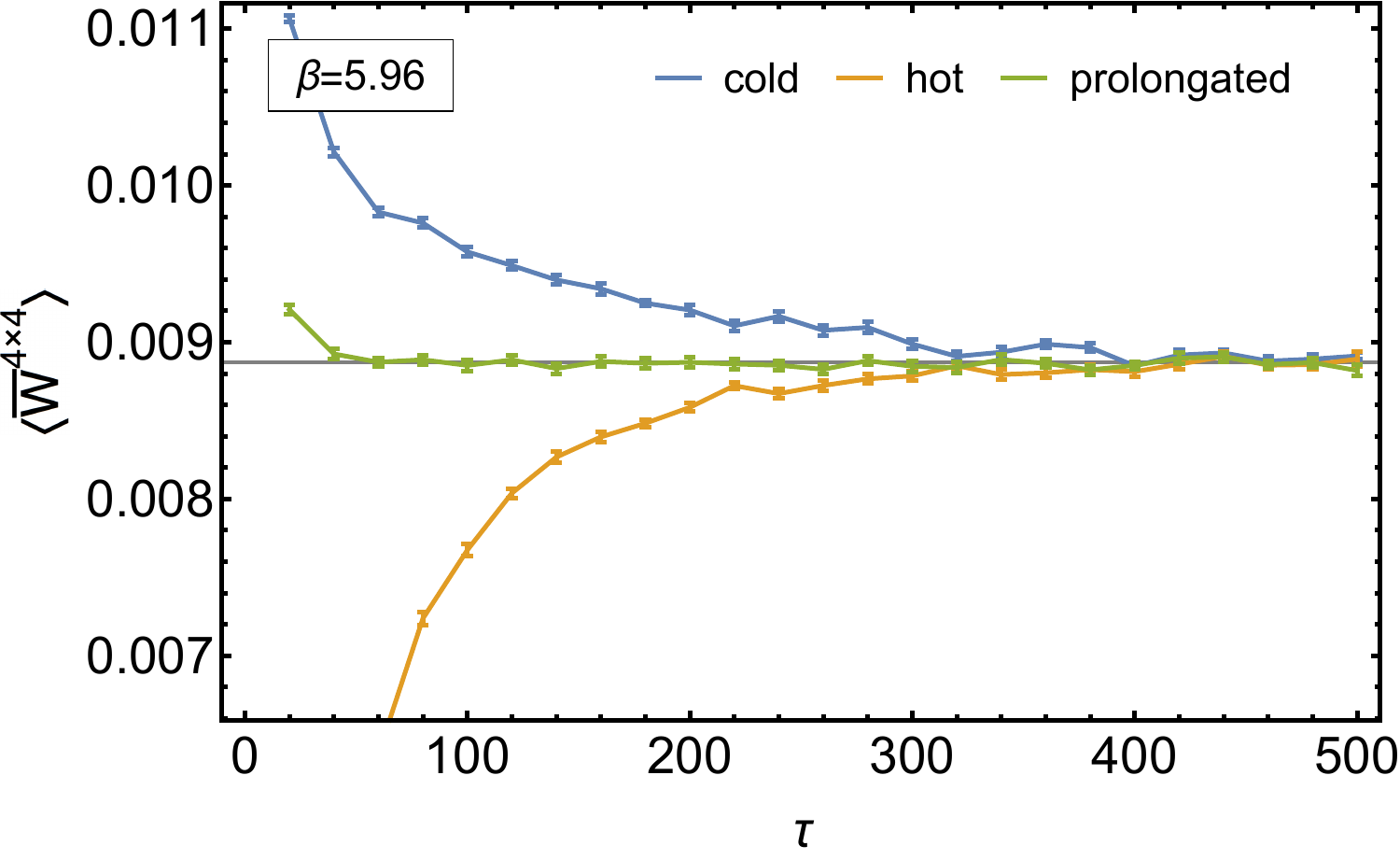}
\caption{\label{fig:intro_figs}%
Left: Correlations in topology between an ensemble before (fine) and after restriction and prolongation (refined), as a function of the coupling, $\beta$ (see \Sec{topological_charge} for details).
Right: Average Wilson loop (approximately $0.4\,\textrm{fm}\times 0.4\,\textrm{fm}$ in dimension) as a function of rethermalization time for an initial ensemble obtained via prolongation of a matched coarse ensemble, and corresponding estimates as a function of the thermalization time for cold and hot starts; (re)thermalization time is measured in terms of the number of unit length HMC trajectories, $\tau$ (see \Sec{rethermalization} and  \Appendix{wilson_loops} for details).
The horizontal band indicates a high precision determination of the Wilson loop obtained from a large decorrelated ensemble.
}
\end{figure}

To test the viability of our strategy, we study a variety of observables that probe long distance scales in pure $SU(3)$ gauge theory.
To facilitate our studies, we utilize restriction as a device for preparing coarse ensembles corresponding to a renormalized coarse action.
An appealing feature of our restriction and prolongation operations is that they well preserve the topological charge distribution of the ensembles to which they are applied.
As a consequence, the ensembles obtained by prolongation will possess properly distributed topology up to lattice artifacts which are inherited from the coarse action.
The prolongator in fact satisfies the stronger property of preserving the topological charge for individual configurations at sufficiently small, but presently accessible, lattice spacings.
This property is demonstrated numerically by studying the growth in correlations in topology between ensembles before and after restriction and prolongation as a function of the inverse lattice spacing, as shown in \Fig{intro_figs} (left).
These features are important, since they enable us to achieve thermalized ensembles in time frames which are far shorter than the decorrelation time for fine evolution, providing the latter is controlled by topology.

Finally, a key measure for establishing the success of our approach is the requisite rethermalization time for an ensemble prepared via prolongation to return to equilibrium under standard updating procedures.
This time is to be compared with the thermalization time for a typical ordered (``cold'') or disordered (``hot'') start, as well as the decorrelation time for fine evolution.
To address this, we monitor the (re)thermalization times for a variety of observables, including the topological susceptibility and rectangular Wilson loops.
In light of the fact that our prolongator preserves topology, we emphasize the study of rethermalization times for nontopological long-distance quantities and demonstrate that they are significantly shorter than the thermalization times and decorrelation times of topological quantities in conventional evolution.
In \Fig{intro_figs} (right), we provide an illustrative comparison of the rethermalization time for a representative prolongated ensemble of size $N_s=24$, and the corresponding thermalization times for hot and cold initial ensembles, as probed by a $0.4\,\textrm{fm}\times 0.4\,\textrm{fm}$ Wilson loop (see \Sec{rethermalization} and  \Appendix{wilson_loops} for details).
For this observable and a large range of other quantities that we investigate, we see that the rethermalization time for the prolongated ensemble is dramatically shorter than the thermalization times measured for hot and cold starts.
It should be emphasized that our choice of prolongator is designed to preserve a large class of Wilson loops on all scales, and it is likely this feature that enables the rapid thermalization seen in this example.

In the remainder of this paper, we elaborate and expand on the results highlighted above.
Before doing so, we first review some basic concepts and known results relating to Markov processes, which provide a theoretical basis for our strategy.
Following an overview of notation and definitions, we then introduce the specific choice of restriction and prolongation operations used throughout this work.
The latter is carefully chosen so as to retain an imprint of the long distance correlations and topological charge of the coarse configurations.
Then, we demonstrate the viability of the multiscale approach through numerical studies of pure Yang-Mills gauge theory in two parts: first, by showing that the proposed prolongation procedure preserves the topological charge on a configuration by configuration basis for sufficiently fine lattice spacing, and second by demonstrating that the rethermalization time required to correct the distribution obtained by prolongation of a coarse ensemble is shorter than the decorrelation time for fine evolution.
For the second studies we consider two commonly used algorithms, namely, HB and HMC.
The latter case is of greater interest, since it is the algorithm used in state-of-the-art QCD simulations with dynamical fermions.
Finally, we conclude with a detailed discussion of the potential applications and pitfalls of these approaches, as well as an outlook on future directions.
Appendices are devoted to technical details of the simulation and prolongation algorithms as well as technical aspects of the data analysis that are required to extract the (re)thermalization and evolution time scales in this work.

\section{Preliminary considerations}
\label{sec:preliminaries}

We begin by reviewing the basic aspects of Markov processes and their use in Monte Carlo importance sampling.
A sequence of configurations
\begin{eqnarray}
s_1 \to s_2 \to s_3 \to \cdots \to s_\tau\ ,
\label{eq:markov_chain}
\end{eqnarray}
labeled for simplicity by a discrete index $\tau$, is generated in a Markov process described by the transition matrix $\calM$.
For simplicity, we assume $\calM$ acts on a discrete configuration space $\Sigma$.
The matrix elements $\calM(s^\prime,s)$ give the transition probabilities for the configuration $s$ to go to the configuration $s^\prime$.
Under adequate conditions (see, e.g.,~\cite{Luscher:2010ae}), there exists a stationary distribution $\chi_0(s) = \calP(s)$, which is a right eigenstate of the transition matrix, i.e., satisfying  $\calP(s^\prime) = \sum_s \calM(s^\prime,s) \calP(s) $, normalized to $\sum_s \calP(s) = 1$ with eigenvalue $\lambda_0=1$.
The left eigenstate is $\tilde\chi_0(s) = 1$ as a consequence of probability conservation: $\sum_{s^\prime} \calM(s^\prime,s) = 1$.
Expectation values of operators $\calO(s)$ averaged over the stationary distribution are given by the inner product $\langle \calO \rangle = \sum_s \calO(s)\calP(s)$.
Although it is not required for a valid algorithm, let us assume that $\calM$ satisfies detailed balance, such that
\begin{eqnarray}
\calM(s^\prime,s) \calP(s) = \calM(s,s^\prime) \calP(s^\prime) \ .
\end{eqnarray}
In this case, the spectrum of $\calM$ consists of real eigenvalues $\lambda_n$, with $|\lambda_n| \equiv e^{-1/\tau_n}$ ordered such that $|\lambda_n| \le |\lambda_{n-1}|$; note that $|\lambda_n|\le1$ for all $n$.
The corresponding right and left eigenvectors are given by $\chi_n(s)$ and $\tilde \chi_n(s) = \chi_n(s)/\calP(s)$, respectively, and are chosen to be real and mutually orthonormal, satisfying $\sum_s\tilde\chi_n(s) \chi_{n^\prime}(s) = \delta_{nn^\prime}$.
The spectral decomposition of $\tau$ applications of the transition matrix is given by
\begin{eqnarray}
\calM^\tau(s^\prime, s) = \sum_{n\ge0} \chi_n(s^\prime) \lambda_n^\tau \tilde\chi_n(s) \ .
\label{eq:M_decomp}
\end{eqnarray}
It follows that the $\tau$th configuration in \Eq{markov_chain} is drawn from the distribution
\begin{eqnarray}
\calP_\tau(s^\prime) = \calP(s^\prime) + \sum_{n>0} \chi_n(s^\prime) \lambda_n^\tau \left[\sum_s \tilde\chi_n(s) \calP_1(s) \right] \ ,
\label{eq:therm_dist}
\end{eqnarray}
where $\calP_1(s)$ is an initial probability distribution.
The expectation of an operator at this point in the Markov process is given by
\begin{eqnarray}
\langle \calO \rangle_\tau = \langle \calO \rangle + \sum_{n>0} \left[\sum_{s} \chi_n(s)\calO(s) \right] \left[\sum_s \tilde\chi_n(s)\calP_1(s) \right] \lambda_n^\tau \ .
\label{eq:therm_obs}
\end{eqnarray}

At late times in the Markov process, both the distribution and the observables converge to their stationary values exponentially, assuming the existence of a gap in the spectrum of $\calM$.
The rate of this convergence (i.e., thermalization time) is dominated by the exponential correlation time $\tau_\textrm{exp} \equiv \tau_1$.
In addition to the time scale, $\tau_\textrm{exp}$, the rate of thermalization is influenced by the overlap between the initial distribution $\calP_1(s)$ and left eigenvectors $\tilde \chi_n(s)$, for $n>0$; in particular, if the overlap vanishes or is exponentially small for $n=1$, then the relevant thermalization time scale would be governed by the shorter time scale, $\tau_2$.
It is observationally established in many examples that near a phase transition, $\tau_\textrm{exp} \sim \xi^z$ where $\xi$ is the largest correlation length of the system (in dimensionless units) and $z$ is a dynamical critical exponent.
Local updating processes are diffusive by nature, implying an exponent $z\sim 2$.
However in some cases, such as lattice gauge theories, the scaling can be far worse (e.g., $z\sim 5$ for topological quantities~\cite{Schaefer:2010hu}).
For critical systems of spatial extent $L$, $\xi\sim L$, and the scaling becomes $\tau_\textrm{exp} \sim L^z$; this kind of volume scaling is a hallmark property of critical slowing down.

A second time scale (or set of scales) of interest is the integrated autocorrelation time $\tau_\textrm{int}(\calO)$, which characterizes the correlations in measurements of an observable after thermalization due to the sequential nature of the Markov process.
In contrast with $\tau_\textrm{exp}$, this time scale depends not only on the algorithmic details (e.g., the eigenvalues and eigenvectors of $\calM$) but also on how well the observable in question couples to the various modes of the stochastic process.
Because of the presence of such correlations, the estimated uncertainties on a given quantity $\langle \calO \rangle$, are enhanced by a factor $\sqrt{2 \tau_\textrm{int}(\calO)}$ compared to those obtained under the assumption that the ensemble is decorrelated.
The integrated autocorrelation time is defined by
\begin{eqnarray}
\tau_\textrm{int}(\calO) = \frac{1}{2} + \sum_{\Delta>0} \frac{ \Gamma_\Delta(\calO)}{\Gamma_0(\calO)} \ ,
\end{eqnarray}
where
\begin{eqnarray}
\Gamma_\Delta(\calO) = \sum_{s^\prime s} \delta\calO(s^\prime) \calM^\Delta(s^\prime,s) \delta\calO(s) \calP(s) 
\end{eqnarray}
is the lag-$\Delta$ autocovariance function, and $\delta\calO(s) =\calO(s) - \langle \calO \rangle$.
Using \Eq{M_decomp}, this expression may be written as
\begin{eqnarray}
\Gamma_\Delta(\calO) = \sum_{n>0} a_n({\cal O}) \, \lambda_n^\Delta \ ,\qquad a_n({\cal O}) = \left[ \sum_s \delta\calO(s) \chi_n(s) \right]^2
\end{eqnarray}
and consequently the integrated correlation time may be expressed as
\begin{eqnarray}
\tau_\textrm{int}({\cal O}) = \left[\sum_{n>0} a_n({\cal O}) \right]^{-1} \sum_{n>0} a_n({\cal O}) \, \eta_n \ ,\qquad \eta_n = \frac{1}{2} + \frac{\lambda_n}{1-\lambda_n} \ ,
\end{eqnarray}
where $\eta_n >  0$ for all $n>0$.
Under the assumption that $\cal O$ is real (or is the real part of an observable), then $a_n({\cal O}) \ge 0$ for all $n>0$, and one can establish the bound $\tau_\textrm{int}(\calO) \le \hat \tau_\textrm{int}$, where
\begin{eqnarray}
\hat\tau_\textrm{int} \le \frac{1}{2} + \frac{|\lambda_1|}{1-|\lambda_1|} \le \tau_\textrm{exp} + \frac{1}{12} \frac{1}{\tau_\textrm{exp}}\ .
\end{eqnarray}
It follows that integrated autcorrelation times are at worst on the order of $\tau_\textrm{exp}$ when the latter is large. 
Interestingly, this bound does not preclude the possibility that $\hat\tau_\textrm{int}\ll\tau_\textrm{exp}$.

\begin{figure} 
\includegraphics[width=0.8\textwidth]{\figdir 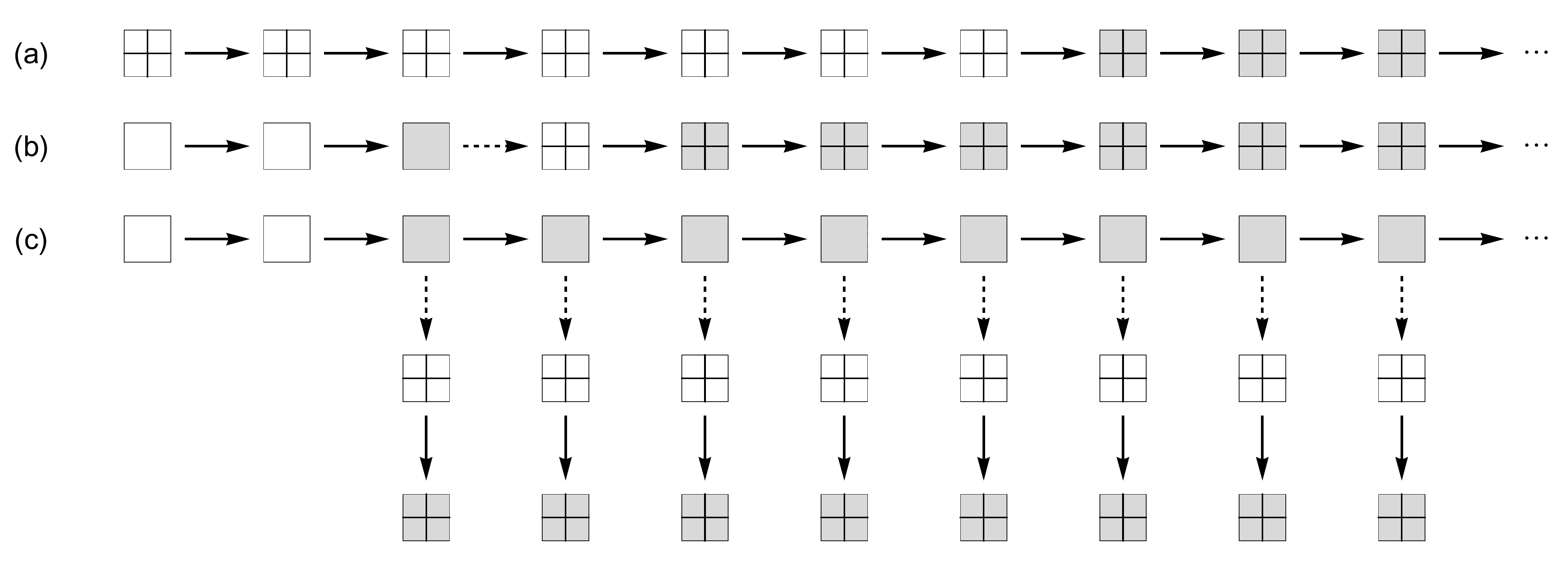}
\caption{\label{fig:mcDiagram}%
Ensemble generation strategies: single fine lattice stream (a), single coarse lattice stream, followed by refinement followed by a single fine lattice stream (b), and a single coarse lattice stream, followed by parallel refinement and rethermalization of refined lattices (c).
In all cases, $\boxplus$ represents a fine configuration, $\square$ represents a coarse configuration, unshaded shapes correspond to unthermalized configurations and shaded shapes correspond to thermalized configurations.
For each simulation strategy, ensemble averages are performed over shaded (fine) configurations, either generated from a single stream (a,b) or in parallel (c).
}
\end{figure}

In a standard Markov chain Monte Carlo simulation, represented schematically by \Fig{mcDiagram} (a), there are two relevant time scales associated with the algorithm: the equilibration or thermalization time $\tau_\textrm{therm} \propto \tau_\textrm{exp}$, and the decorrelation time for observables, which is bounded by $2\hat\tau_\textrm{int}$.
The former will depend to some extent on the initial configuration, drawn from the probability distribution $\calP_1(s)$; if the initial configuration is drawn from the stationary distribution $\calP(s)$, then the thermalization time will vanish.\footnote{Strictly speaking, it does not make sense to talk about a thermalized configuration, but rather a configuration that is drawn from a thermalized distribution.}

Next, let us introduce operators that map probability distributions between fine and coarse configuration spaces.
Borrowing the terminology of multigrid, we refer to these as {\it restriction} operators, $\calR$, when mapping from the fine to coarse configuration space and {\it prolongation} operators, $\calQ$, when mapping from the coarse to fine configuration space.
To facilitate the discussion, we adorn all coarse and fine quantities with the labels ($c$) and ($f$), respectively.
For example, fine and coarse configurations are labeled as $s^f\in \Sigma^f$ and $s^c \in \Sigma^c$, where $\Sigma^f$ and $\Sigma^c$ represent the fine and coarse configuration spaces, respectively.
The restrictor and prolongator can be represented by the matrices $\calR(s^c,s^f)$ and $\calQ(s^f,s^c)$ which act on fine and coarse configuration spaces, respectively.
The restrictor and prolongator should be probability preserving, and therefore must satisfy $\sum_{s^f} \calQ(s^f,s^c) = \sum_{s^c} \calR(s^c,s^f) = 1$.
Such transformations can be one to one, in which case the rectangular matrices $\calR$ and $\calQ$ have at most one nonzero entry per row and column, or they can be probabilistic.
Both restriction and prolongation operations are nonunique, need not satisfy $\calR \calQ = 1$, and cannot satisfy $\calQ \calR=1$ since the rank of $\calQ$ and $\calR$ is that of $\textrm{dim}(\Sigma^c)$ and not $\textrm{dim}(\Sigma^f)$.
Explicitly, the restriction operation acting on a fine probability distribution $\calP^f$ produces a coarse probability distribution, given by
\begin{eqnarray}
\calP^c(s^c) = \sum_{s^f} \calR(s^c,s^f) \calP^f(s^f)\ ,
\end{eqnarray}
and can be interpreted as a renormalization group transformation (e.g., decimation or block spin averaging in a simple implementation).
This can be seen by noting the equality of partition functions $\sum_{s^c}\calP^c(s^c) = \sum_{s^f} \calP^f(s^f)$.
On the other hand, the prolongation operation maps a coarse probability distribution to a fine distribution, given by
\begin{eqnarray}
\calP^f(s^f) = \sum_{s^c} \calQ(s^f,s^c) \calP^c(s^c)\ ,
\end{eqnarray}
and can be interpreted as a kind of inverse RG transformation.

With the concepts of restriction and prolongation in hand, consider a simulation represented schematically by \Fig{mcDiagram} (b), corresponding to the scenario in which $N_s=1$ and $N_e \gg 1$.
Here, evolution is first performed on a coarse lattice using an algorithm represented by the coarse transition matrix $\calM^c$ (which implicitly depends on a coarse action) until it is thermalized.
Subsequently the lattice is prolongated, and finally rethermalized using an algorithm represented by the fine transition matrix $\calM^f$.
Note that the subsequent rethermalization is needed to correct the prolongated configuration at the scale of the fine cutoff.
In this example, there are now three relevant time scales associated with the algorithm in its entirety: the coarse thermalization time $\tau^c_\textrm{therm}$, the rethermalization time $\tau^f_\textrm{retherm}$, and the decorrelation time of the fine evolution, bounded by $2\hat\tau^f_\textrm{int}$.
The procedure represented by \Fig{mcDiagram} (b) will be computationally less costly than that shown in \Fig{mcDiagram} (a) provided $\tau_\textrm{therm}^c + \tau^f_\textrm{retherm} < \tau^f_\textrm{therm}$.
Nevertheless, the improvements that can be found here are attenuated by the cost of the generation of a large ensemble since $\tau_\textrm{therm}^f/(N_e 2\hat\tau_\textrm{int}^f)\to 0$ as $N_e\to \infty$. 

As previously discussed, the rethermalization time of the prolongated configuration is at worst governed by the time scale $\tau^f_\textrm{exp}$, which is algorithm dependent, and overlap factors, which depend in part on the initial refined distribution and are thereby controllable.
In light of \Eq{therm_dist} and \Eq{therm_obs}, we can in principle accelerate the approach to equilibrium of the fine ensemble, by setting to zero the overlap of our refined ensemble with a fixed set of the slowest modes,
\begin{eqnarray}
\sum_{s^f s^c}  \tilde \chi_n^f(s^f) \calQ(s^f,s^c) \calP^c(s^c) = 0 \ .
\label{eq:weak_cond}
\end{eqnarray}
Removal of the lowest mode in this fashion, for example, would imply that the rethermalization time is no longer governed by $\tau^f_\textrm{exp}$, but rather by the shorter time scale, $\tau^f_2$.
In practice this is difficult to achieve, but by judicious choices, one seeks to  approximate this condition for as wide a range of slow modes as possible.
Note that this condition depends on the prolongator, on the Markov process used for fine evolution, and implicitly on the renormalized coarse action.
All of these factors are therefore important in maximizing the efficiency of our algorithm.

Finally, let us consider a simulation represented schematically by \Fig{mcDiagram} (c), corresponding to the scenario in which $N_s\gg 1$, and $N_e=1$. 
In this case, the evolution is first performed on a coarse lattice using an algorithm represented by the transition matrix $\calM^c$ until an ensemble of decorrelated configurations are generated.
The ensemble of decorrelated coarse configurations are subsequently prolongated, and finally rethermalized using an algorithm represented by the transition matrix $\calM^f$.
This procedure has three time scales associated with it: the coarse thermalization time $\tau^c_\textrm{therm}$, the decorrelation time for coarse evolution, bounded by $2\hat\tau^c_\textrm{int}$, and the rethermalization time $\tau^f_\textrm{retherm}$.
The procedure is computationally less costly than \Fig{mcDiagram} (a) provided that $\tau^c_\textrm{therm} + N_s \hat 2 \tau^c_\textrm{int} + N_s \tau^f_\textrm{retherm} < \tau^f_\textrm{therm} + N_s 2\hat \tau^f_\textrm{int}$, where $N_s$ is the size of the target ensemble being generated.
For large $N_s$, the condition reduces to  $2 \hat \tau^c_\textrm{int} + \tau^f_\textrm{retherm} < 2 \hat \tau^f_\textrm{int}$.
Since the decorrelation time for coarse evolution is usually negligible compared to that for fine evolution, the approach will be less computationally costly when the decorrelation time for fine evolution exceeds the rethermalization time for the prolongated ensemble.
Note that the computationally most intensive component of this algorithm, namely rethermalization, is embarrassingly parallel, and so each stream can be generated with maximal efficiency on available computing resources.

Our goal for the remainder of this paper is to explore choices of $\calP^c$ and $\calQ$, given a fine transition matrix $\calM^f$, such that \Fig{mcDiagram} (c) becomes a viable simulation strategy in pure gauge theory, and to investigate the time scales associated with the various approaches described above.
Provided that the fine evolution is ergodic, the proposed algorithm as a whole will also be ergodic.
Given that the prolongated ensemble inherits the long-distance properties of the coarse ensemble, the relevant parts of the fine configurations space $\Sigma^f$ are in some sense evenly populated.
Since the rethermalization merely corrects the missing short-distance part of the prolongated distribution, our underlying assumption is that all of the fine configuration space can be covered within the rethermalization time.

\section{Actions, observables, restriction, and prolongation}
\label{sec:notation}

Here, we provide explicit details pertaining to the implementation of our multiscale thermalization algorithm.
We begin by considering a $D$-dimensional hypercubic lattice, with lattice spacing $a$ and periodic boundary conditions.
Let us label the sites of the lattice with $D$-vectors $\bfn$, with components $n_\mu = \bfn\cdot\bfe_\mu$, where $\bfe_\mu$ is a unit basis vector in the $\mu$ direction and $\mu = 0,\cdots, D-1$.
Note that a $D$-dimensional hypercubic lattice comprises $q$-dimensional unit ``$q$-cells,'' where $q = 0,\cdots,D$.
For example, with $D=4$, the lattice comprises sites ($q=0$), bonds ($q=1$), and plaquettes ($q=2$), and so on.
The total number of such cells is given by
\begin{eqnarray}
N_q = N_0 \binom{D}{q}\ ,
\end{eqnarray}
where $N_0$ is the total number of lattice sites.

The Wilson action~\cite{PhysRevD.10.2445} for pure lattice Yang-Mills gauge theory is given by
\begin{eqnarray}
S = \beta \sum_{\bfn}\sum_{\mu<\nu} \left[ 1 - W^{1\times 1}_{\mu\nu}(\bfn) \right]\ ,
\end{eqnarray}
where $U_{\mu}(\bfn) \in SU(N_c)$ are variables associated with the bonds of the lattice, $W_{\mu\nu}^{1\times 1}(\bfn)$ are $1\times 1$ Wilson loops associated with the plaquettes of the lattice, and $\beta = 2N_c/g^2$ is the coupling.
Note that if Wilson lines are given by
\begin{eqnarray}
L^m_\mu(\bfn) = \calP \prod_{n=0}^{m-1} U_\mu(\bfn+n\bfe_\mu)\ ,
\end{eqnarray}
where $\calP$ is the path-ordering symbol, then a rectangular $m\times n$ Wilson loop in the $\mu$-$\nu$ plane is given by
\begin{eqnarray}
W^{m\times n}_{\mu\nu}(\bfn) = \frac{1}{N_c} \Re\, \Tr\, L^m_\mu(\bfn) L^n_\nu(\bfn+m\bfe_\mu) L^m_\mu(\bfn+n\bfe_\nu)^\dagger L_\nu^n(\bfn)^\dagger\ ,
\end{eqnarray}
and the corresponding space-time averaged Wilson loop is given by
\begin{eqnarray}
\bar W^{m\times n} = \frac{1}{N_2} \sum_{\bfn}\sum_{\mu<\nu} W^{m\times n}_{\mu\nu}(\bfn)\ .
\end{eqnarray}

\begin{figure} 
\includegraphics[width=\figWidthQuarters]{\figdir 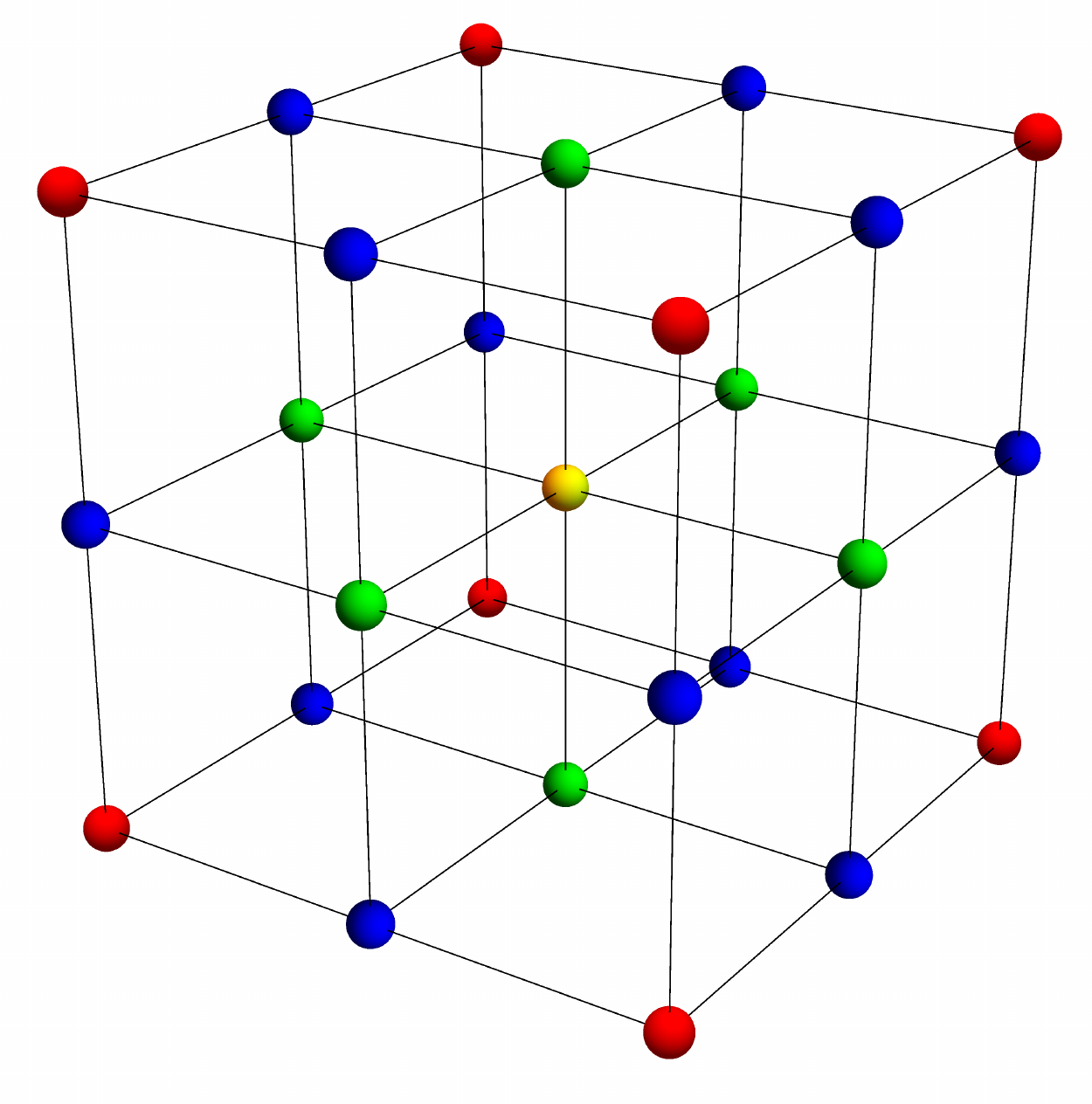}
\hspace{25pt}
\includegraphics[width=\figWidthQuarters]{\figdir 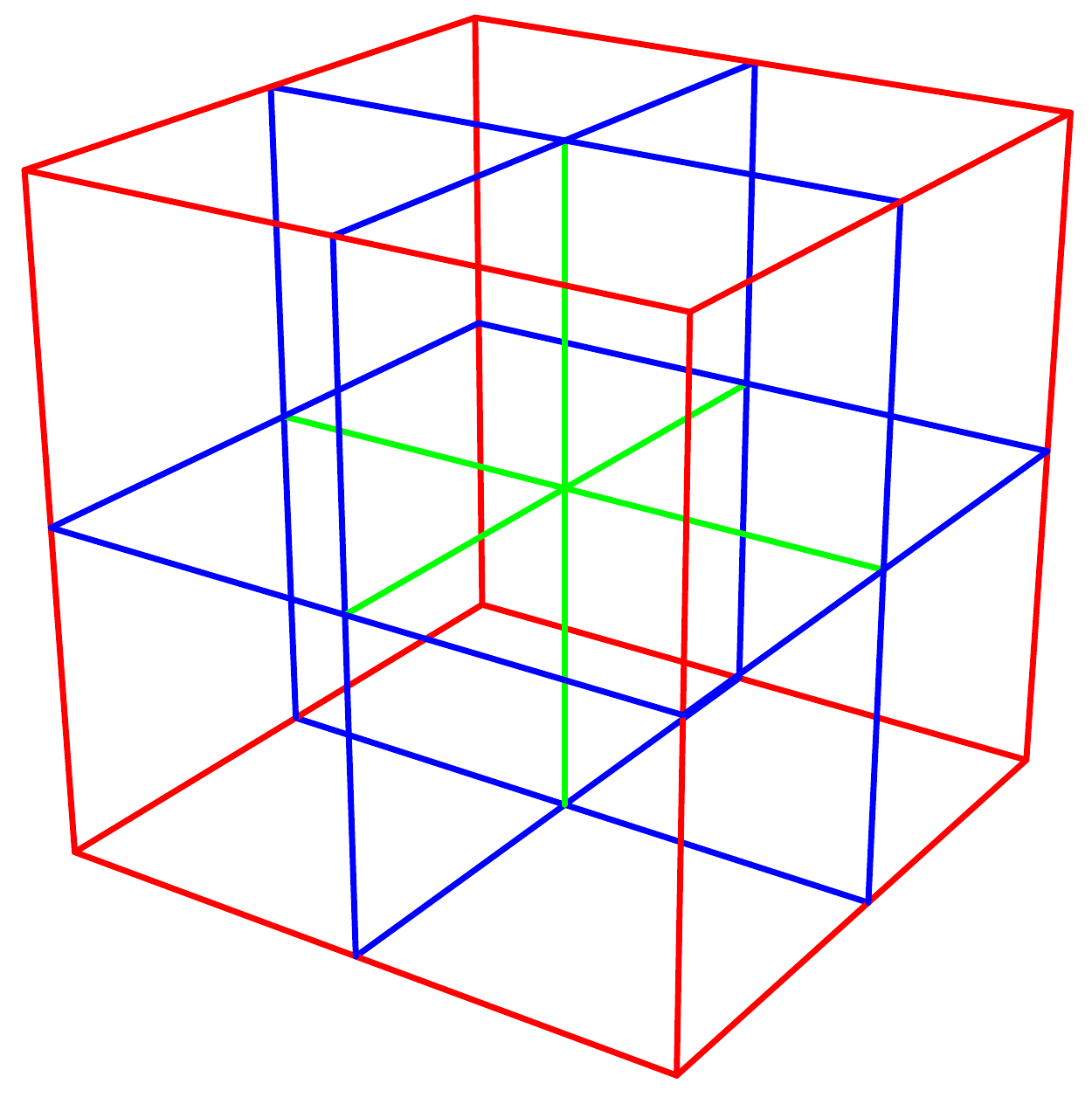}
\hspace{25pt}
\includegraphics[width=\figWidthQuarters]{\figdir 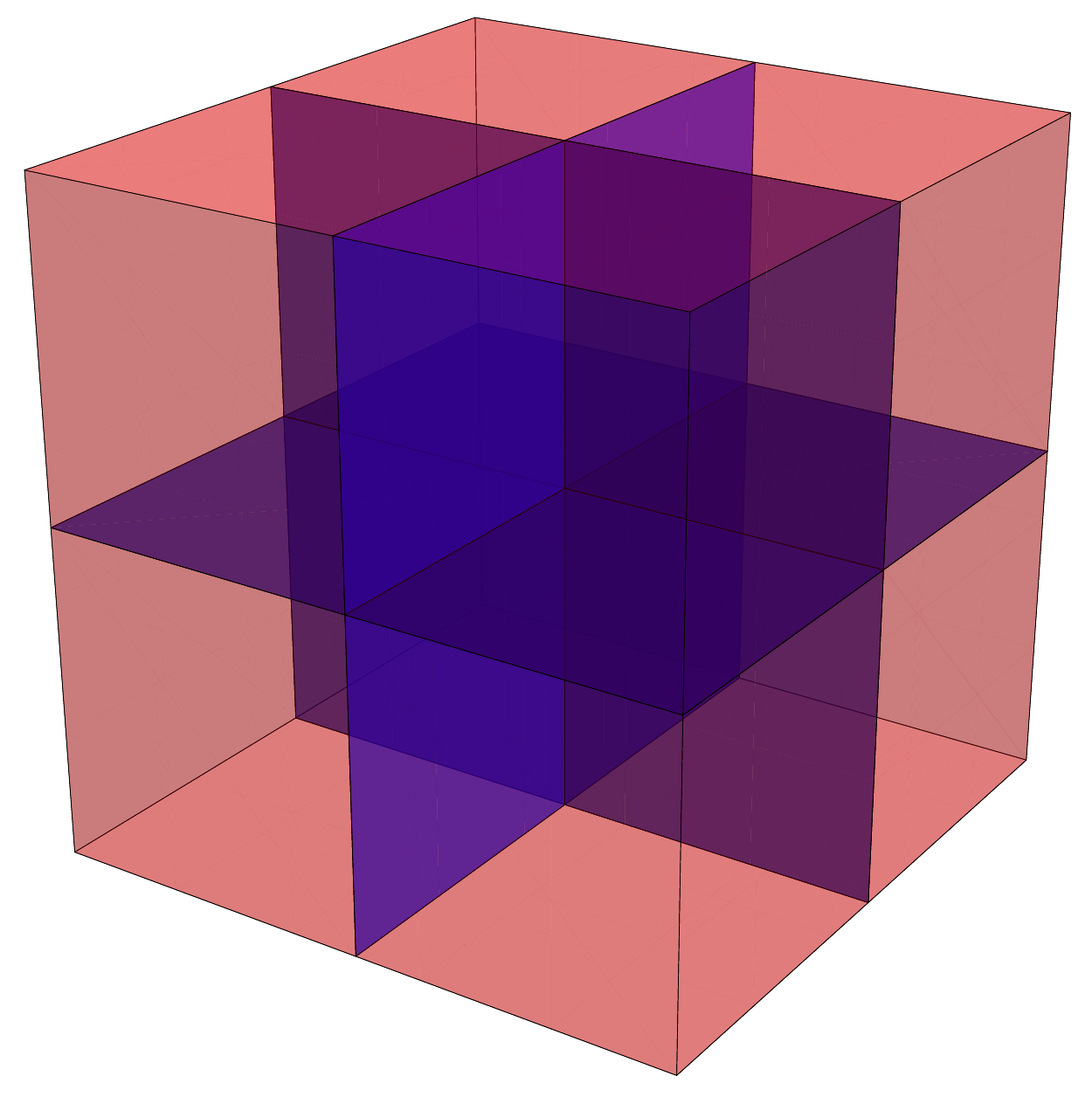}
\caption{\label{fig:classification}%
Classification of lattice cells, labeled by the integers $0$ (red), $1$ (blue), $2$ (green), and $3$ (yellow).
Left: $\chi(\bfn)$.
Center: $\chi_\mu(\bfn)$.
Right: $\chi_{\mu\nu}(\bfn)$.
}
\end{figure}

For the purpose of this study, we consider restriction and prolongation operations that take an ensemble associated with a ``fine'' lattice with spacing $a$ to an ensemble associated with a ``coarse'' lattice with spacing $2a$, and back.
To facilitate the discussion, we begin by classifying the various $q$-cells of the fine lattice according to their positions with respect to the $2^D$ hypercubes which define the coarse lattice.
We define the function 
\begin{eqnarray}
\chi(\bfn) = \sum_\mu ( \textrm{$n_\mu$ mod $2$} ) \ ,
\end{eqnarray}
which allows us to associate integers $0,\cdots, D$ to the sites of the $2^D$ hypercubes, as shown in \Fig{classification} (a) for $D=3$ space-time dimensions.
The subset of sites associated with the coarse lattice satisfy $\chi(\bfn)=0$ and are consequently given by $\bfn/2\in {\mathbb Z}^D$.
Note that this convention is but one of $2^D$ possibilities for the alignment of the coarse lattice with respect to the fine; since the fine lattice theory is invariant under lattice translations, any choice is acceptable without loss of generality.
Similarly, we may define the quantities
\begin{eqnarray}
\chi_\mu(\bfn) &=& \chi(\bfn-n_\mu \bfe_\mu ) \cr
\chi_{\mu\nu}(\bfn) &=& \chi(\bfn-n_\mu \bfe_\mu-n_\nu\bfe_\nu) \cr
\chi_{\mu\nu\sigma}(\bfn) &=& \chi(\bfn-n_\mu\bfe_\mu-n_\nu\bfe_\nu-n_\sigma\bfe_\sigma) \cr
\chi_{\mu\nu\sigma\rho}(\bfn) &=& \chi(\bfn-n_\mu\bfe_\mu-n_\nu\bfe_\nu-n_\sigma\bfe_\sigma-n_\rho\bfe_\rho)\ , 
\end{eqnarray}
which associate the integers $0,\cdots,D-q$ with the remaining $q$-cells of the lattice, where $q=0,\cdots,D$.
The classification of bonds is shown in \Fig{classification} (b) and the classification of plaquettes is shown in \Fig{classification} (c) for $D=3$ space-time dimensions.

For this study, we consider the simplest restriction (or blocking) procedure, which proceeds by assigning products of fine bond variables $U^f_\mu$ to the coarse bond variables $U^c_\mu$,
\begin{eqnarray}
U^c_\mu(\bfn/2) = U^f_\mu(\bfn) U^f_\mu(\bfn+\bfe_\mu) \ ,
\label{eq:restriction}
\end{eqnarray}
for all values of $\bfn$ satisfying $\chi(\bfn)=0$.
Other schemes are also possible; however, the specific choice plays a minor role in the present implementation of our algorithm.
In all cases, information is lost in the restriction operation.

Prolongation proceeds in two steps: first, the transfer of the coarse lattice variables to an appropriate subset of bonds on the fine lattice, and second, interpolation of the transferred variables to the remaining undefined bonds of the fine lattice.
The prolongator is designed to preserve the long-distance structure of the theory (as it is encoded by the configurations), including correlation lengths and topological charge.
As a consequence, if the coarse configurations to which prolongation is applied are thermalized, the resulting fine configurations will also be thermalized except for short-distance defects at the scale of the cutoff.
Fine evolution can correct for such short-distance defects, and it is reasonable to expect that the evolution time required to bring the entire prolongated ensemble into thermal equilibrium will be short.
Since topology freezing is one of the major issues in present-day simulations that we aim to address, it is advantageous for the restriction and prolongation procedures to preserve the topological charge either with configuration by configuration within an ensemble or in terms of its distribution over the ensemble.\footnote{Note that the topological charge depends on the ultraviolet regulator and a particular definition will be discussed below. It is only in the limit of weak coupling where configurations satisfy an admissibility criterion~\cite{Luscher:1981zq} that the definition becomes unique.}

\begin{figure} 
\includegraphics[width=\figWidthFifths]{\figdir 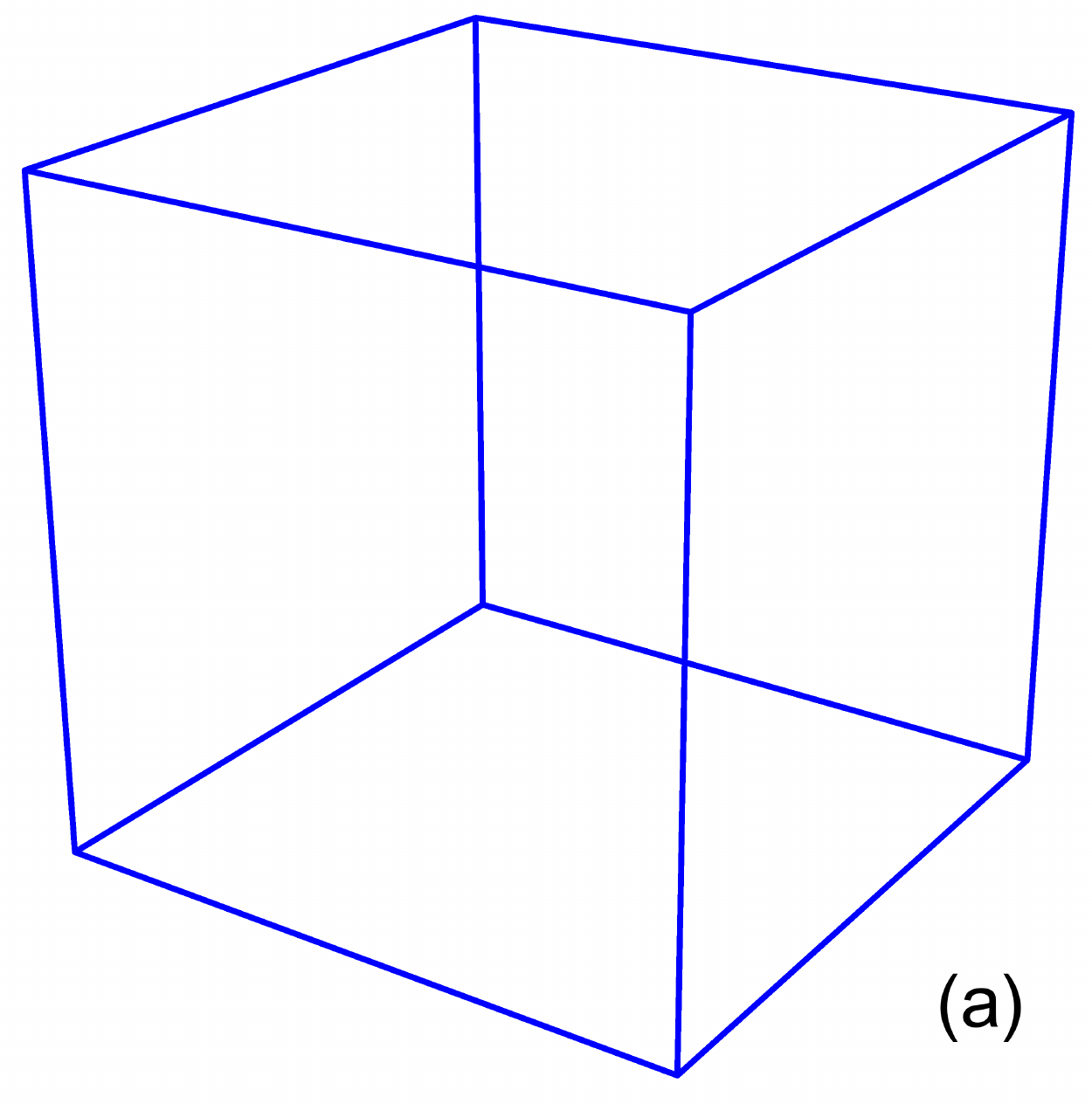}
\hspace{25pt}
\includegraphics[width=\figWidthFifths]{\figdir 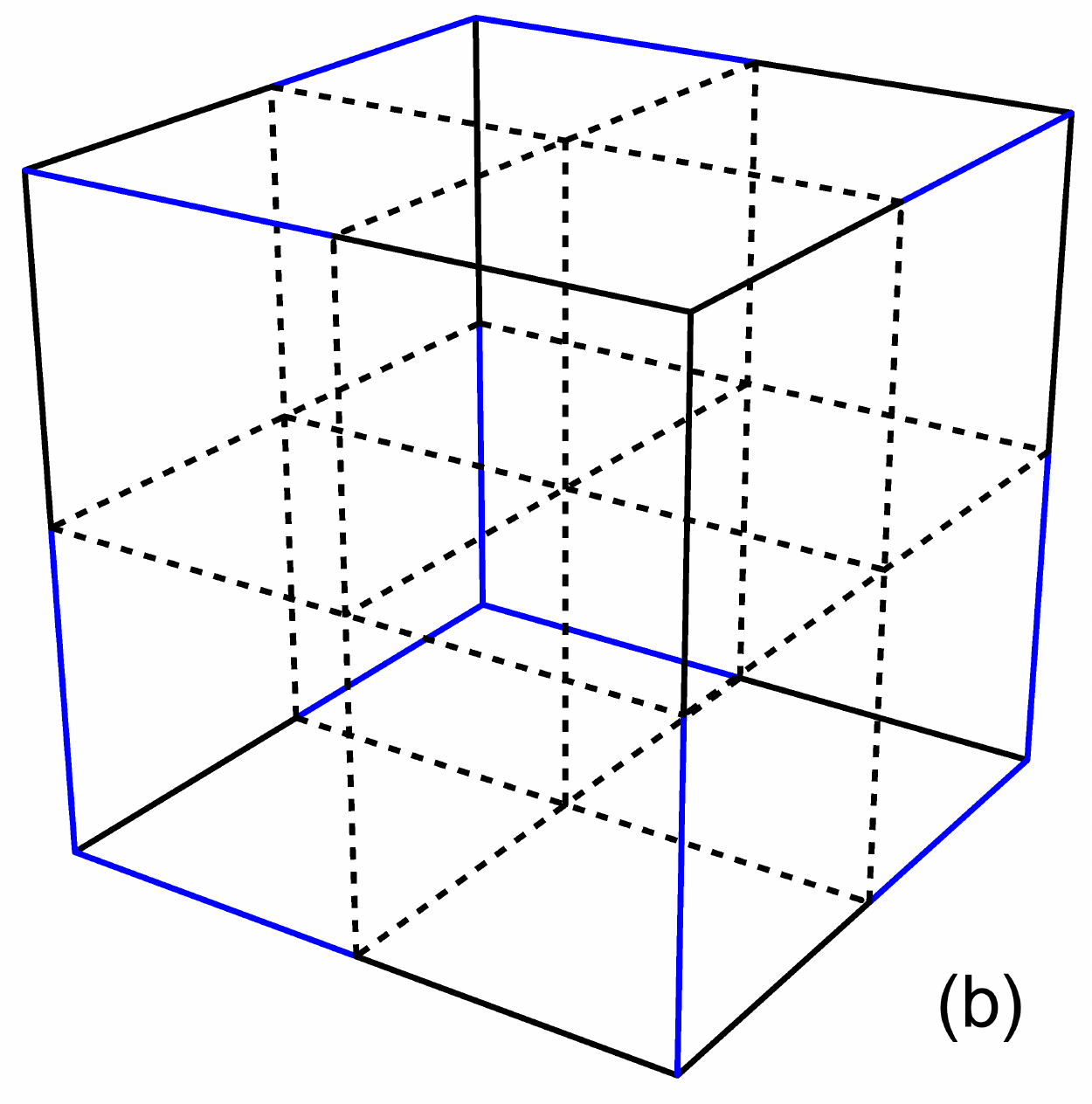}
\hspace{25pt}
\includegraphics[width=\figWidthFifths]{\figdir 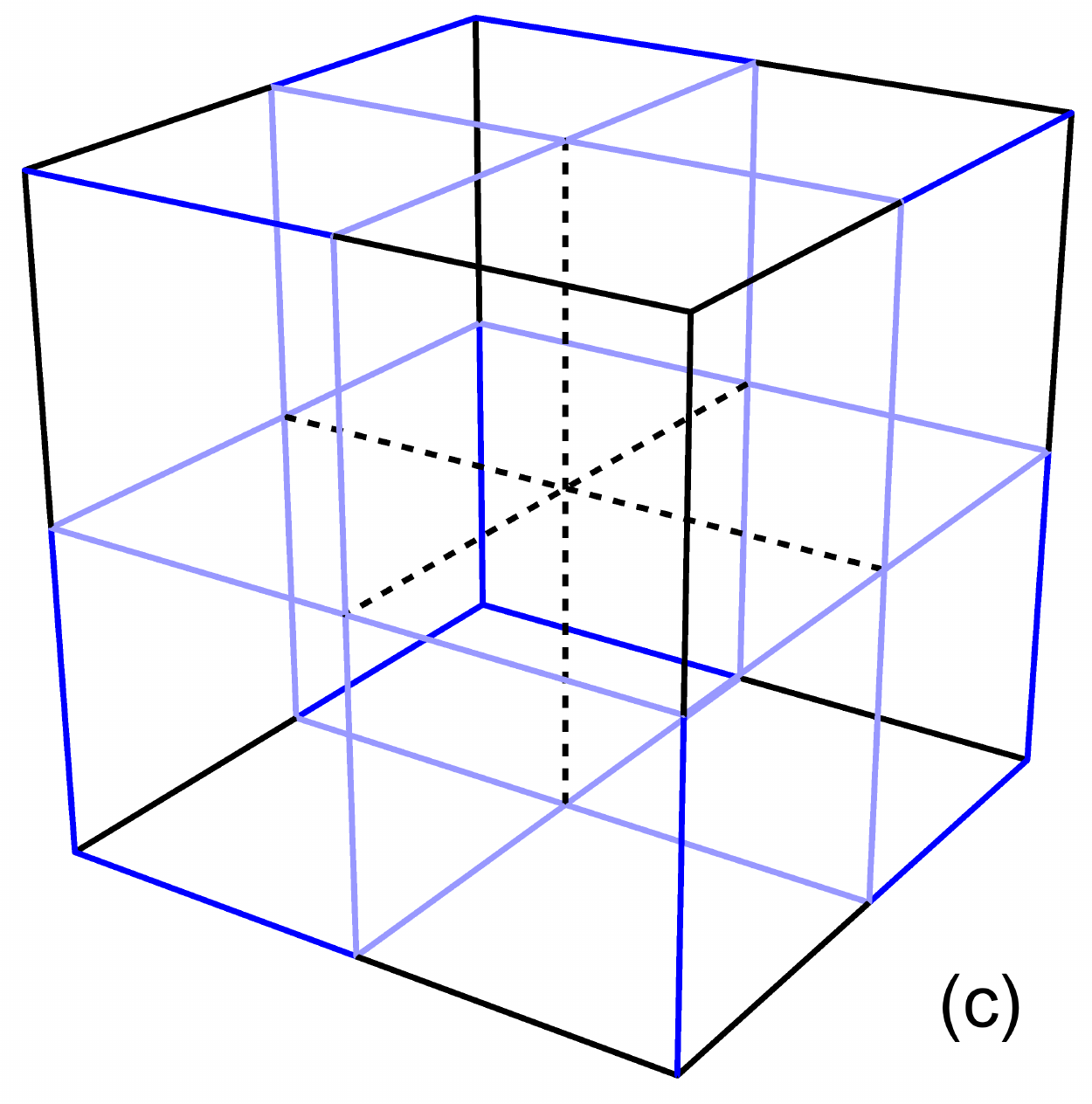}
\hspace{25pt}
\includegraphics[width=\figWidthFifths]{\figdir 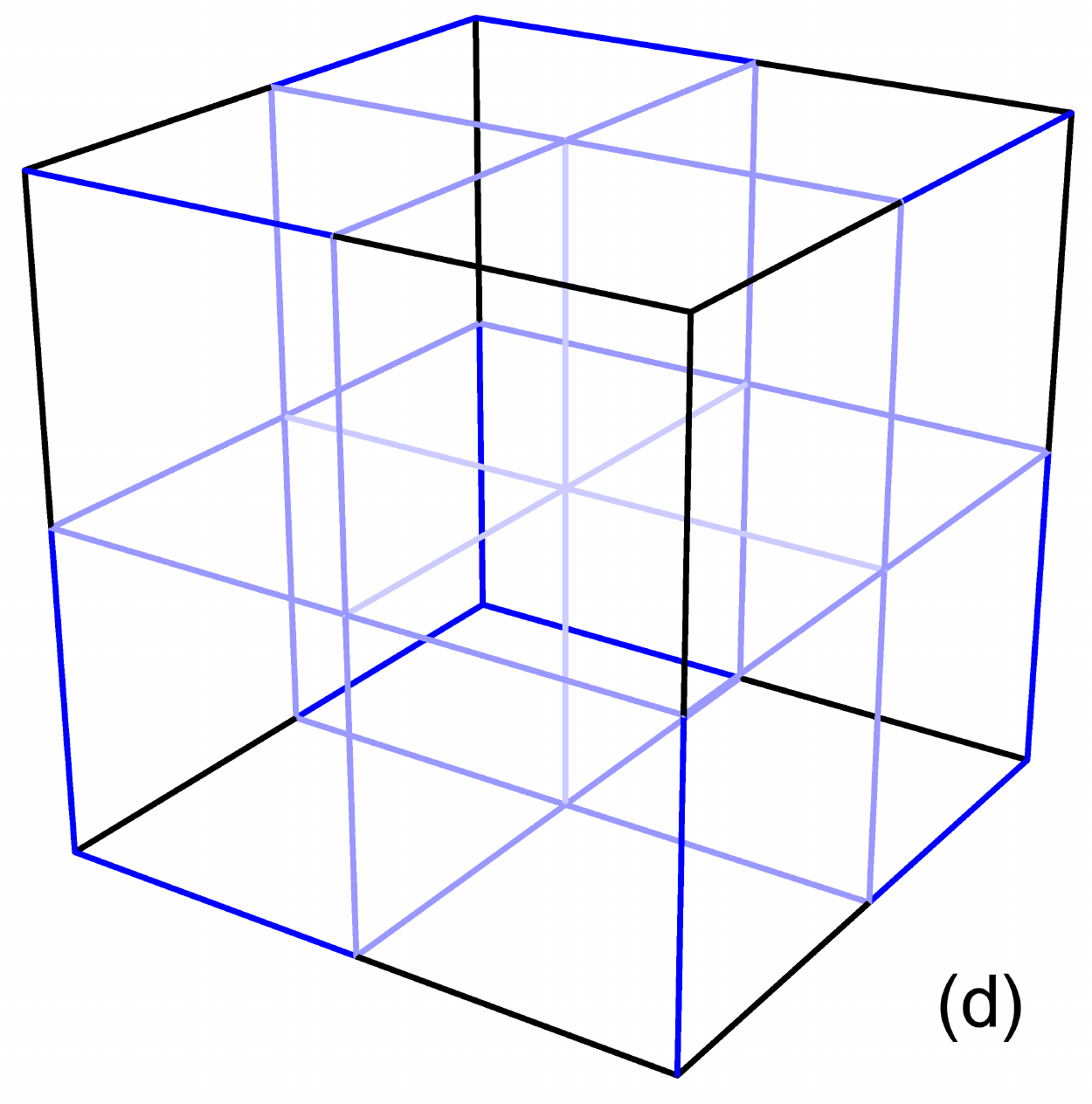}
\caption{\label{fig:refinement}%
Schematic description of prolongation: (a) coarse bond variables; (b) transfer of coarse bond variables to a fine lattice; (c) interpolation performed on $2\times2$ plaquettes; (d) interpolation performed on $2\times2\times2$ cubes; the procedure is continued for the remaining stages (not shown).
Black bond variables are set to unity, as allowed by gauge freedom, and dashed bond variables are undefined at intermediate stages of the refinement; dark blue bond variables are transferred from the coarse lattice, whereas the lighter blue bond variables are determined via the interpolation.
}
\end{figure}

The first step of the prolongation procedure is to associate the coarse bond variables with the fine lattice.
Since there are more bonds on the fine lattice than the coarse, there is no unique prescription for doing this.
However, a simple choice is to demand that
\begin{eqnarray}
U^f_\mu(\bfn) = U^c_\mu(\bfn/2) \ ,\qquad U^f_\mu(\bfn+\bfe_\mu) = 1\ ,
\label{eq:prolongation}
\end{eqnarray}
for all $\bfn$ satisfying $\chi(\bfn)=0$.
Note that the gauge freedom of the fine action allows us to set one of the two fine bond variables above to unity; in \Fig{refinement} (b) we show an example of this assignment for a $2\times2\times2$ cell, having transferred bond variables from the coarse unit cell shown in  \Fig{refinement} (a).
The remaining bond variables are undefined and may be set to arbitrary values; in this study we initially set them to unity.
Note that $2\times2$ Wilson loops originating from the even sites of the fine lattice [i.e., where $\chi(\bfn)=0$] are exactly equal to the plaquettes of the coarse lattice.
Furthermore, all even length Wilson loops originating from the even sites are exactly preserved by the map.
This is the key to our construction, and it implies that there is a set of long distance loops from which the renormalization group invariant area law can be computed (i.e., using Creutz ratios constructed from even-sided loops).

The second step, interpolation, is designed to remove the most damaging ultraviolet defects induced by the first step.
There are a number of ways to carry out the interpolation of gauge fields (see, e.g.,~\cite{Luscher:1981zq,Phillips:1986qd,'tHooft1995491}).
Following the approach of 't Hooft~\cite{'tHooft1995491}, we use an interpolation which respects the L{\"u}scher bound for sufficiently smooth configurations~\cite{Luscher:1981zq}, and as a consequence, exactly preserves the topological charge for those configurations.
The gauge field interpolation is carried out by sequentially minimizing the partial actions,
\begin{eqnarray}
S_d = \beta \sum_{\bfn}\sum_{\mu<\nu} \delta_{d,\chi_{\mu\nu}(\bfn)}  \left[ 1 - W^{1\times 1}_{\mu\nu}(\bfn) \right]
\label{eq:smoothing}
\end{eqnarray}
with respect to ``active'' bond variables which satisfy $\chi_\mu(\bfn)=d+1$, for $d=0,\cdots,D-2$.
The interpolation proceeds starting from low dimensional to high dimensional cells. 
A useful property of this prescription is that at each stage of the interpolation, the active bond variables in one $2^{d+2}$ cell are completely decoupled from those in neighboring $2^{d+2}$ cells.
Thus the interpolation can be performed locally at each stage.

At stage $d=0$, the minimization can be performed analytically, following~\cite{'tHooft1995491}; however, the analytic forms become complicated for $d>0$.
In this study, we followed a numerically simpler procedure for performing the minimization that is valid for all stages.
Specifically, repeated applications of APE smearing~\cite{Falcioni1985624,Albanese1987163} of the form
\begin{eqnarray}
U_\mu(\bfn) \to U_\mu^\prime(\bfn) = {\mathbb P}_{SU(N_c)} \left[  U_\mu(\bfn) + c \sum_{\sigma=\pm}\sum_\nu \delta_{d,\chi_{\mu\nu}(\bfn)} T^\sigma_{\mu\nu}(\bfn) \right]
\label{eq:APE}
\end{eqnarray}
were performed on the active bonds at a given stage, where
\begin{eqnarray}
T^+_{\mu\nu}(\bfn) = U_\nu(\bfn) U_\mu(\bfn+\bfe_\nu) U_\nu(\bfn+\bfe_\mu)^\dagger
\end{eqnarray}
and
\begin{eqnarray}
T^-_{\mu\nu}(\bfn) = U_\nu(\bfn-\bfe_\nu)^\dagger U_\mu(\bfn-\bfe_\nu) U_\nu(\bfn-\bfe_\nu+\bfe_\mu)^\dagger
\end{eqnarray}
are forward and backward oriented staple operators, ${\mathbb P}_{SU(N_c)}$ is a projection operator onto $SU(N_c)$, and $c$ is a small parameter to be specified later.
The number of times this smearing is applied to the gauge fields will also be specified later.

Before moving on to numerical studies, we define several additional quantities, which will prove useful later on:
partially space-time averaged plaquettes, associated with the different plaquettes subsets,
\begin{eqnarray}
\bar W^{1\times 1}_d = \frac{1}{N^{1\times1}_d}  \sum_{\bfn}\sum_{\mu<\nu}  \delta_{d,\chi_{\mu\nu}(\bfn)} W^{1\times 1}_{\mu\nu}(\bfn) \ ,
\label{eq:partial_1x1}
\end{eqnarray}
and average displaced $2\times 2$ Wilson loops, given by
\begin{eqnarray}
\bar W^{2\times 2}_d = \frac{1}{N^{2\times2}_d }  \sum_{\bfn}\sum_{\mu<\nu}  \delta_{d,\chi(\bfn)} W^{2\times 2}_{\mu\nu}(\bfn) \ .
\label{eq:displaced_2x2}
\end{eqnarray}
The normalization for these quantities are given by
\begin{eqnarray}
N^{1\times1}_d =  2 (D-d)(D-d-1) \binom{D}{d} \frac{N_0}{2^D}
\end{eqnarray}
and
\begin{eqnarray}
N^{2\times2}_d  = \binom{D}{2}  \binom{D}{d}  \frac{N_0}{2^D}\ ,
\end{eqnarray}
respectively.
Note that
\begin{eqnarray}
\sum_{d=0}^D N^{1\times1}_d = \sum_{d=0}^D N^{2\times2}_d = \binom{D}{2} N_s\ ,
\end{eqnarray}
which is just the total number of plaquettes on the lattice.

\section{Simulations}
\label{sec:simulations}

\subsection{Target ensembles}
\label{sec:ensembles}

In the remainder of this work we consider pure $SU(3)$ gauge theory in $D=4$ dimensions and make use of four decorrelated target ensembles of size $N$, generated in a standard way.
Physical observables on these ensembles will serve as benchmarks that the multiscale thermalization algorithm should reproduce.
The ensembles are described in \Tab{target_ensembles} and have lattice spacings ranging from approximately $0.07$ to $0.2$ fm, separated by multiples of $\sqrt{2}$.
The lattice spacings were determined from empirical formulas relating the Sommer scale in lattice units ($r_0/a$) to the coupling~\cite{Edwards1998377,Guagnelli1998389}, taking $r_0 = 0.5$ fm.
The spatial extents of the lattices were chosen to be approximately $2.25$ to $2.40$ fm; the temporal extents were chosen to be twice the spatial extents in order to minimize thermal effects.
Standard boundary conditions, periodic in all directions, were used throughout.
Ensembles were generated with the Cabibbo-Marinari HB algorithm~\cite{Cabibbo:1982zn} combined with overrelaxation~\cite{Brown:1987rra}.
Each HB sweep was performed on a checkerboard sweep schedule with $N_{hb}=1$ attempted updates to each $SU(2)$ subgroup per bond variable via the method of Creutz~\cite{Creutz:1979dw}.
Each HB sweep was followed by $N_{ov}=10$ overrelaxation sweeps following the same checkerboard sweep schedule.
For all ensembles, 1500 HB sweeps were initially performed for thermalization starting from a weak field configuration; subsequent configurations were saved after every $100$ sweeps for future use.

\begin{table}
\caption{%
\label{tab:target_ensembles}%
Decorrelated target ensembles of size $N$, generated using HB with 10 overrelaxation sweeps.
Lattice spacing is set via the Sommer scale $r_0 = 0.5$ fm, based on the works~\cite{Edwards1998377} (coarsest) and~\cite{Guagnelli1998389} (finer).
The reference scale $t_0$ is defined in \Sec{wilson_flow}.
}
\begin{ruledtabular}
\begin{tabular}{ccccc}
Lattice & $\beta$ & $a$ [fm] & $N$ & $t_0/a^2$ \\
\hline
$12^3\times 24$   & 5.626           & 0.1995(20) & 385  & 0.72966(69)  \\
$16^3\times 36$   & 5.78\phantom{0} & 0.1423(5)  & 385  & 1.3858(15)  \\
$24^3\times 48$   & 5.96\phantom{0} & 0.0999(4)  & 185  & 2.7891(45)  \\
$32^3\times 72$   & 6.17\phantom{0} & 0.0710(3)  & 185  & 5.5007(83)  \\
\end{tabular}
\end{ruledtabular}
\end{table}

\subsection{Wilson flow}
\label{sec:wilson_flow}

Wilson flow~\cite{Narayanan:2006rf,Luscher:2010iy,Lohmayer:2011si} was used to define a number of the observables studied in this work.
The diffusive nature of the flow allows us to consider a series of observables, which probe different length scales at different flow times, $t$.
Wilson flow was applied to the target ensembles described in \Tab{target_ensembles} using both a fixed step size algorithm~\cite{Luscher:2010iy} and an adaptive step size algorithm~\cite{Fritzsch:2013je}.
The accuracy of the integration along the flow is controlled by the size of the step in the former case and a tolerance level in the latter case (see~\cite{Fritzsch:2013je} for an explicit definition of this tolerance).
The adaptive approach is more efficient because the flow has a smoothing effect on the fields.
Consequently, the forces that drive the flow become smaller with flow time, thus enabling the use of larger step sizes at later times.
We have established the validity of our implementation of the adaptive step Wilson flow by direct comparison with fixed step size Wilson flow for the target ensembles in \Tab{target_ensembles}.
For the autocorrelation time and (re)thermalization studies performed later in this work, the adaptive step size algorithm was used, due to its higher efficiency.

\begin{figure} 
\includegraphics[width=\figWidthHalf]{\figdir 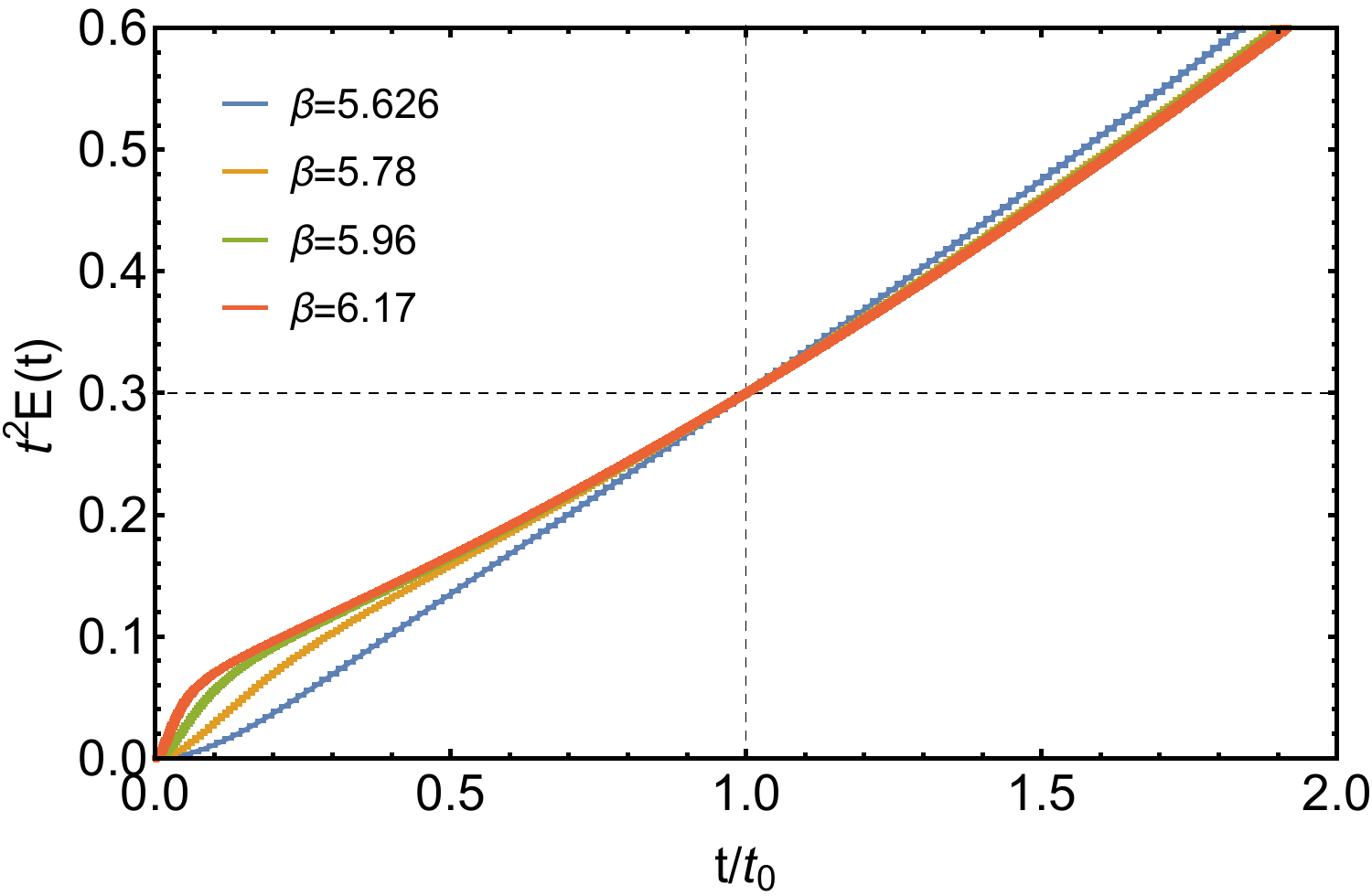}
\hspace{10pt}
\includegraphics[width=\figWidthHalf]{\figdir 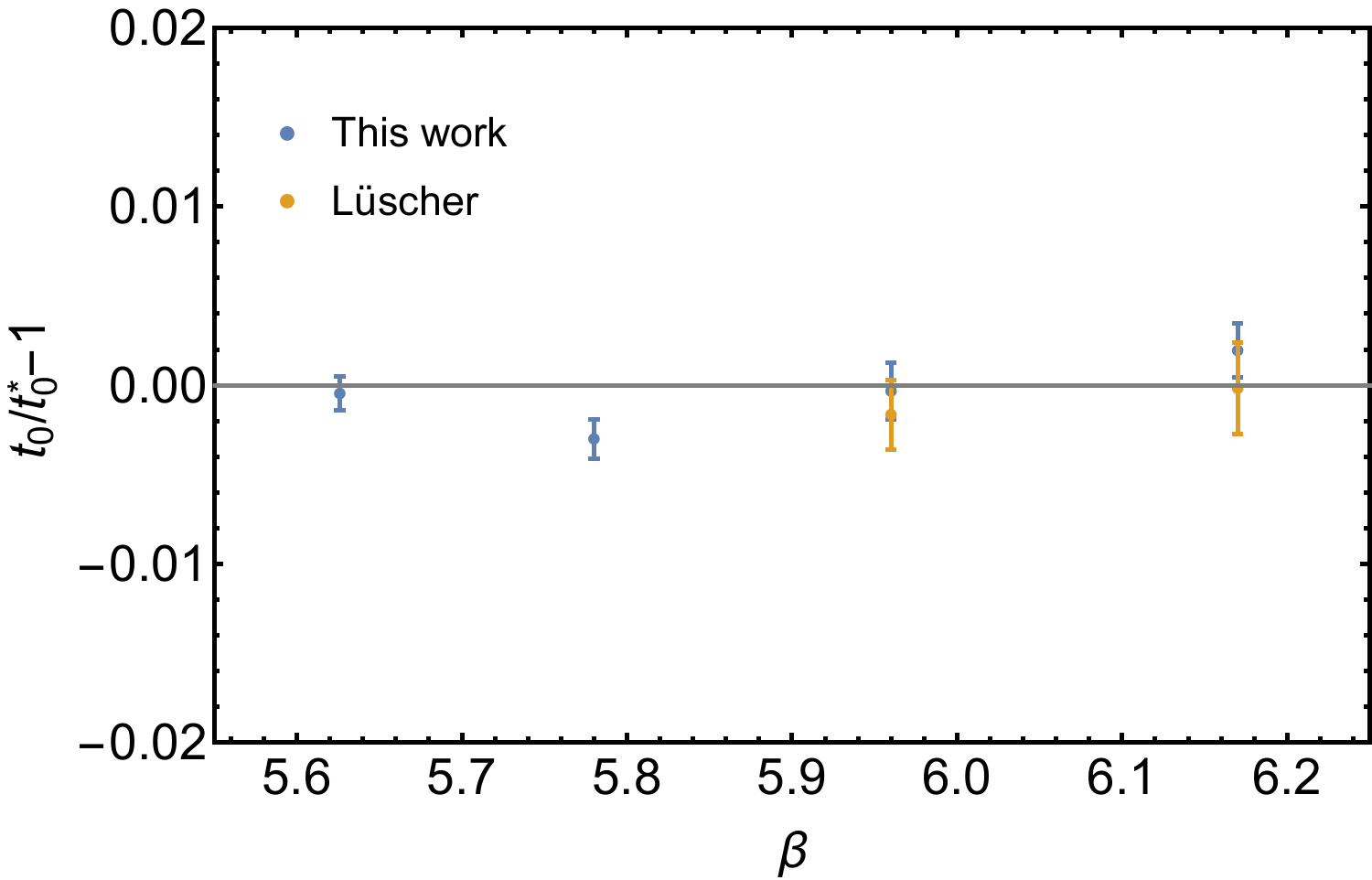}
\caption{\label{fig:wflow}%
Left: $t^2 E(t)$ as a function of $t/t_0$; determined values of $t_0$ are provided in \Tab{target_ensembles}.
Right: Deviation between $t_0$ obtained using Wilson flow with a fixed step size (this work) and the nominal values $t^\star_0$ introduced in \Tab{wilson_flow}.
Corresponding results obtained in~\cite{Luscher:2010iy} (L{\"u}scher) are shown for comparison.
}
\end{figure}

For the target ensembles, Wilson flow measurements were performed using a fixed step size of 0.01; results for the quantity $t^2 E(t)$ (for this study, we use the clover-leaf definition) are provided in \Fig{wflow} as a function of $t/t_0$.
The Wilson flow scale, $t_0$, is defined by $t_0^2 E(t_0) = 0.3$; values of this scale and corresponding statistical errors were obtained by linearly interpolating the nearest estimates of $t^2 E(t)$ to $0.3$.
The results from this analysis are provided in \Tab{target_ensembles}; for the $24^3$ and $32^3$ ensembles, we obtained estimates of $t_0$ which are consistent with~\cite{Luscher:2010iy}.
For the same ensembles, the adaptive step size algorithm was used with a tolerance of 0.01.
Measurements along the flow were made in multiples of $t_0^\star/4$, where $t_0^\star$ is introduced in \Tab{wilson_flow} as a nominal value for $t_0$ [this parameter was chosen to be sufficiently close to $t_0$, but also a multiple of $0.01$ so that we may directly compare estimates of $E(t)$ using both methods].
For the $32^3$ ensemble, we found a maximum deviation of about $0.02\%$ in the estimates of $E(t)$ for $t\in[0, 5t_0^\star]$.
For $12^3$, $16^3$, and $24^3$ lattices, the deviation was less than $0.001\%$ on the same interval.
The good agreement for these ensembles not only validates our implementation, but also indicates that our choice of tolerance level is adequate for the studies we are pursuing.
Note that for $t\gtrsim5t_0$, the flow radius $\sqrt{8t}$ exceeds the size of our lattices, and beyond that, the Wilson flow was used primarily for its smoothing properties in determining the topological charge.
For all ensembles, the topological charge was found to be consistent between fixed and adaptive step algorithms over the entire range of flow times on a per configuration basis.

\subsection{Restriction and prolongation}
\label{sec:prolongation}

\begin{figure} 
\includegraphics[width=\figWidthHalf]{\figdir 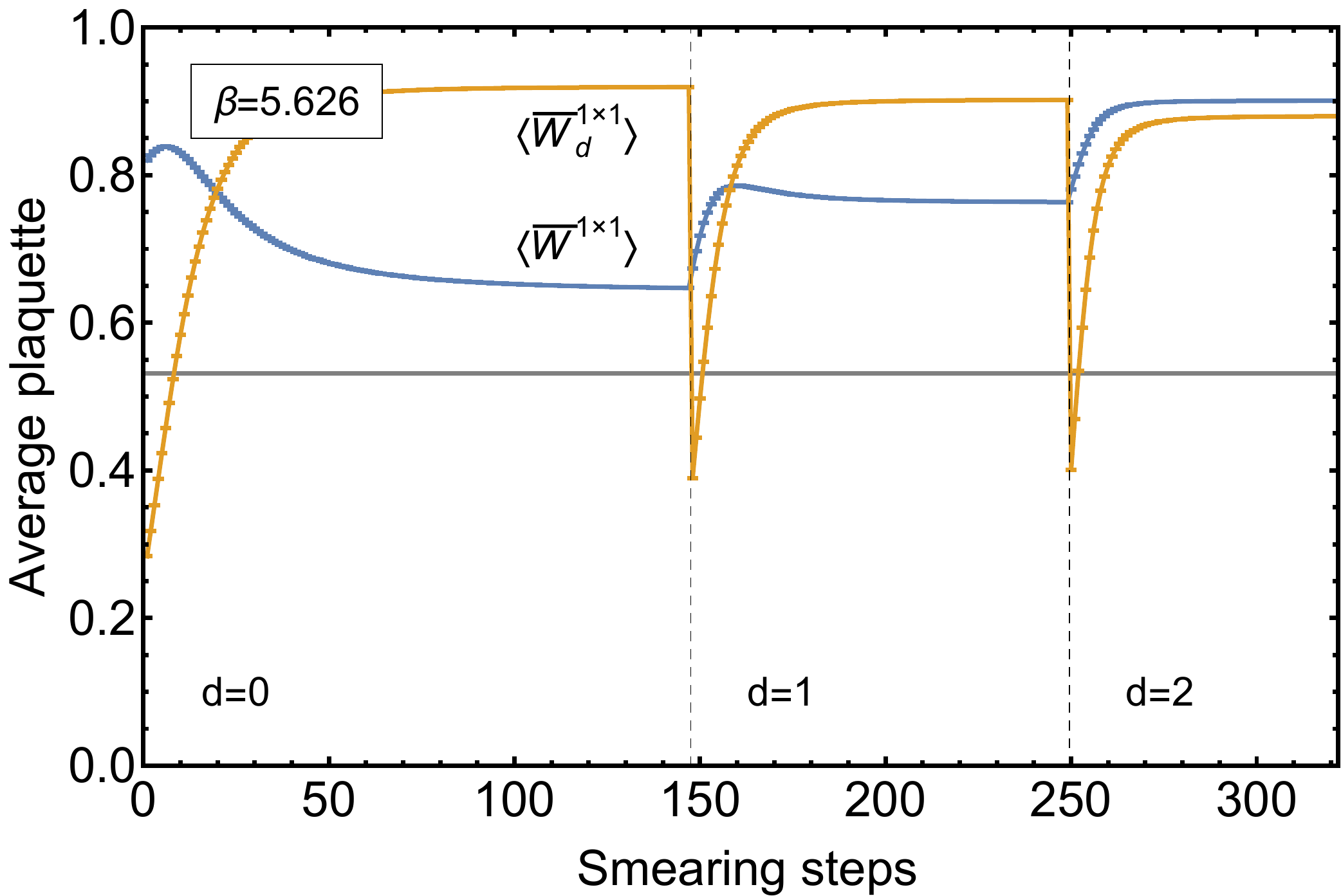}
\includegraphics[width=\figWidthHalf]{\figdir 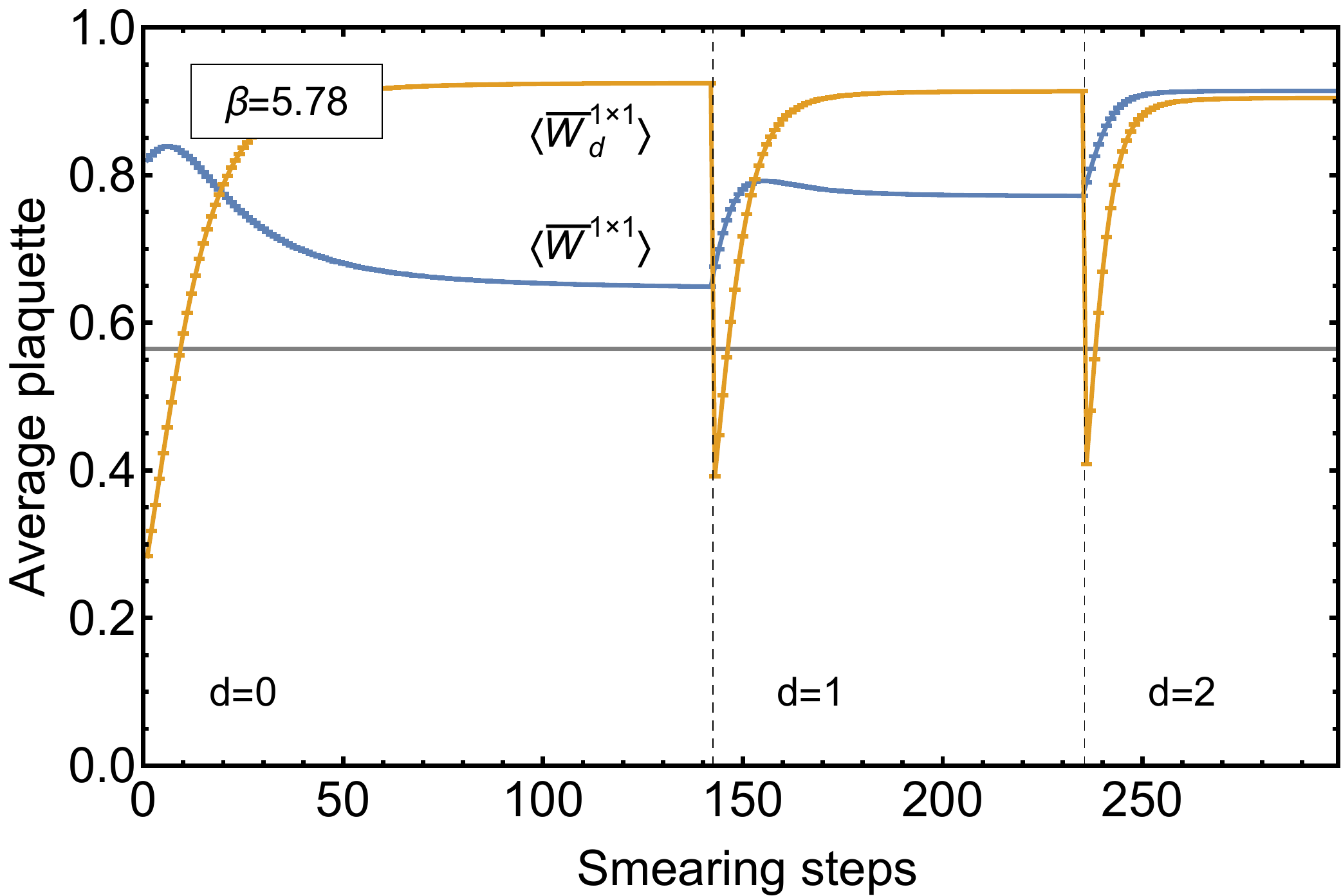} \\
\includegraphics[width=\figWidthHalf]{\figdir 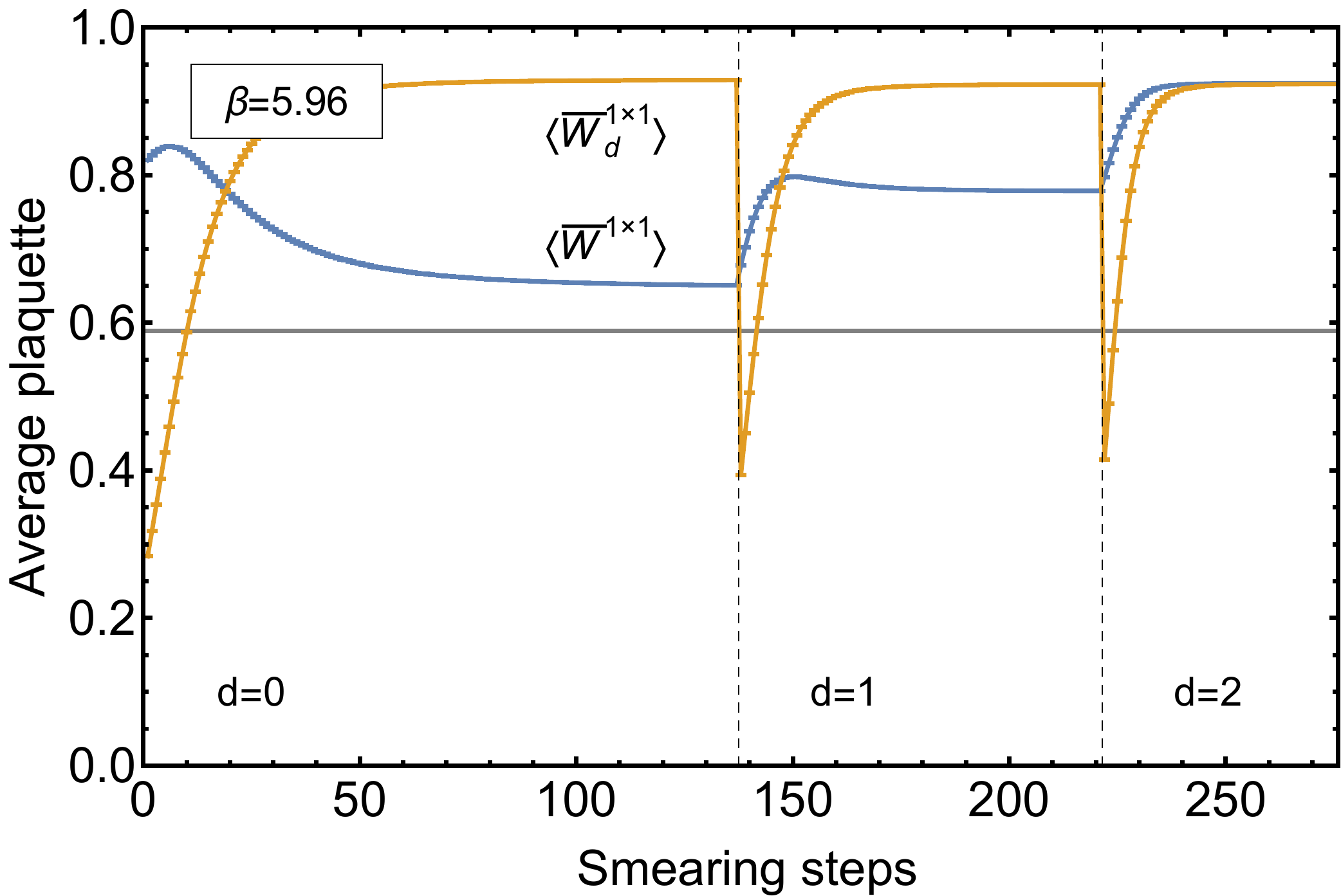}
\includegraphics[width=\figWidthHalf]{\figdir 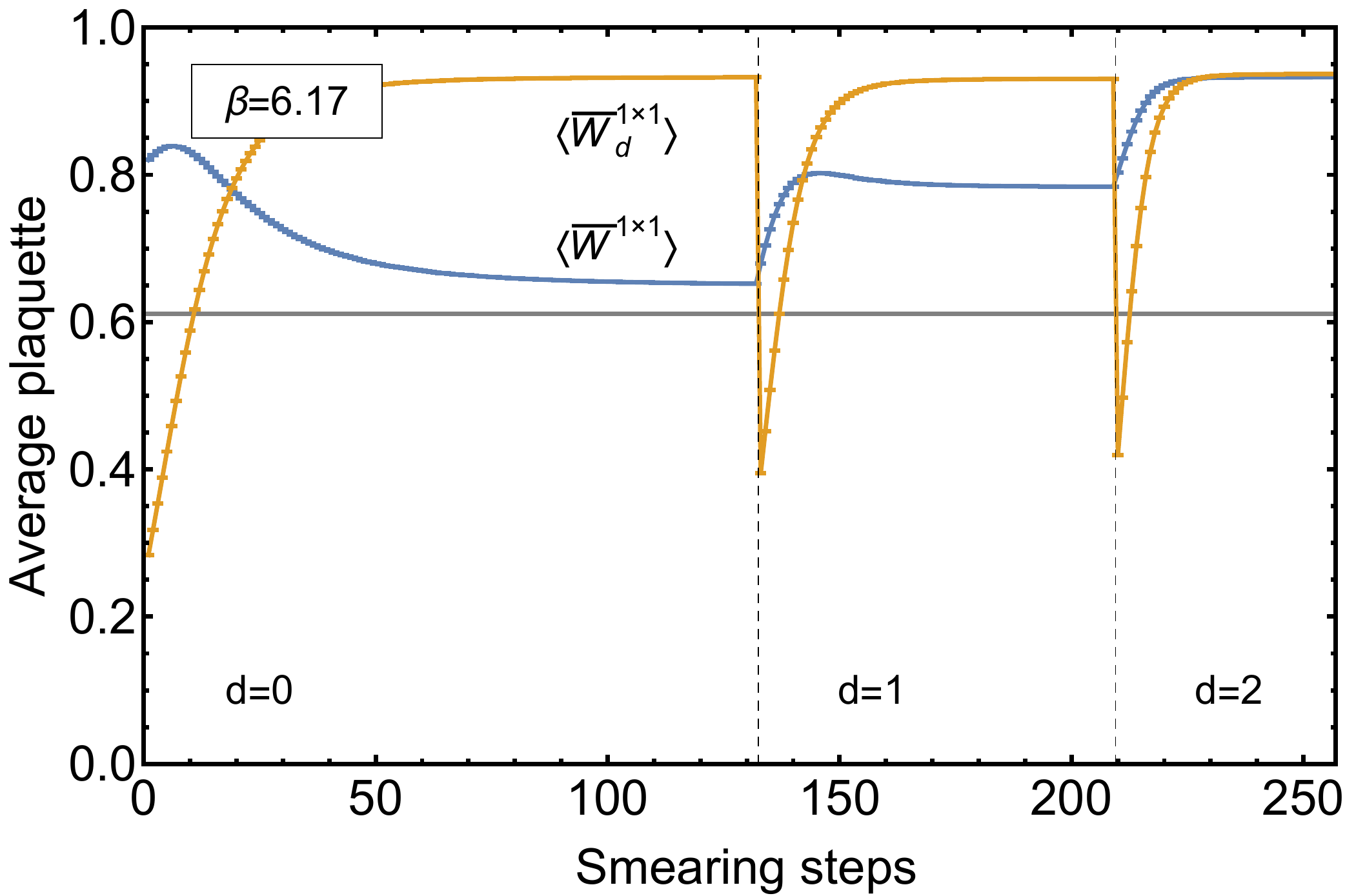}
\caption{\label{fig:interpolation}%
Average plaquette as a function of the number of cooling sweeps at each level of interpolation.
}
\end{figure}

As a first test of the prolongation algorithm described in \Sec{notation}, we investigate how the target fine ensembles described in \Tab{target_ensembles} are modified by the application of restriction using \Eq{restriction}, followed immediately by prolongation using \Eq{prolongation} and \Eq{smoothing}.
The gauge field interpolation was performed by sequentially minimizing the partial actions $S_d$ until at each stage, $d$, the relative change in the partial action reached $0.001$\%.
For the ensembles considered, the action minimization required by our interpolation procedure was performed using repeated applications of APE smearing, using \Eq{APE} with $c=0.05$.
The variation of the average plaquette, $\langle \bar W^{1\times1}\rangle$,  and partially averaged plaquette $\langle \bar W^{1\times 1}_d \rangle$ are shown in \Fig{interpolation} for each ensemble as a function of the number of smearing applications, beginning with undefined bonds set to unity.
Notice from the results that the average plaquette $\langle \bar W^{1\times1} \rangle$, which is proportional to the action, $S$, up to an overall additive constant, is not a monotonically increasing function of the number of cooling sweeps.
This is due to the fact that it is not the total action that is being minimized at each stage of the interpolation, but rather the partial action $S_d$.

\begin{figure} 
\includegraphics[width=\figWidthHalf]{\figdir 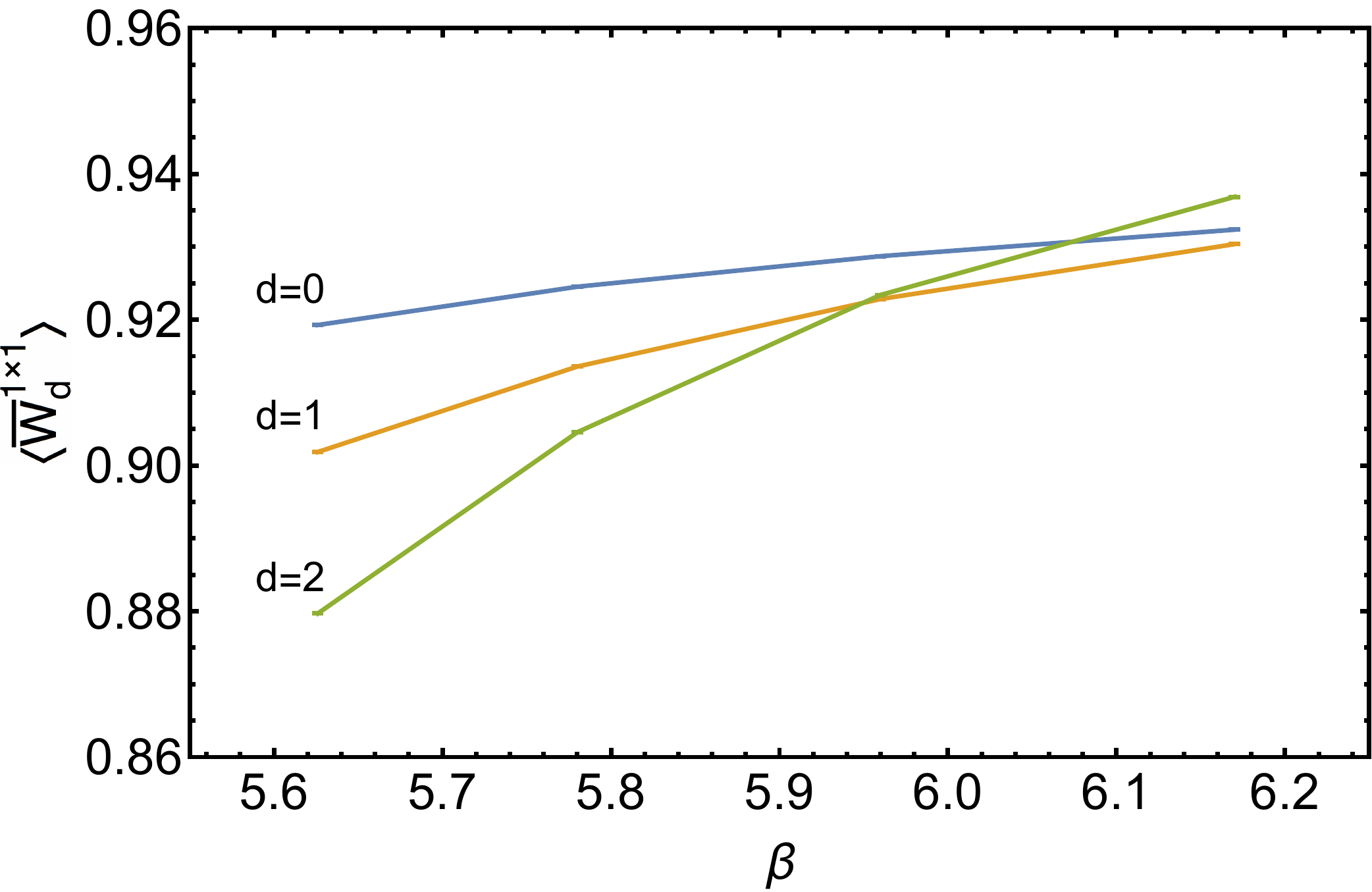}
\hspace{10pt}
\includegraphics[width=\figWidthHalf]{\figdir 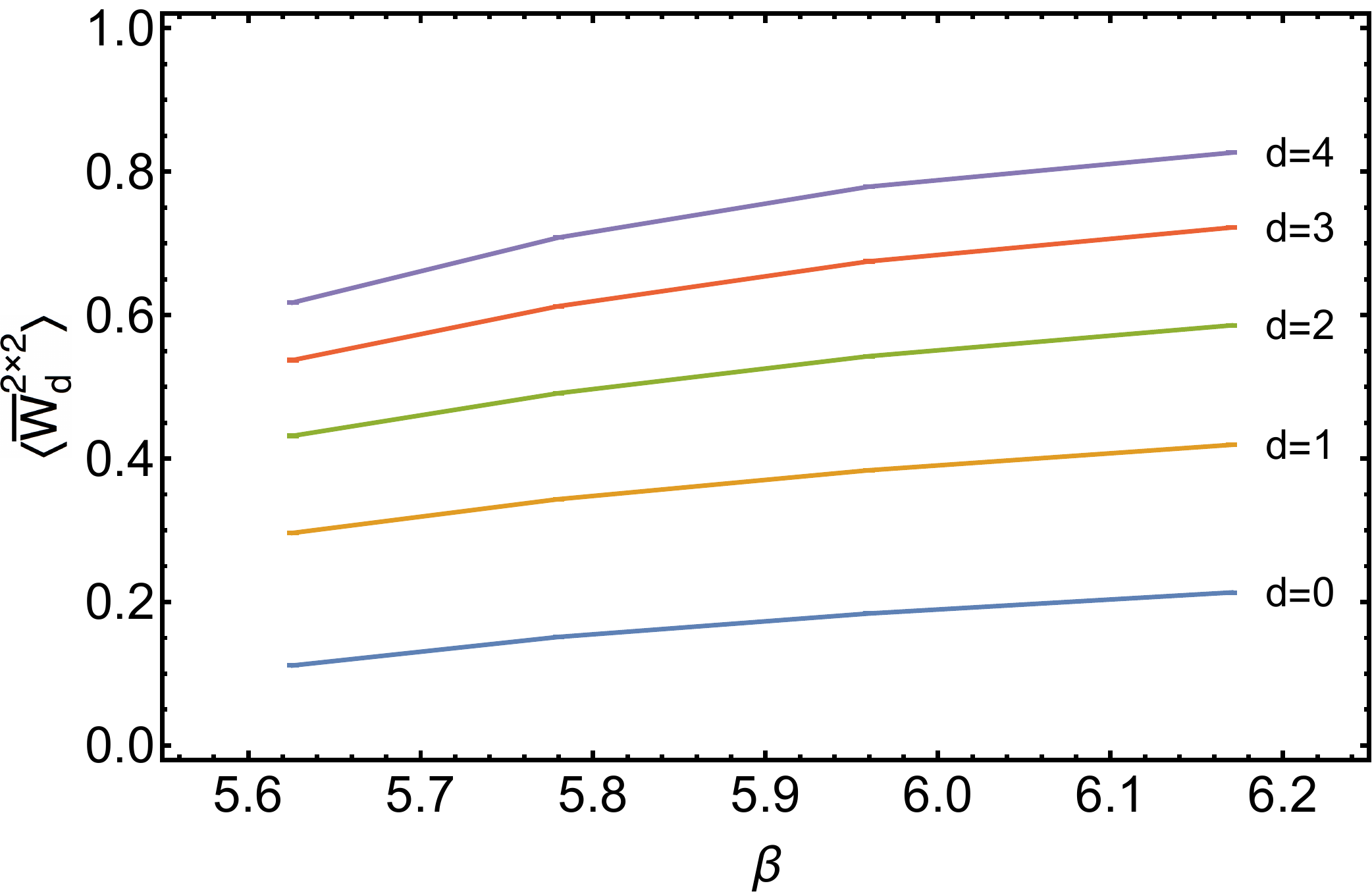}
\caption{\label{fig:partial_displaced_plaq}%
Partially averaged plaquette (left) and average displaced $2\times 2$ Wilson loop (right) as a function of $\beta$, measured on the four target ensembles after restriction and prolongation.
}
\end{figure}

The average partial plaquettes $\langle \bar W^{1\times 1}_d \rangle$ and displaced $2\times 2$ Wilson loops $\langle \bar W^{2\times 2}_d \rangle$ are shown in \Fig{partial_displaced_plaq} as a function of the target ensemble coupling for each ensemble after restriction and prolongation.
The former demonstrates that although the prolongated configurations retain an imprint of the coarse lattice, the configurations are nonetheless smooth by comparison to configurations from the associated target ensemble.
The latter observable provides a measure of the reduced translational symmetry of the restriction/prolongation operators.
Note that $\langle \bar W^{2\times 2}_0 \rangle$ is just the average plaquette measured on the coarse lattice, whereas $\langle \bar W^{2\times 2}_4 \rangle$ corresponds to fully displaced plaquettes.
Clear signals of the reduced translational symmetry is evident with approximately a factor of 4 difference between the two.
Later, we explore the rate at which displaced $2\times 2$ Wilson loops converge to the same value as a function of Monte Carlo evolution time, since this provides a measure of how quickly the full translational symmetry of the fine theory is restored.

\subsection{Topological charge}
\label{sec:topological_charge}

\begin{figure} 
\includegraphics[width=\figWidthHalf]{\figdir 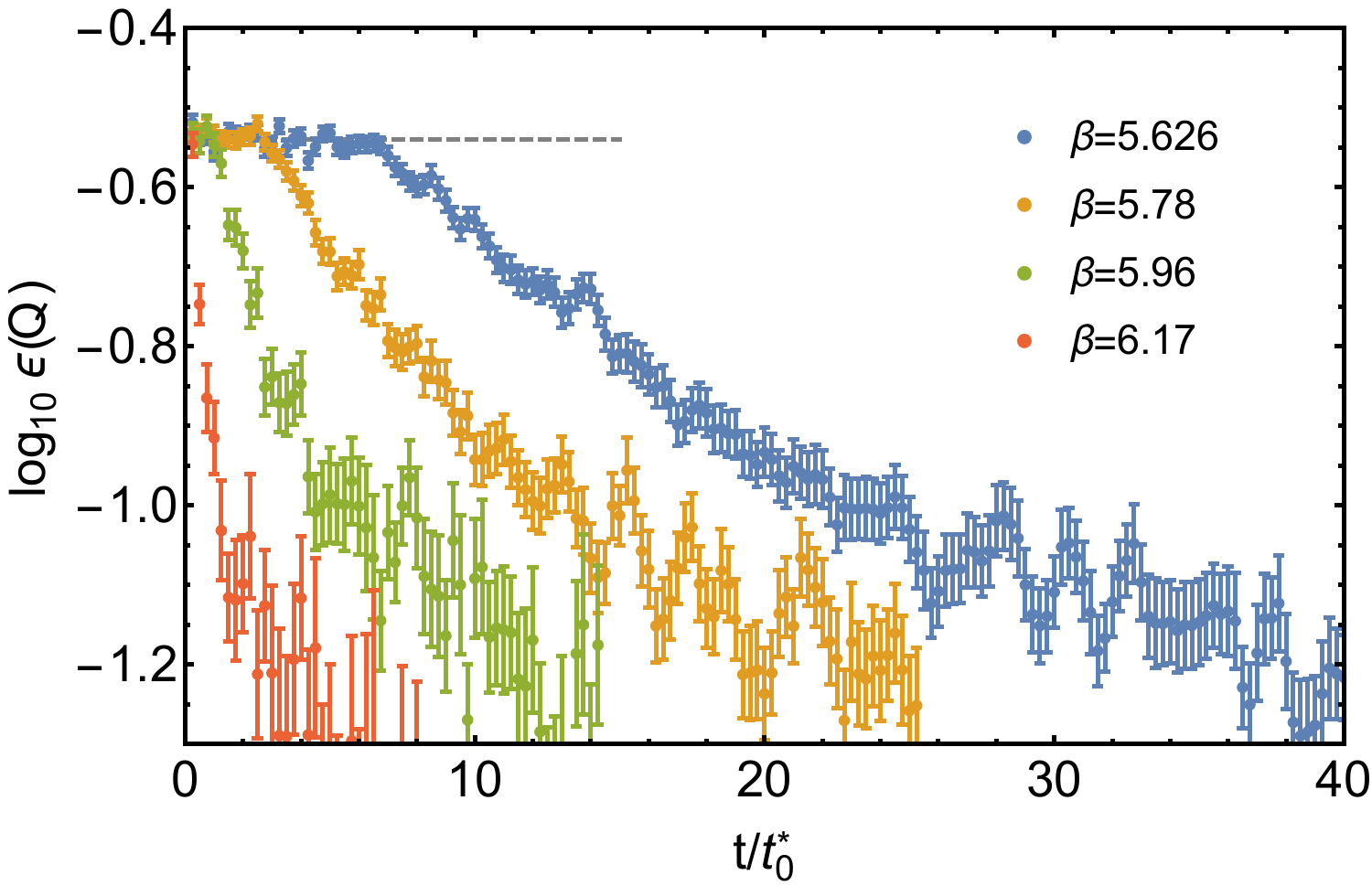}
\hspace{10pt}
\includegraphics[width=\figWidthHalf]{\figdir 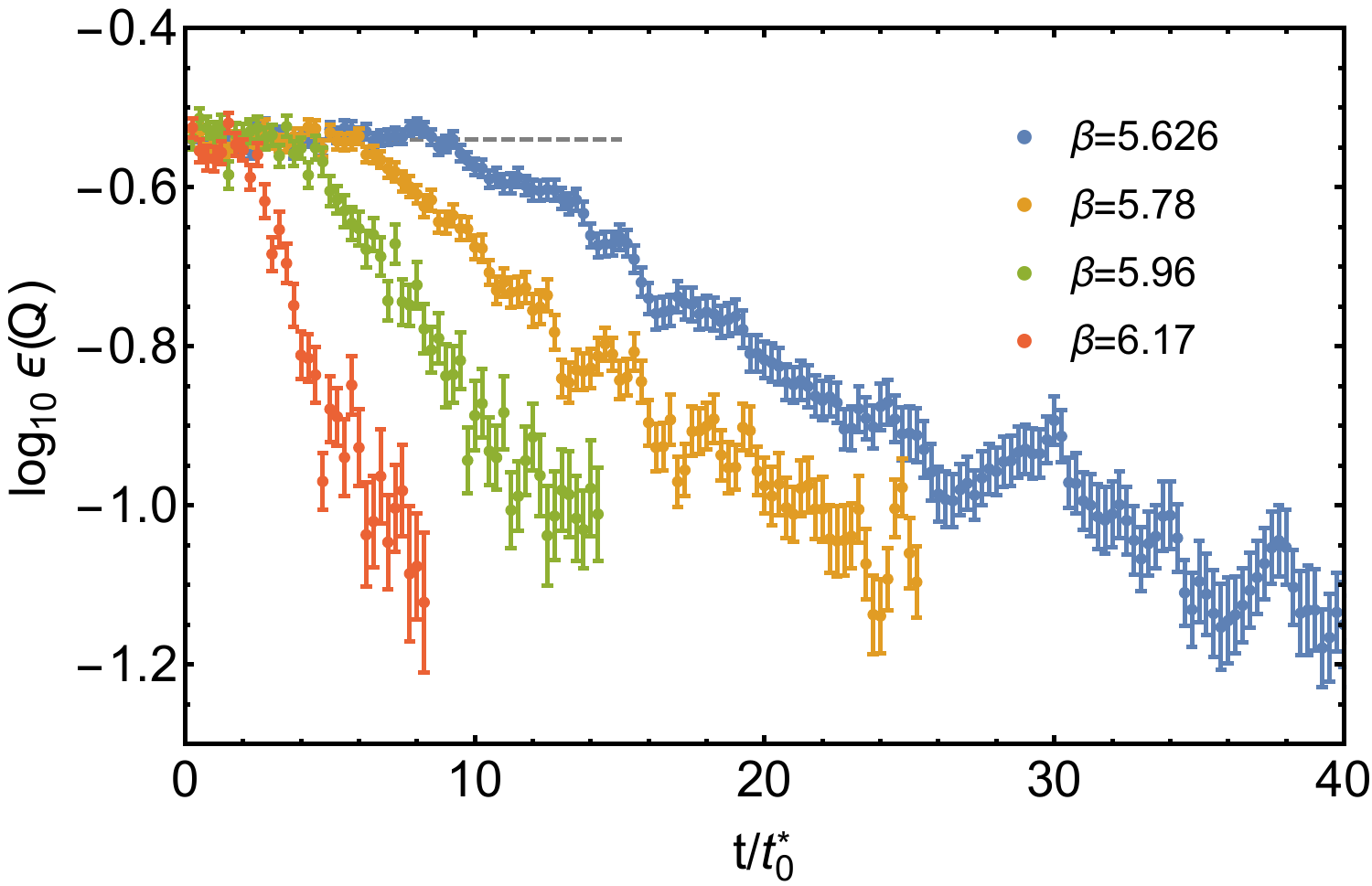}
\caption{\label{fig:eps}%
Plots of $\epsilon(Q)$ as a function of $t/t^\star_0$ for fine (left) and refined (right) ensembles; dashed line corresponds to $1/\sqrt{12}$.
}
\end{figure}

The topological charge, $Q(t)$, and susceptibility, $\chi(t)$ [defined as the square of $Q(t)$ divided by the spacetime volume], is determined as a function of the Wilson flow time using the three-loop improved gluonic definition of the topological charge operator~\cite{deForcrand1997409}. 
As a function of flow time, the topological charge is expected to approach integer values.
A measure of the deviation of the topological charge from integer values over the ensemble is given by
\begin{eqnarray}
\epsilon(Q)  = \sqrt{ \frac{1}{N} \sum_{i=1}^N \left( Q_i - [Q_i] \right)^2 } \ ,
\end{eqnarray}
where $[Q_i]$ is defined as the nearest integer to $Q_i$ for each decorrelated configuration, labeled by $i$.
Note that a locally uniform distribution for $Q_i$ about integer values (e.g., when the distribution for $Q$ is broad and smooth on scales much larger than unity) yields $\epsilon(Q) = 1/\sqrt{12}$.
This measure is expected to approach zero for the gluonic definition of the topological charge at late flow times.
Plots of this quantity are provided in \Fig{eps} for the target (fine) ensembles and ensembles obtained from restriction and prolongation (refined), as described in the previous section.
In units of the nominal scale $t_0^\star$, we find that the topological charge approaches integer values faster as the lattice spacing is decreased.
This can be understood in terms of smoothness of the gauge field configurations.
Following~\cite{Luscher:2010iy}, we may consider the quantity
\begin{eqnarray}
s_p = N_c \left[1 - W_{\mu\nu}^{1\times 1}(\bfn) \right]\ ,
\end{eqnarray}
measured on field configurations at flow time $t_0$, where $p$ is a plaquette associated with site $\bfn$ and basis vectors $\bfe_\mu$ and $\bfe_\nu$.
Given the measure of the configurations' smoothness, $h = \max_p(s_p)$, gauge configurations satisfying the admissibility criterion $h< 0.067$ will fall into distinct topological sectors~\cite{Luscher:2010iy}.
Configurations that violate the bound due to lattice artifacts, on the other hand, will not.
In this study, we find that 0\% of the configurations satisfy the criterion for $a \gtrsim0.1$ fm, whereas only 9\% of the configurations satisfy the criterion at $a \sim0.07$ fm; these results appear consistent with~\cite{Luscher:2010iy}.
Note that according to that study, at $a\sim 0.05$ fm, this percentage increases to about 70\%.
For equal flow times in lattice units, we find that the fine ensemble is more likely to have configurations that satisfy the admissibility condition than its refined counterpart; this result is counterintuitive, is due to the fact that the refined ensemble is smoother at the scale of the lattice spacing, and thus undergoes diffusion under Wilson flow at a rate slower than the fine ensemble.
As a by-product of this, we see in \Fig{eps} (b) that the topological charge attains integer values at a somewhat slower rate than that in \Fig{eps} (a).

\begin{table}
\caption{%
\label{tab:wilson_flow}%
Wilson flow parameters used for comparative study of $Q$ on the fine and refined lattices, autocorrelation time studies, and (re)thermalization time studies; all flows were performed using an adaptive step size~\cite{Fritzsch:2013je} with a maximum allowed step of $0.2$.
}
\begin{ruledtabular}
\begin{tabular}{ccccc}
Lattice & $t^\star_0/a^2$ & Flow time extent [$t^\star_0/a^2$] & Measurement frequency [$t^\star_0/a^2$] & Tolerance \\
\hline
$12^3\times 24$   & 0.73\phantom{0}  & 40 & 1/4 & 0.01 \\
$16^3\times 36$   & 1.39\phantom{0}  & 25 & 1/4 & 0.01 \\
$24^3\times 48$   & 2.79\phantom{0}  & 14 & 1/4 & 0.01 \\
$32^3\times 72$   & 5.49\phantom{0}  & 8  & 1/4 & 0.01 \\
\end{tabular}
\end{ruledtabular}
\end{table}

\begin{figure} 
\includegraphics[width=\figWidthHalf]{\figdir 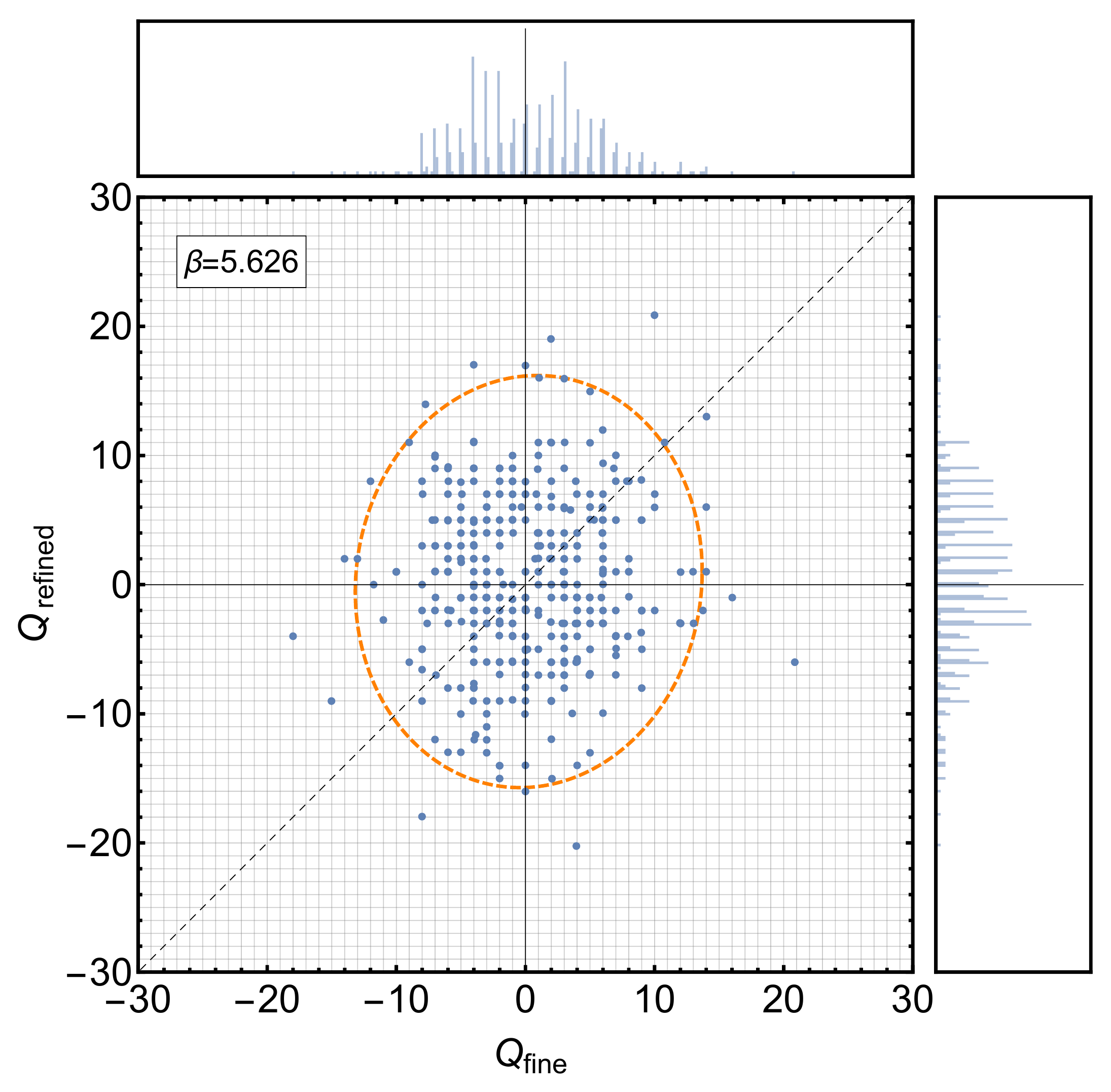}
\hspace{10pt}
\includegraphics[width=\figWidthHalf]{\figdir 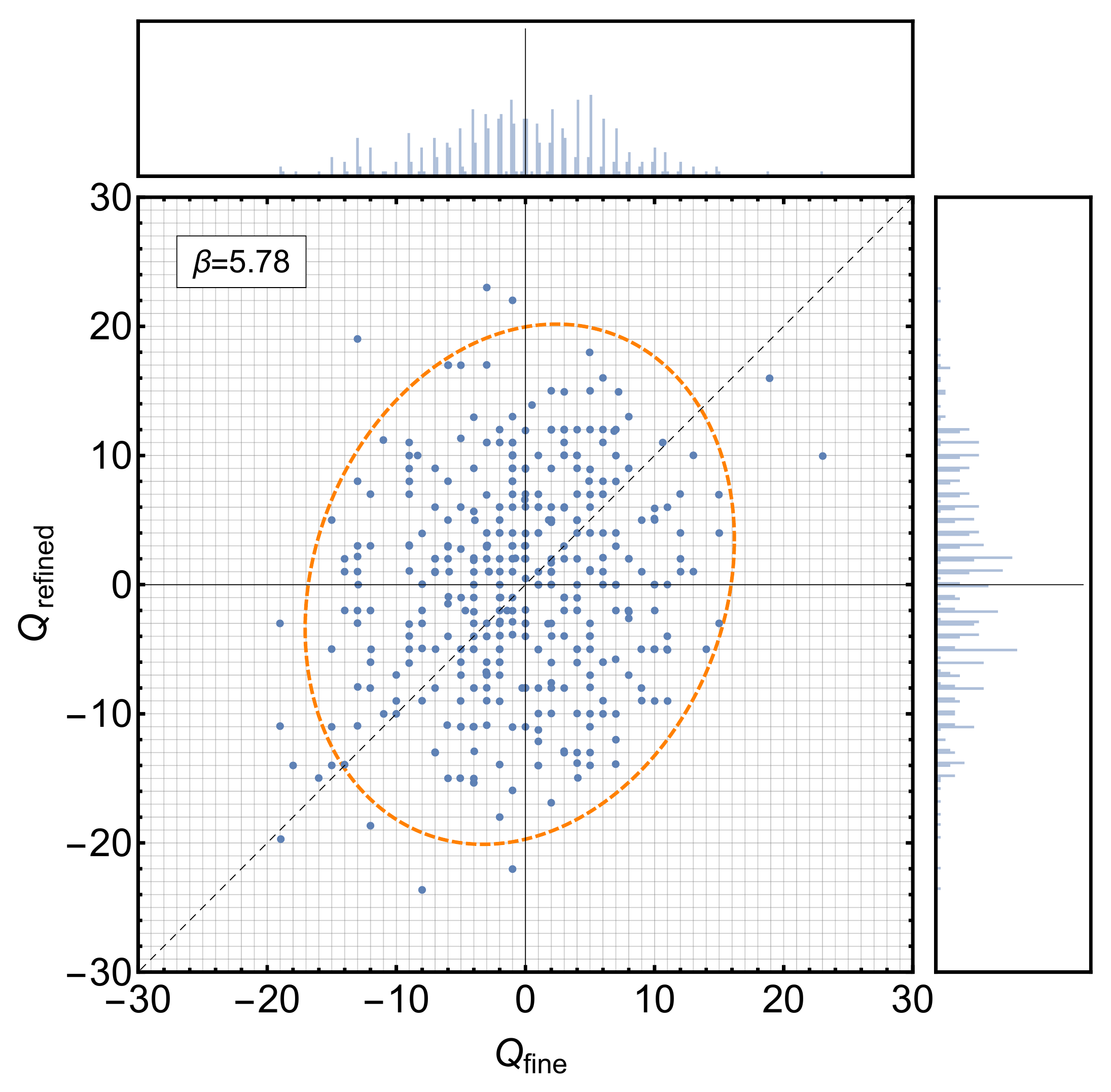}
\\
\includegraphics[width=\figWidthHalf]{\figdir 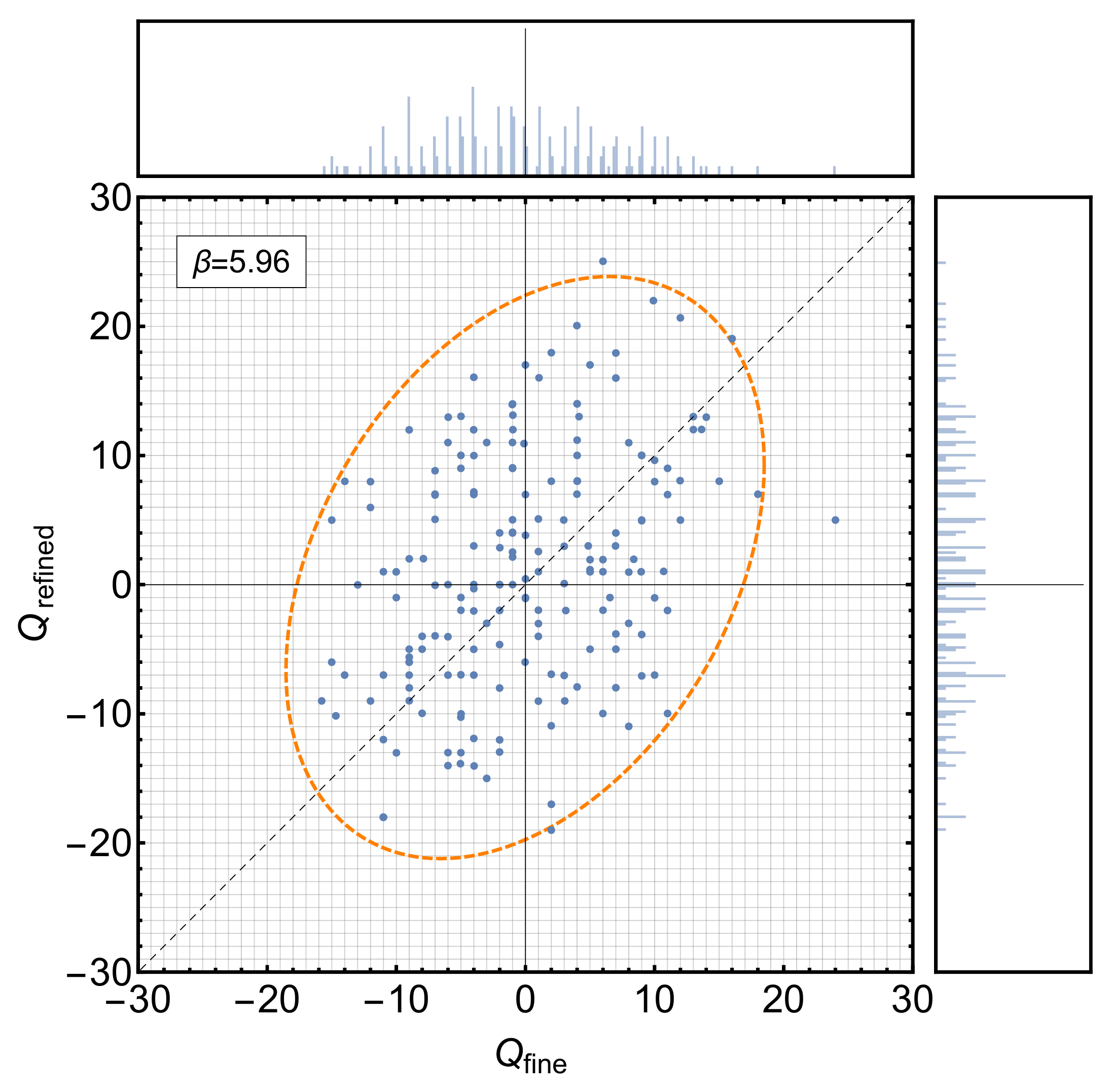}
\hspace{10pt}
\includegraphics[width=\figWidthHalf]{\figdir 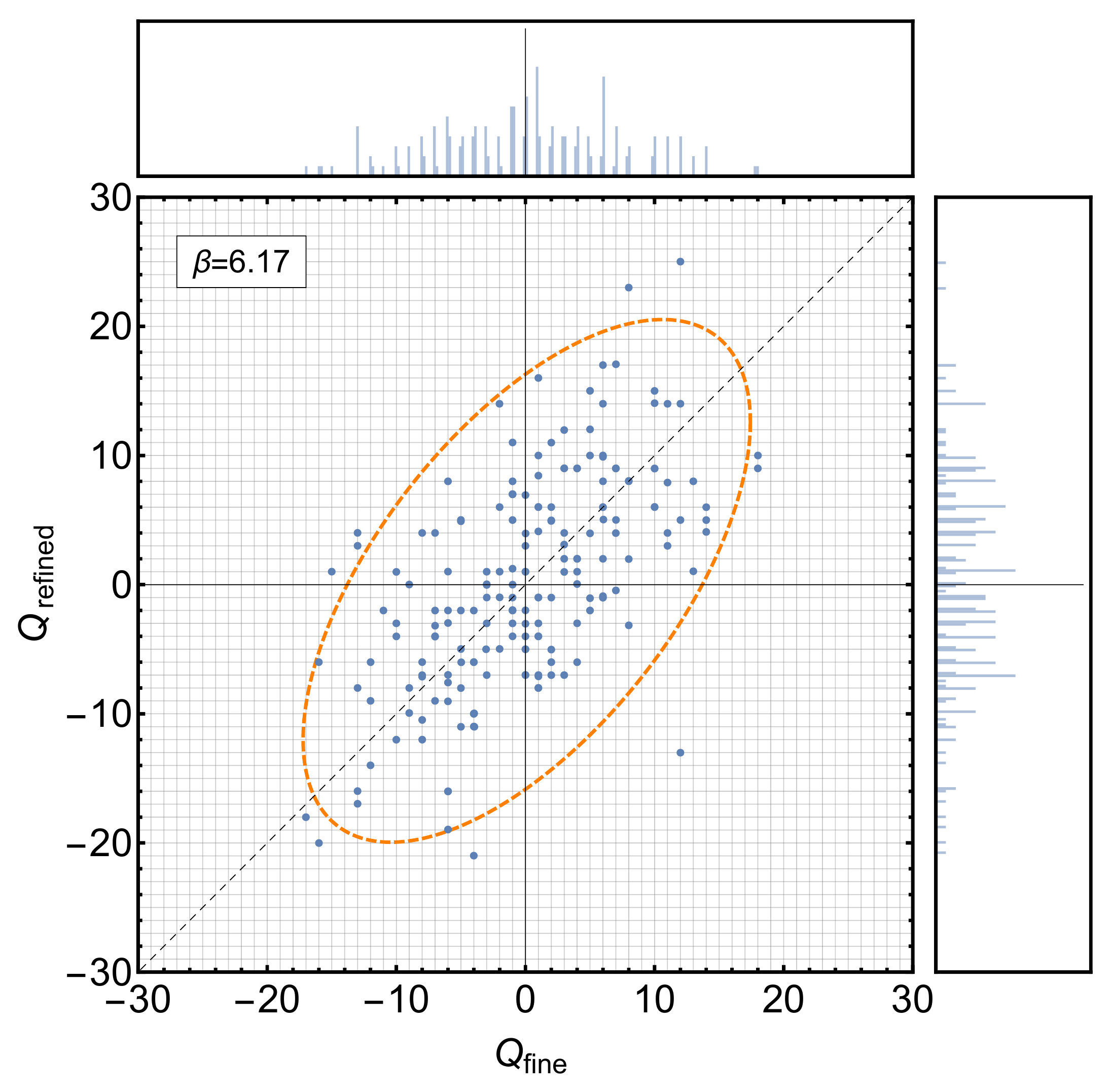}
\caption{\label{fig:qqCorr}%
Plot of $Q_\textrm{fine}$ versus $Q_\textrm{refined}$ determined at the maximum flow time extent for each ensemble pair. Dashed ellipses indicate 95\% confidence intervals centered about the mean.
The skewed nature of the ellipses with respect to the diagonal suggests that the refined topological charge distribution is broader than the fine distribution.
}
\end{figure}

\begin{figure} 
\includegraphics[width=\figWidthHalf]{\figdir 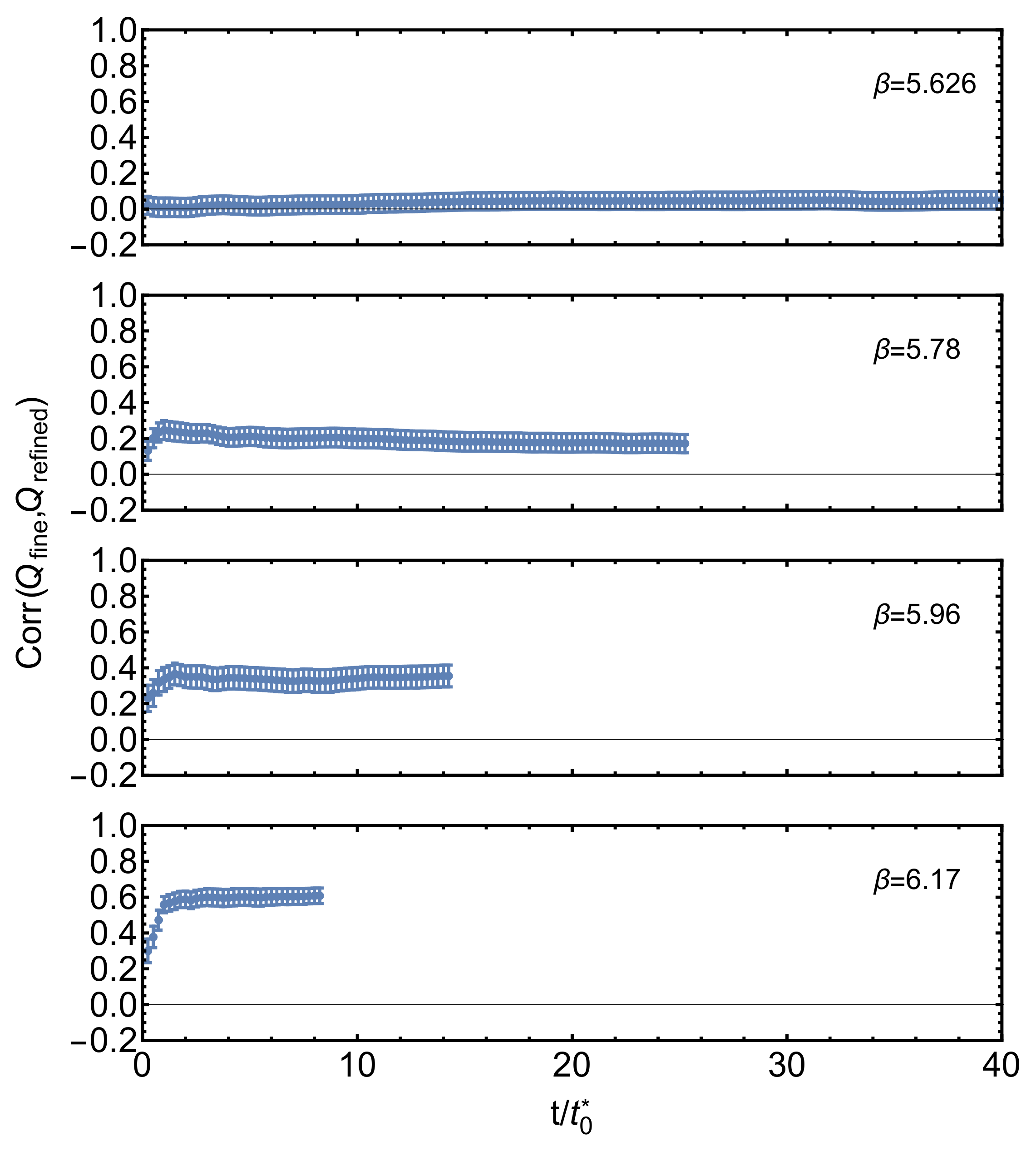}
\includegraphics[width=\figWidthHalf]{\figdir 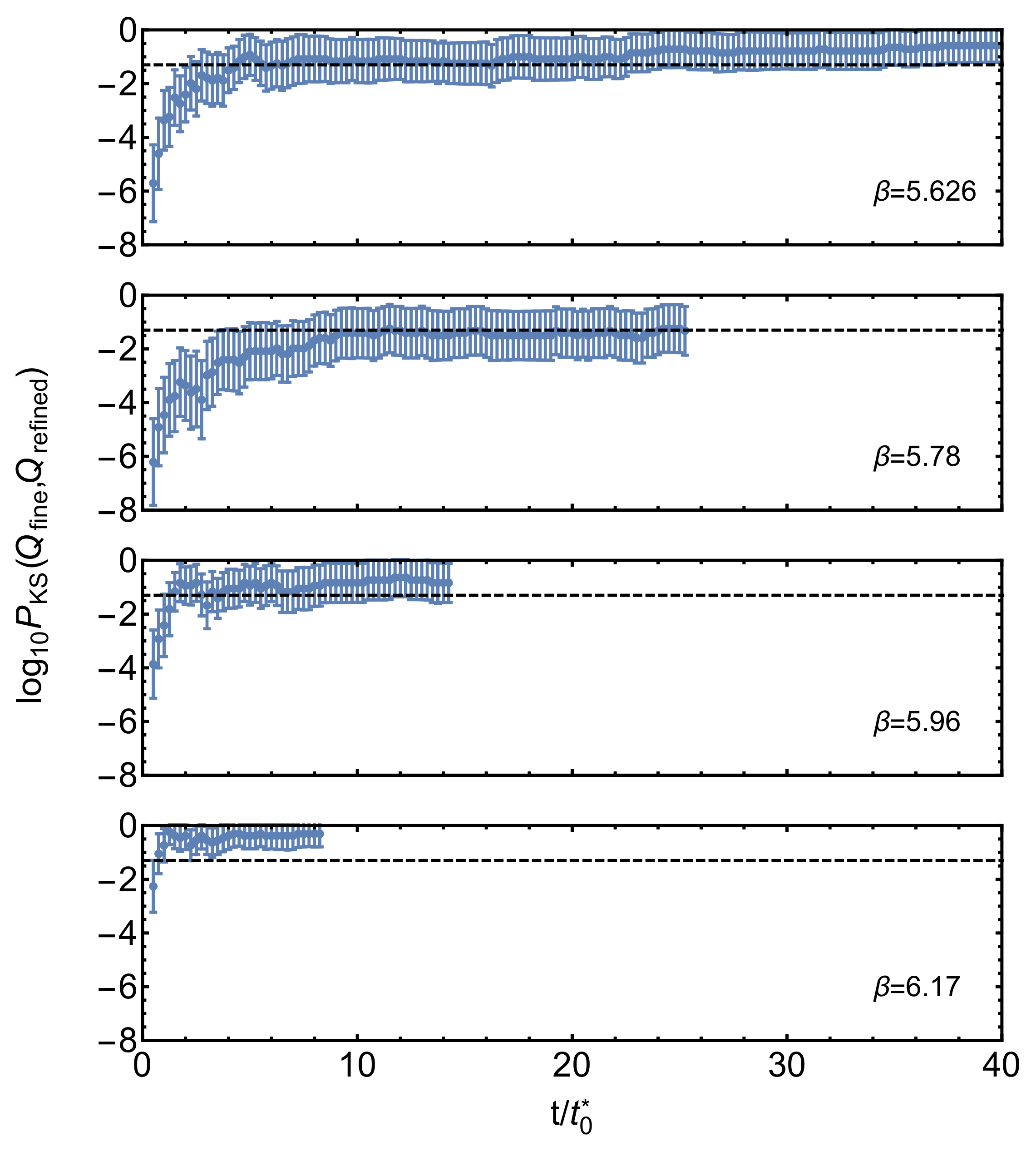}
\caption{\label{fig:qq_corr_ks}%
Left: Correlation between the topological charge on fine and refined ensembles as a function of flow time for the four values of $\beta$ considered in this work.
Right: $P$-values obtained from a two-sample Kolmogorov-Smirnov test that the topological charge distributions on fine and refined lattices come from the same underlying distribution, as a function of the Wilson flow time.
Dashed line corresponds to $0.05$.
}
\end{figure}

In \Fig{qqCorr}, we show scatterplots of the topological charge measured at the longest flow time for each of the target ensembles, and the corresponding topological charge measured after restriction and prolongation.
As was evident in \Fig{eps}, we see that for both fine and refined ensembles, the topological charge takes on integer values at large flow times.
Furthermore, for large $\beta$ we find that the distributions become increasingly skewed, indicating increasing correlation between the topological charge of the original and refined configurations.
In \Fig{qq_corr_ks} (left) we provide plots of this correlation as a function of flow time for each value of $\beta$.
We note that the correlations in the topological charge are largely independent of flow time for $t\gtrsim t_0$; this observation holds even at early times, where the topological charge need not take integer values.
In \Fig{intro_figs} (left), the correlations in the topological charge measured  at the latest flow time are plotted for each value of $\beta$.
Despite the large violations of the admissibility condition at the lattice spacings considered in this work, we nonetheless see a clear increasing trend in the topological charge correlations between fine and refined lattices.
Our expectation is that these correlations will rapidly approach unity as the lattice spacing is further reduced by a factor of $\sqrt{2}-2$.
Finally, in \Fig{qq_corr_ks} (right) we show the $p$-values obtained from a two-sample Kolmogorov-Smirnov test that the topological charge distributions on fine and refined ensembles come from the same underlying distribution.
The $p$-values were obtained as a function of Wilson flow time, and in each case are consistent or exceed $0.05$ after flow times of a few $t_0^\star$.
Already at $\beta = 6.17$ ($a = 0.07$ fm), it is difficult to distinguish between the distributions, and at even finer lattice spacings, the distributions will be exactly preserved.
This is important as it indicates that when we apply prolongation to a coarse ensemble with a well-sampled topological charge distribution, the resulting fine ensemble will continue to have a well sampled topological charge distribution, albeit with lattice artifacts inherited from the coarse level of discretization.
These differences can be corrected by the fine evolution, or by improvement of the coarse action.

\subsection{Algorithms and autocorrelations}
\label{sec:algorithms_autocorrelations}

To establish the viability of our multiscale approach, we must first determine the decorrelation time scales associated with conventional gauge evolution methods and their associated scaling behavior with the lattice spacing.
In this work, we focus on two particular algorithms, namely, HB and HMC.
The former algorithm is described in \Sec{ensembles}; however, for this application we consider the parameter choices $N_{hb}=100$ and $N_{ov}=0$.
The total number of updates is given by $\tau$.
For the later case, we use a PQP-type leapfrog algorithm, with $\tau$ trajectories of unit length and the number of leapfrog steps per trajectory tuned to yield an approximate acceptance rate of 70\% for each coupling.
Full details regarding the tuning of this algorithm are provided in \Appendix{hmc_tuning}.
Note that the strategy we take here of keeping the trajectory length fixed, rather than scaling it inversely with the lattice spacing, differs from that of~\cite{Luscher:2011kk,McGlynn:2014bxa}.
The algorithmic implementations we consider are chosen to provide a relatively simple benchmark for comparison with the multiscale approach.
More sophisticated and efficient implementations exist and could be used in both the traditional and the multiscale approaches, but require significant tuning and optimization.
We do not pursue these directions in this initial quenched investigation.

\begin{figure} 
\includegraphics[width=\figWidthHalf]{\figdir 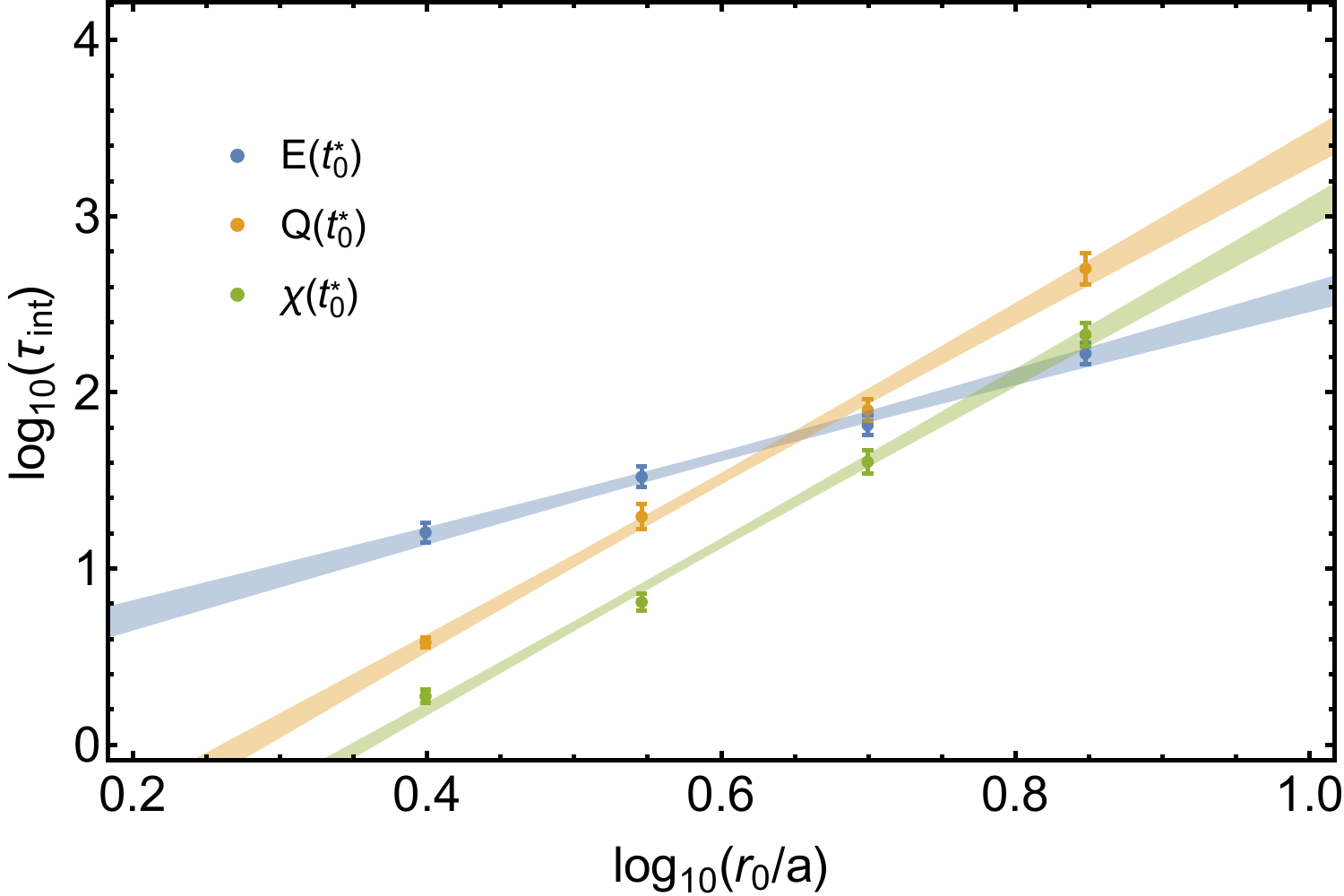}
\hspace{10pt}
\includegraphics[width=\figWidthHalf]{\figdir 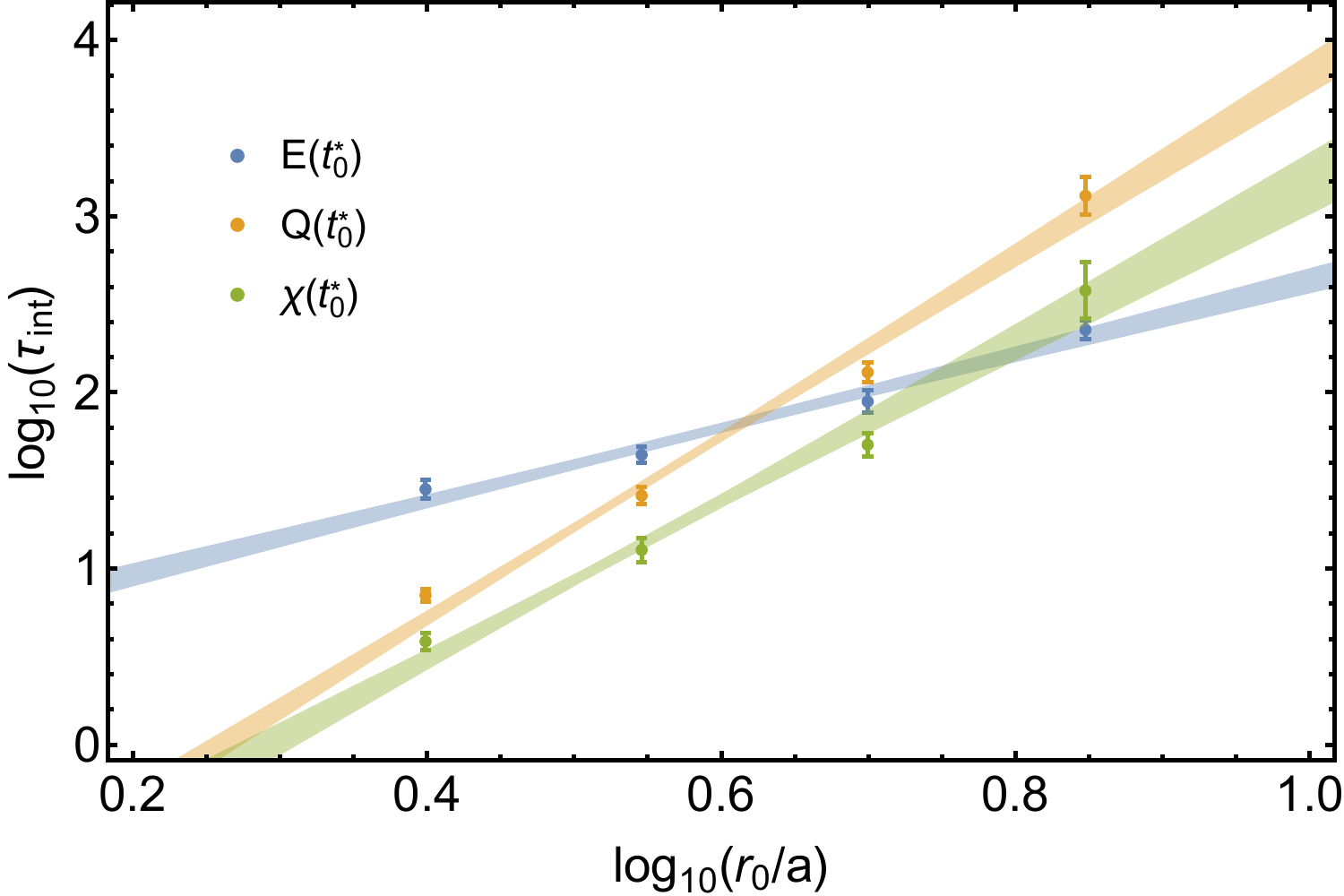}
\caption{\label{fig:tint}%
Integrated autocorrelation times for Wilson flow observables as a function of $r_0/a$ using HB (left) and HMC (right) algorithms.
}
\end{figure}

Integrated autocorrelation times were determined for various observables using the methods described in \Appendix{autocorrelations}.
Errors on the autocorrelation times were estimated using a highly efficient implementation of the jackknife method, with jackknife blocks of size $N_J$.
Obtained errors were consistent with those obtained with analytic approximations described in~\cite{Luscher:2005rx} and based on~\cite{Madras:1988ei,Wolff:2003sm}.
In \Tab{autocorr_ensembles}, we provide details of the ensembles generated for these estimates, including the ensemble size $N$, jackknife block size $N_J$, and measurement frequency of observables $\Delta\tau$.
Note that the total number of trajectories per ensemble is given by $N\Delta\tau$.
In the same table, we report the autocorrelation times for the topological charge, topological susceptibility, and the quantity $t^2 E(t)$ at flow time $t_0^\star$, all of which are long distance observables.
Note that the choice of $N_J \Delta\tau$ exceeds $2\tau_\textrm{int}$ estimated for each observable, suggesting a self-consistency in our error estimates.
For the finer lattice spacings, we find that the integrated autocorrelation time for Wilson loops of all sizes were significantly less than that of the Wilson flow quantities provided in \Tab{autocorr_ensembles}.
For $24^3$ and $32^3$ ensembles, our sampling resolution was insufficient to obtain reliable estimates of the integrated autocorrelation times for Wilson loops, and therefore such estimates for all $\beta$ are omitted.
The integrated autocorrelation times for each observable were fit to the functional form
\begin{eqnarray}
\tau_\textrm{int} = \textrm{const}\times \left( \frac{r_0}{a}\right)^{z_\textrm{int}}\ ,
\end{eqnarray}
and the fit results are provided in \Tab{autocorr_fits}.
Note that at a fixed physical volume, the computational cost to obtain decorrelated measurements of an observable is proportional to $\tau_\textrm{int}\times (r_0/a)^D $ for HB, and $\tau_\textrm{int}\times (r_0/a)^{D+1}$ for HMC, where $D=4$ powers of $r_0/a$ arise from trivial scaling of the number of lattice sites.
The latter has an additional power of $r_0/a$ due to the fact that the number of steps per trajectory needed to attain constant acceptance probability is roughly inversely proportional to the lattice spacing (see \Appendix{hmc_tuning} for details).
\Fig{tint} shows the autocorrelation times as a function of the inverse lattice spacing, exhibiting the expected critical slowing down as the continuum limit is approached at large $\beta$.
Results are shown for both HB and HMC evolution.

\begin{table}
\caption{%
\label{tab:autocorr_ensembles}%
Integrated autocorrelation times for various observables obtained from $N$ measurements performed on every $\Delta \tau$th update.
}
\begin{ruledtabular}
\begin{tabular}{cccccccc}
Algorithm & Lattice & $N$ & $\Delta\tau$ & $N_J$ & $\tau_\textrm{int}\left( E(t^\star_0) \right)$ & $\tau_\textrm{int}\left( Q(t^\star_0) \right)$ & $\tau_\textrm{int}\left( \chi(t^\star_0) \right)$ \\
\hline
HB  & $12^3\times 24$ & 5000   & 1 & 50 & 16.0(2.1)   & 3.8(0.3)      & 1.9(0.2)    \\
    & $16^3\times 36$ & 12000  & 1 & 100 & 33.2(4.4)   & 19.7(3.2)     & 6.5(0.7)    \\
    & $24^3\times 48$ & 8000   & 4 & 100 & 65.2(8.1)   & 79.4(11.1)    & 40.3(6.0)    \\
    & $32^3\times 72$ & 9000   & 8 & 300 & 166.2(22.7) & 504.4(103.9)  & 212.6(32.2)    \\
HMC & $12^3\times 24$ & 9000   & 1 & 50 & 28.2(3.0) & 7.1(0.6)  & 3.9(0.4) \\
    & $16^3\times 36$ & 24000  & 1 & 100 & 44.2(4.7) & 26.0(2.9) & 12.7(2.0) \\
    & $24^3\times 48$ & 12000  & 4 & 100 & 88.8(12.7) & 130.0(16.7) & 50.5(7.8) \\
    & $32^3\times 72$ & 9000   & 8 & 300 & 227.3(28.5) & 1307.5(322.8) & 378.4(139.4) \\
\end{tabular}
\end{ruledtabular}
\end{table}

\begin{table}
\caption{%
\label{tab:autocorr_fits}%
Integrated autocorrelation time fit results for $z_\textrm{int}$ for various observables.
}
\begin{ruledtabular}
\begin{tabular}{ccccc}
Algorithm & $z_\textrm{int}\left( E(t^\star_0) \right)$ & $z_\textrm{int}\left( Q(t^\star_0) \right)$ & $z_\textrm{int}\left( \chi(t^\star_0) \right)$ \\
\hline
HB  & 2.3(2) & 4.7(2) & 4.7(2) \\
HMC & 2.1(2) & 5.2(2) & 4.5(4) \\
\end{tabular}
\end{ruledtabular}
\end{table}

\subsection{Thermalization and rethermalization}
\label{sec:rethermalization}

With autocorrelation times for conventional gauge evolution at hand, we are finally in a position to assess the utility of the multiscale strategy proposed in \Sec{introduction}.
To make the analysis quantitative, we proceed by studying the equilibration properties of both conventionally prepared initial configurations (thermalization) and initial configurations prepared by prolongation (rethermalization).
For these studies, we consider ensembles of size $N_s=24$ under Markov evolution using the same action and algorithms as in the previous section, and the two gauge couplings, $\beta=5.96$ and $\beta=6.17$, corresponding to $a\sim 0.1$ fm and $a\sim 0.07$ fm, respectively.
Thermalization studies were performed using initial configurations drawn from an ordered delta-function distribution (i.e., a cold start) and from a random distribution (i.e., hot start).
Rethermalization studies were performed using initial configurations which had been prepared in two ways, in each case utilizing a subset of the decorrelated target ensembles described in \Tab{target_ensembles}.
In the first case (r-I), configurations were prepared by restriction and prolongation of $\beta=5.96$ and $\beta=6.17$ configurations, similar to the analysis of topological charge correlations in \Sec{topological_charge}.
In the second case (r-II), initial configurations were prepared via prolongation of RG-matched coarse ensembles generated using the Wilson action.
The matching was performed via the Sommer scale, with coarse couplings corresponding to $\beta=5.626$ ($a\sim 0.2$ fm) and $\beta=5.78$ ($a\sim0.14$ fm).

Each (re)thermalization study was performed using both HB and HMC algorithms.
In the latter case, acceptance probabilities were found to be exponentially small at the start of (re)thermalization.
For both lattice spacings, we therefore initially evolved the ensembles for 24 trajectories without an accept/reject step in order to achieve reasonable acceptance probabilities; beyond that, evolution was performed with an accept/reject step.
In all cases, the warm-up period necessary to achieve reasonable acceptance probabilities was significantly shorter than the (re)thermalization time.

\begin{figure} 
\includegraphics[width=\figWidthHalf]{\figdir 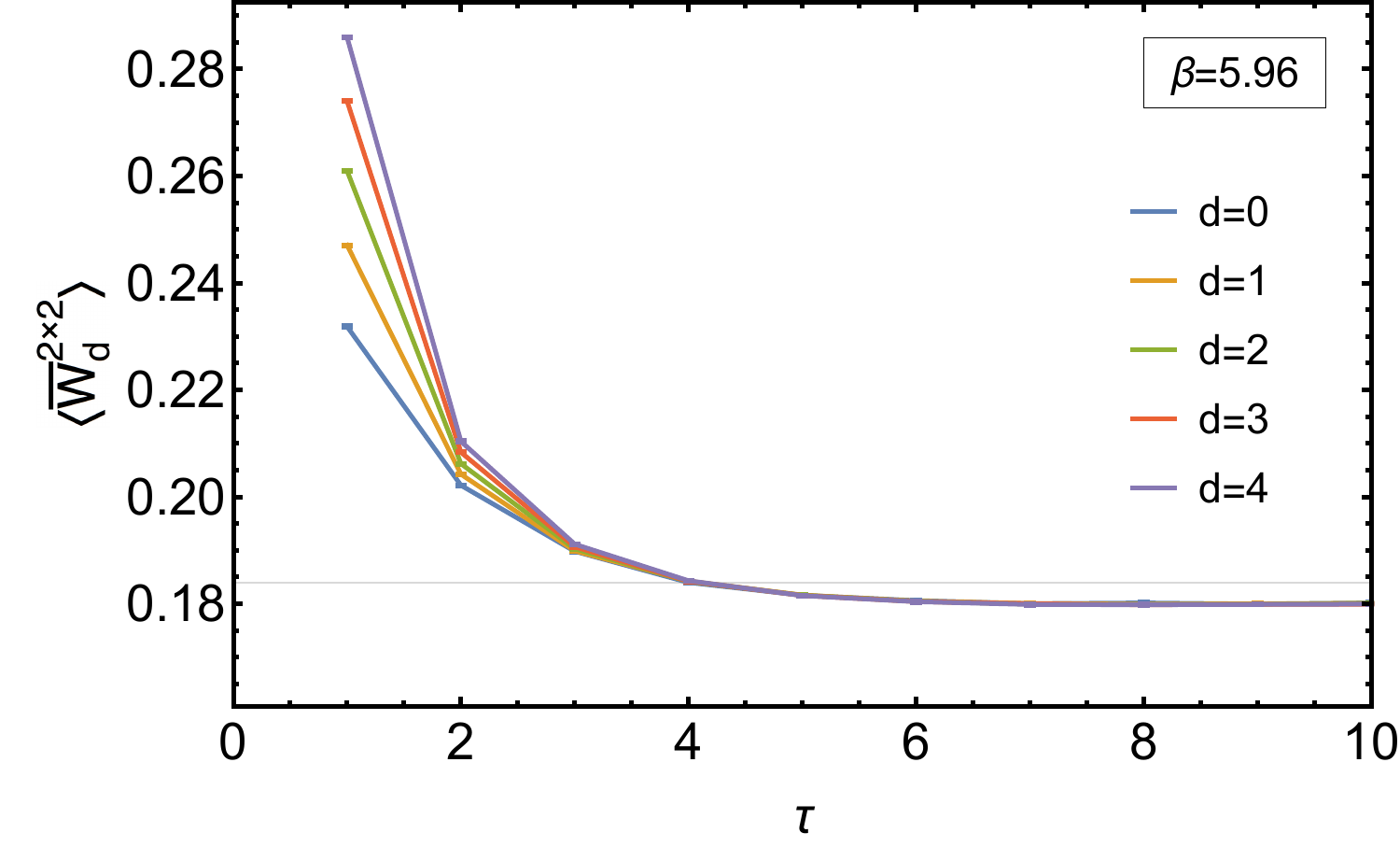}
\hspace{10pt}
\includegraphics[width=\figWidthHalf]{\figdir 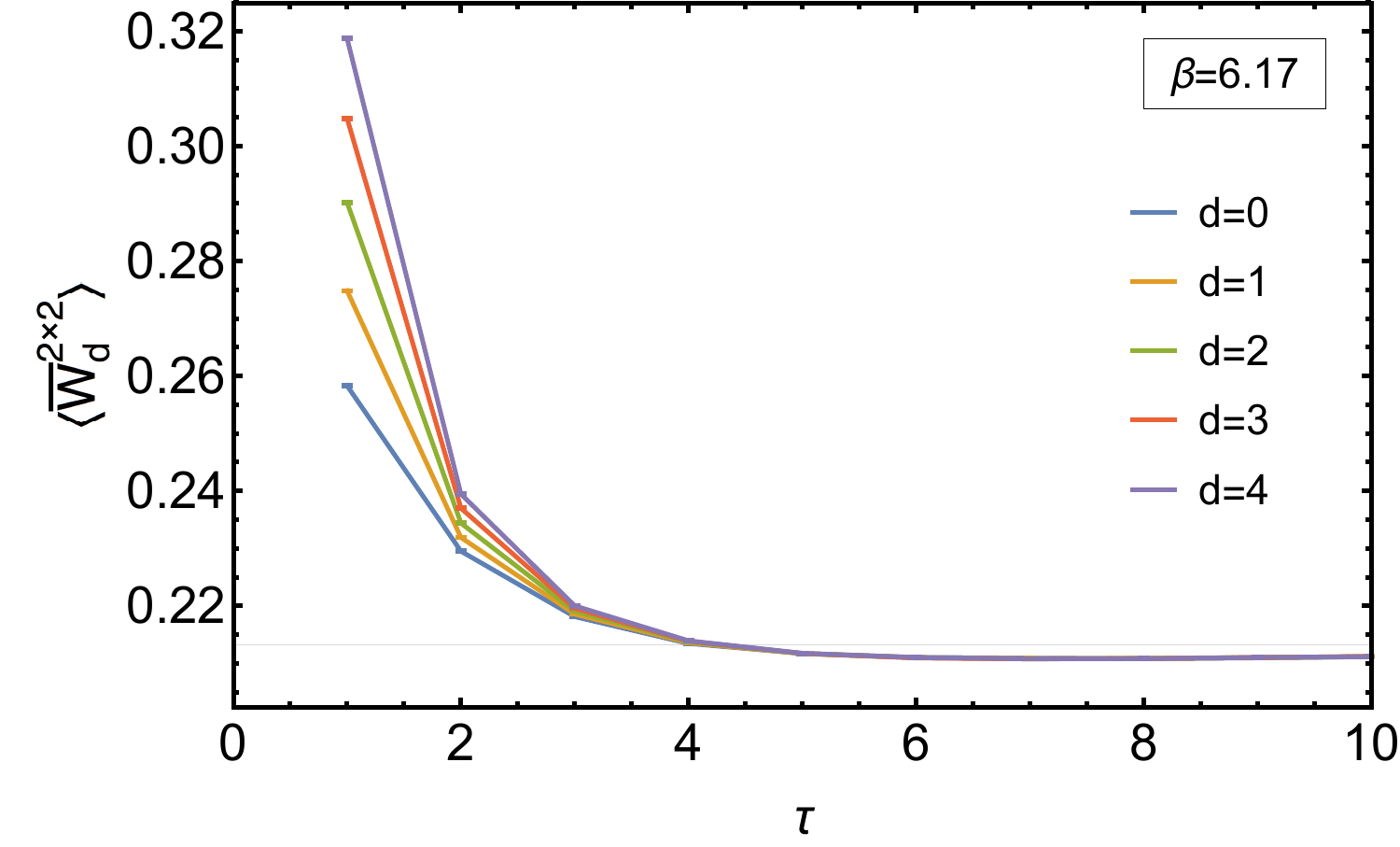} \\
\vspace{10pt}
\includegraphics[width=\figWidthHalf]{\figdir 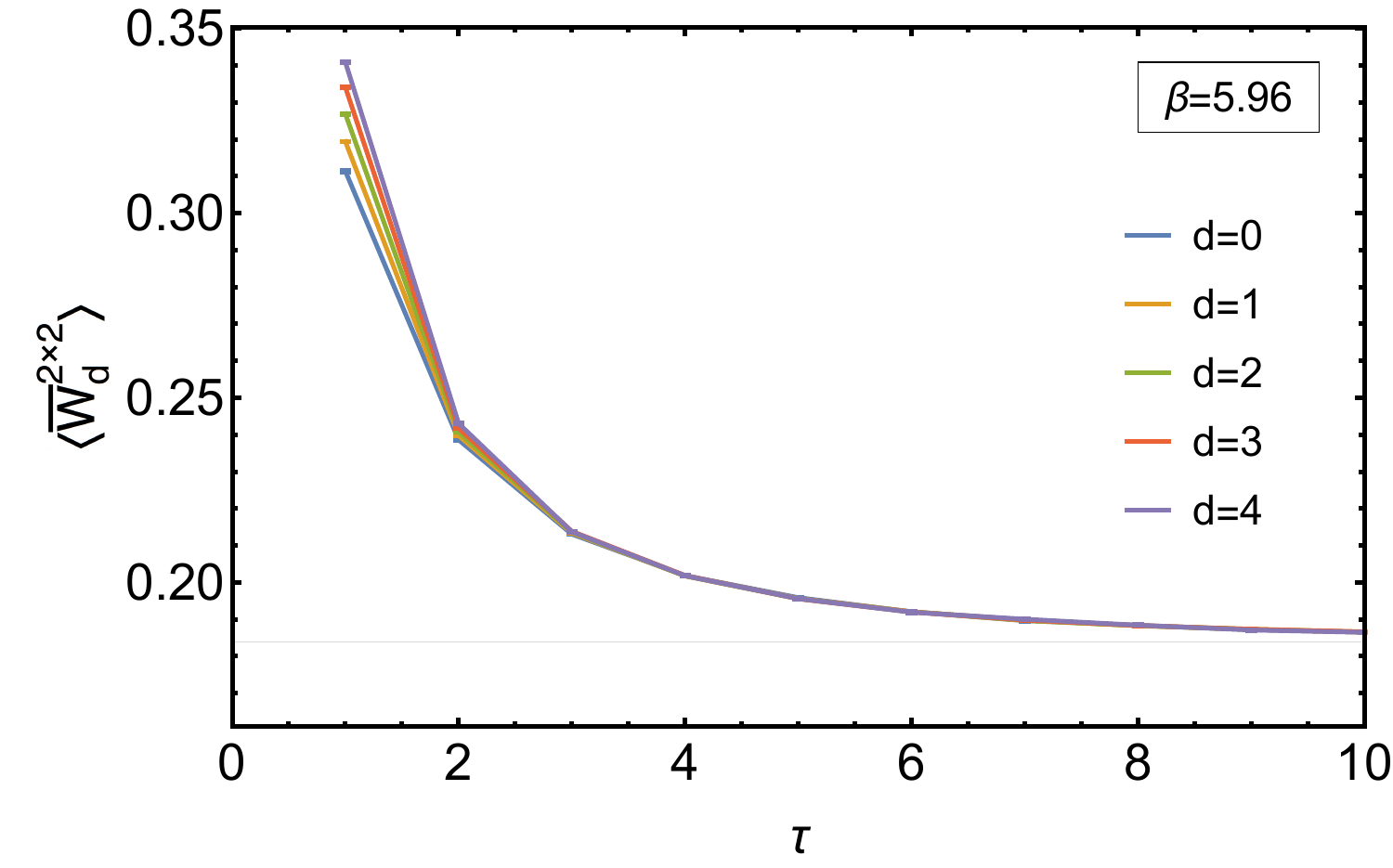}
\hspace{10pt}
\includegraphics[width=\figWidthHalf]{\figdir 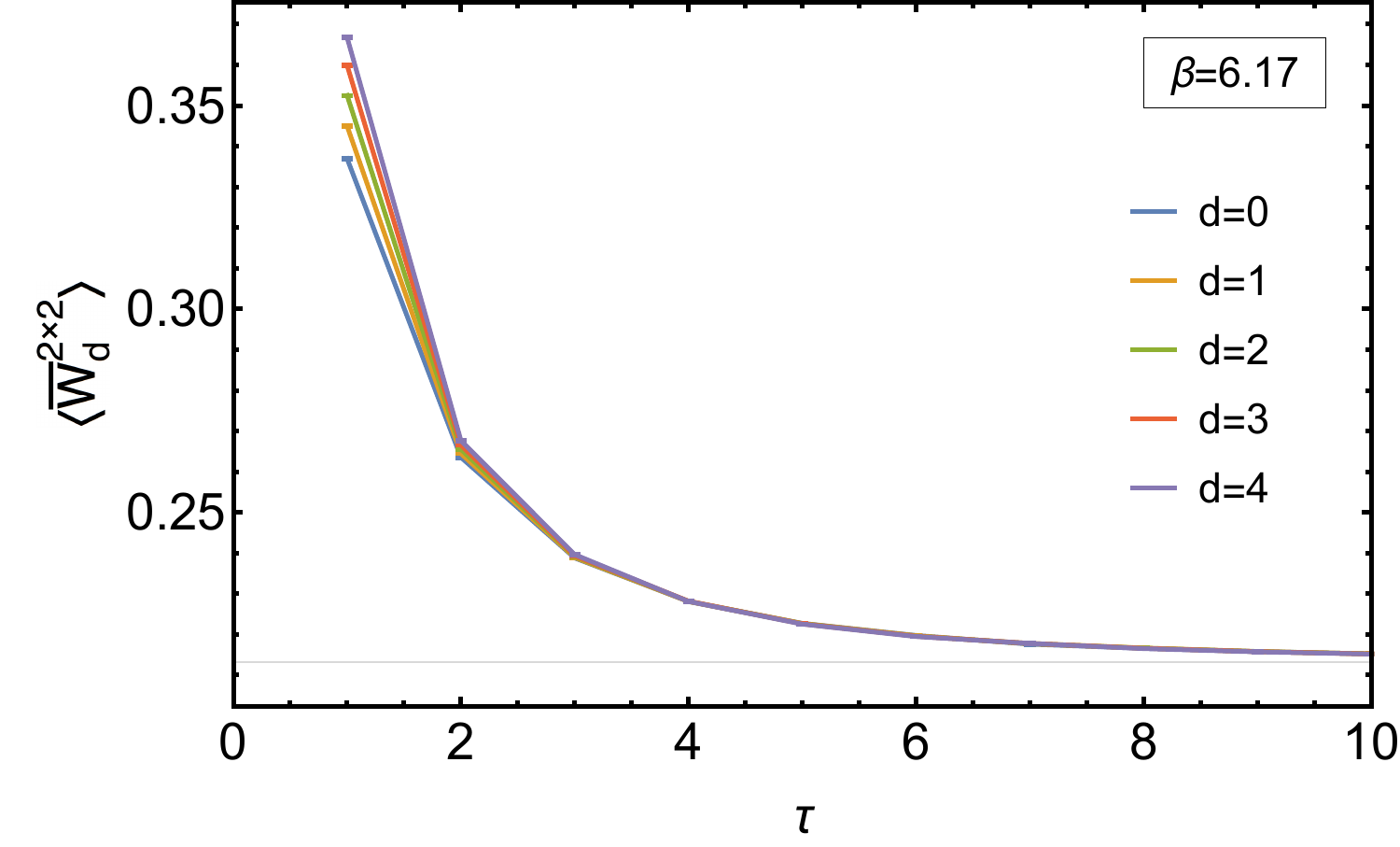} \\
\vspace{10pt}
\includegraphics[width=\figWidthHalf]{\figdir 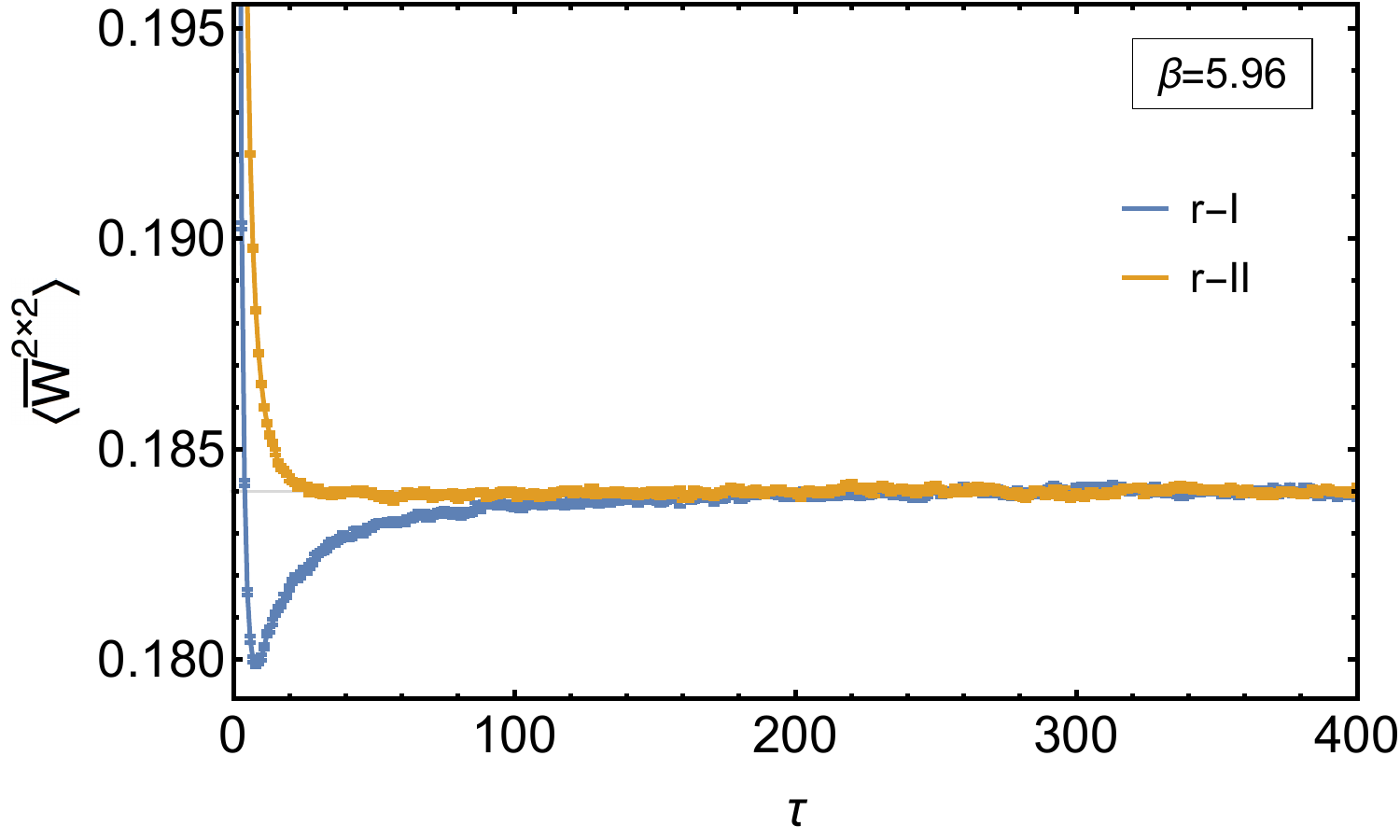}
\hspace{10pt}
\includegraphics[width=\figWidthHalf]{\figdir 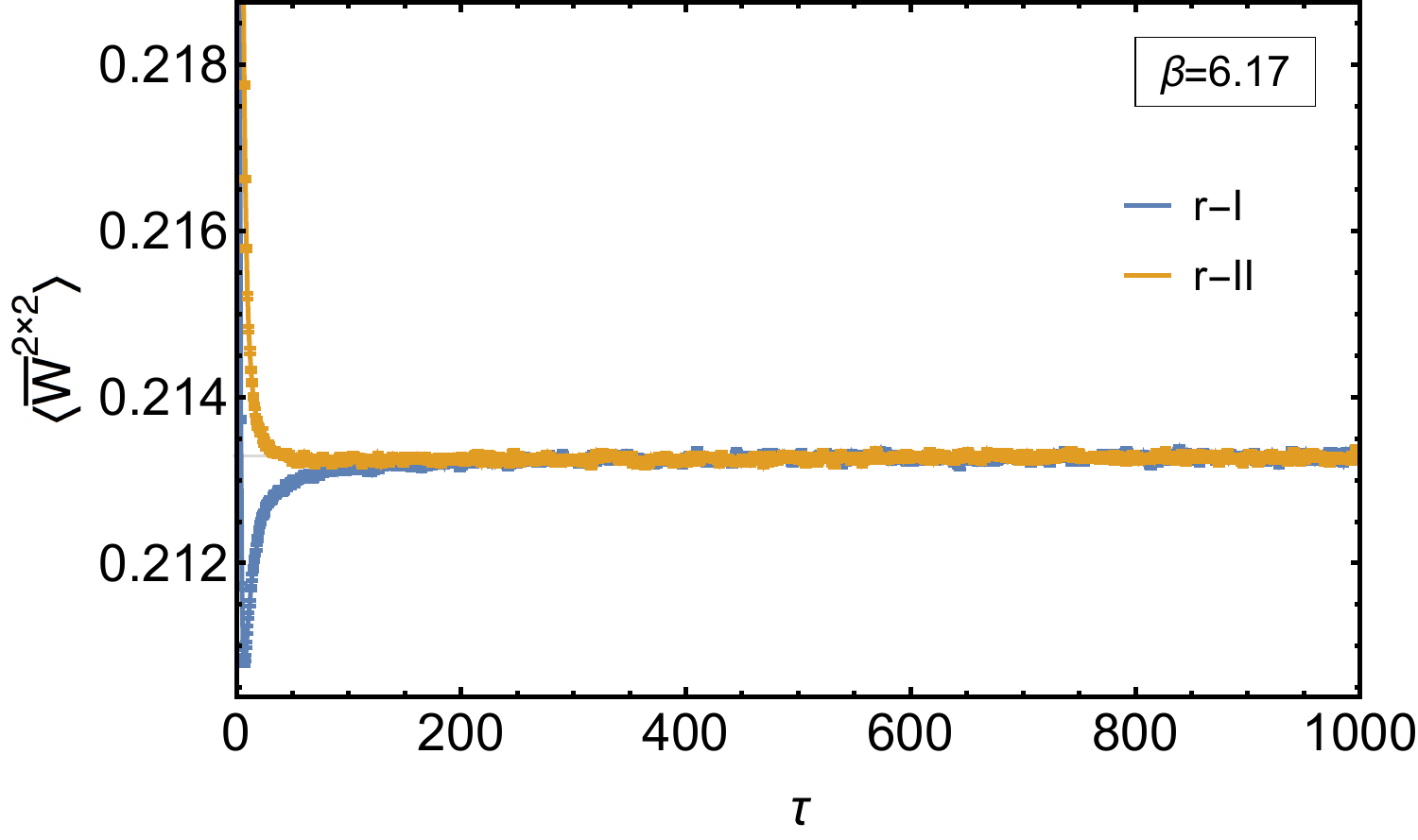}
\caption{\label{fig:hb_retherm_disp}%
Top: HB rethermalization of average displaced $2\times 2$ Wilson loops (r-I).
Center: HB rethermalization of average displaced $2\times 2$ Wilson loops (r-II).
Bottom: HB rethermalization of $2\times 2$ Wilson loops.
}
\end{figure}

\begin{figure} 
\includegraphics[width=\figWidthHalf]{\figdir 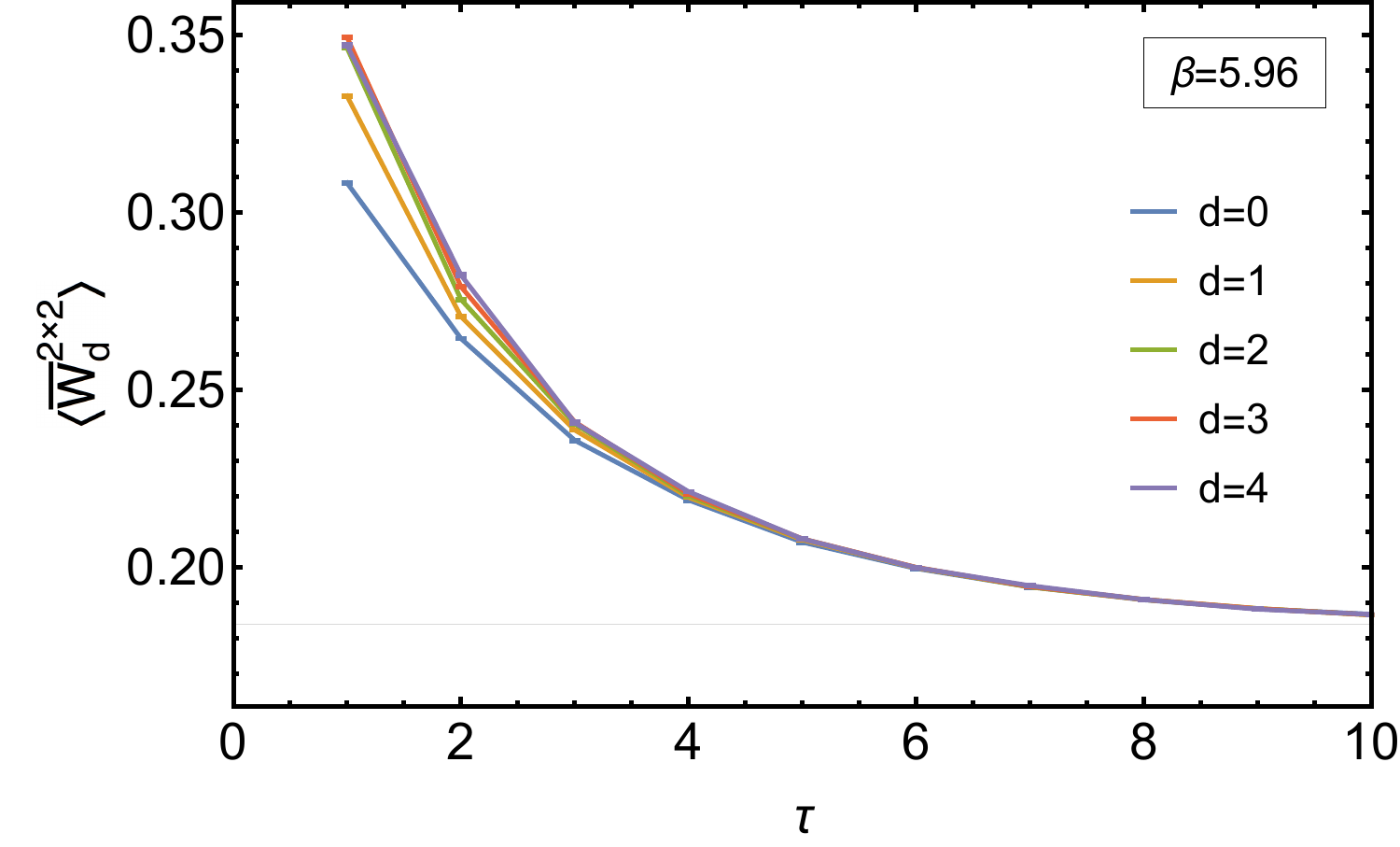}
\hspace{10pt}
\includegraphics[width=\figWidthHalf]{\figdir 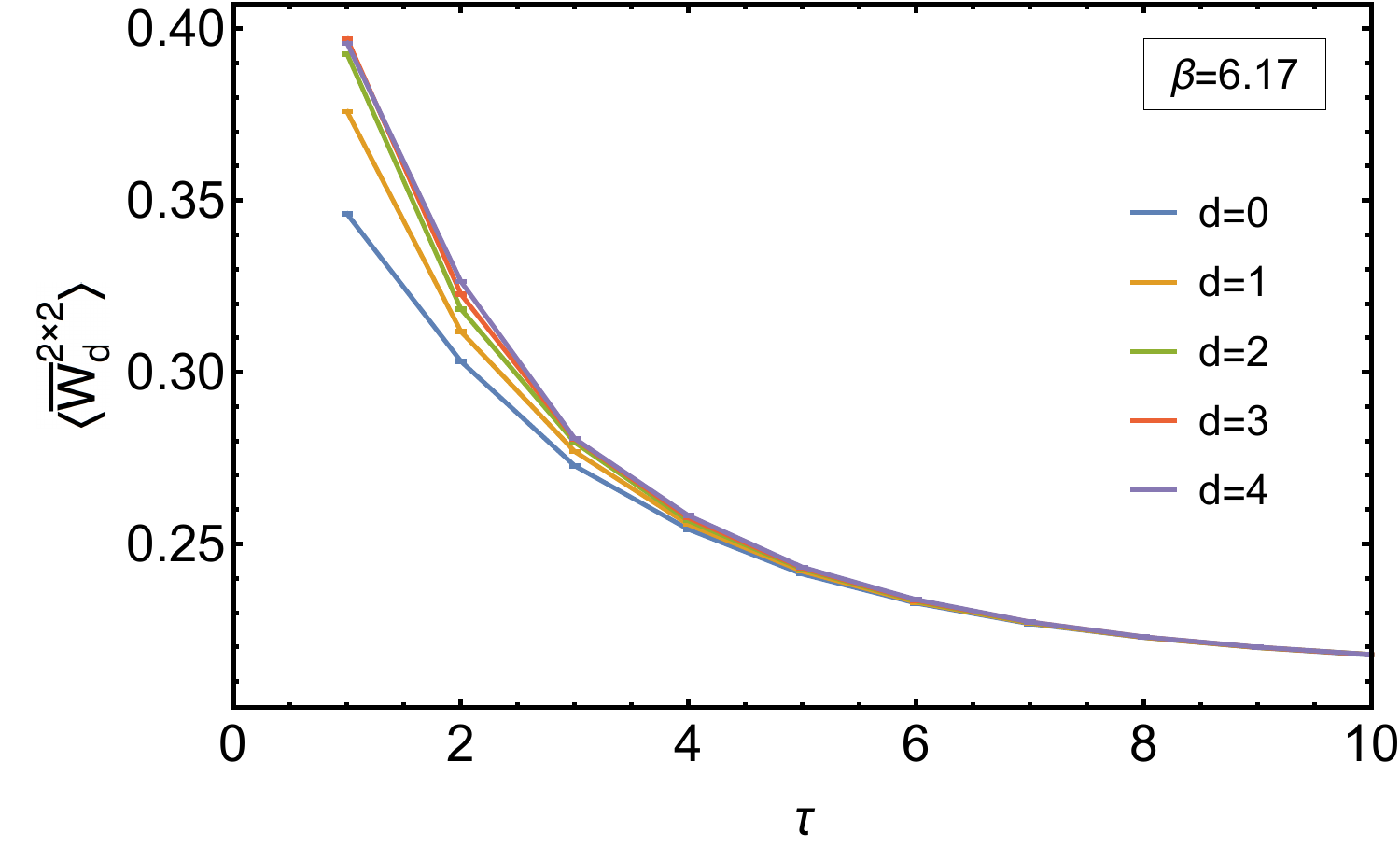} \\
\vspace{10pt}
\includegraphics[width=\figWidthHalf]{\figdir 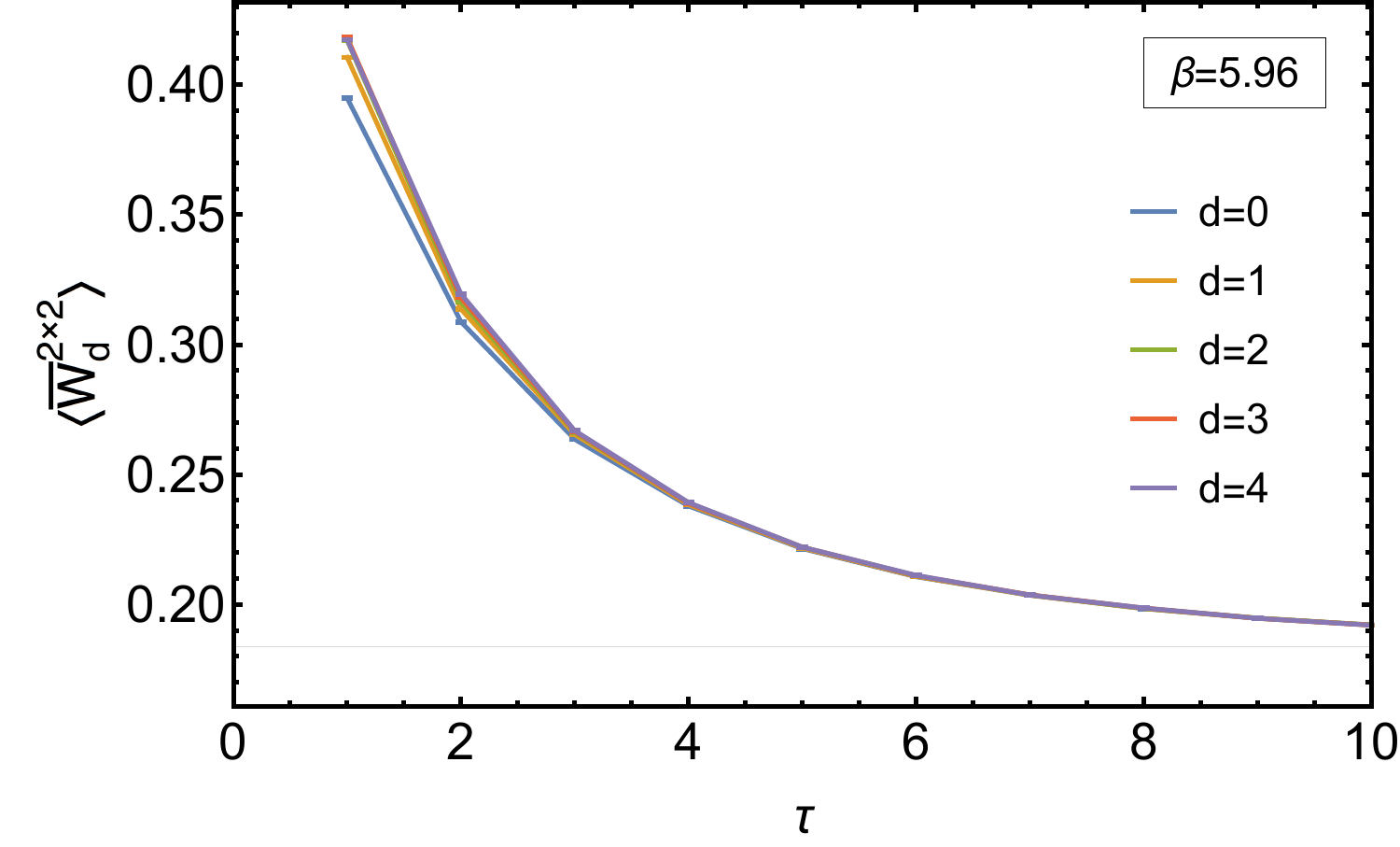}
\hspace{10pt}
\includegraphics[width=\figWidthHalf]{\figdir 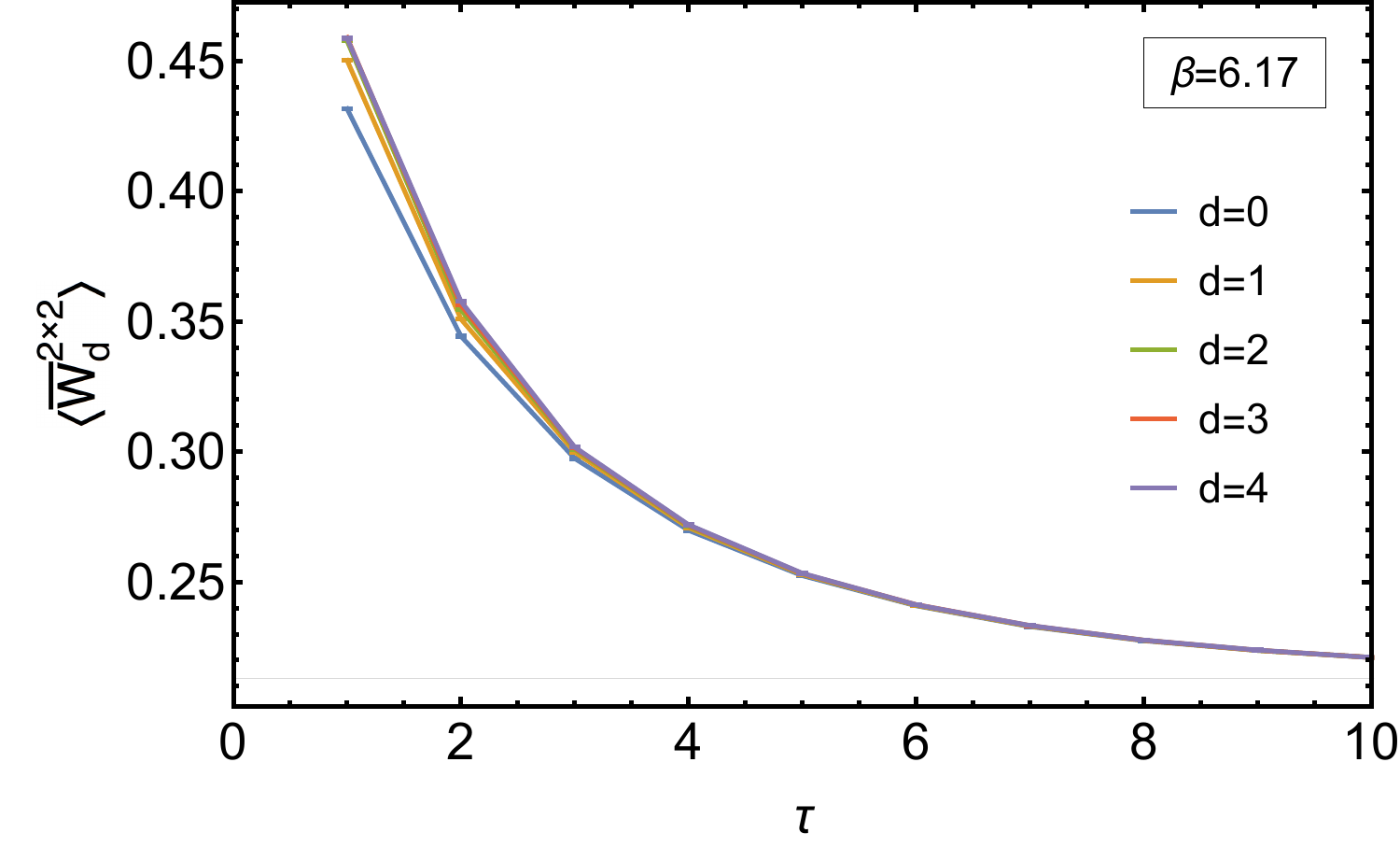} \\
\vspace{10pt}
\includegraphics[width=\figWidthHalf]{\figdir 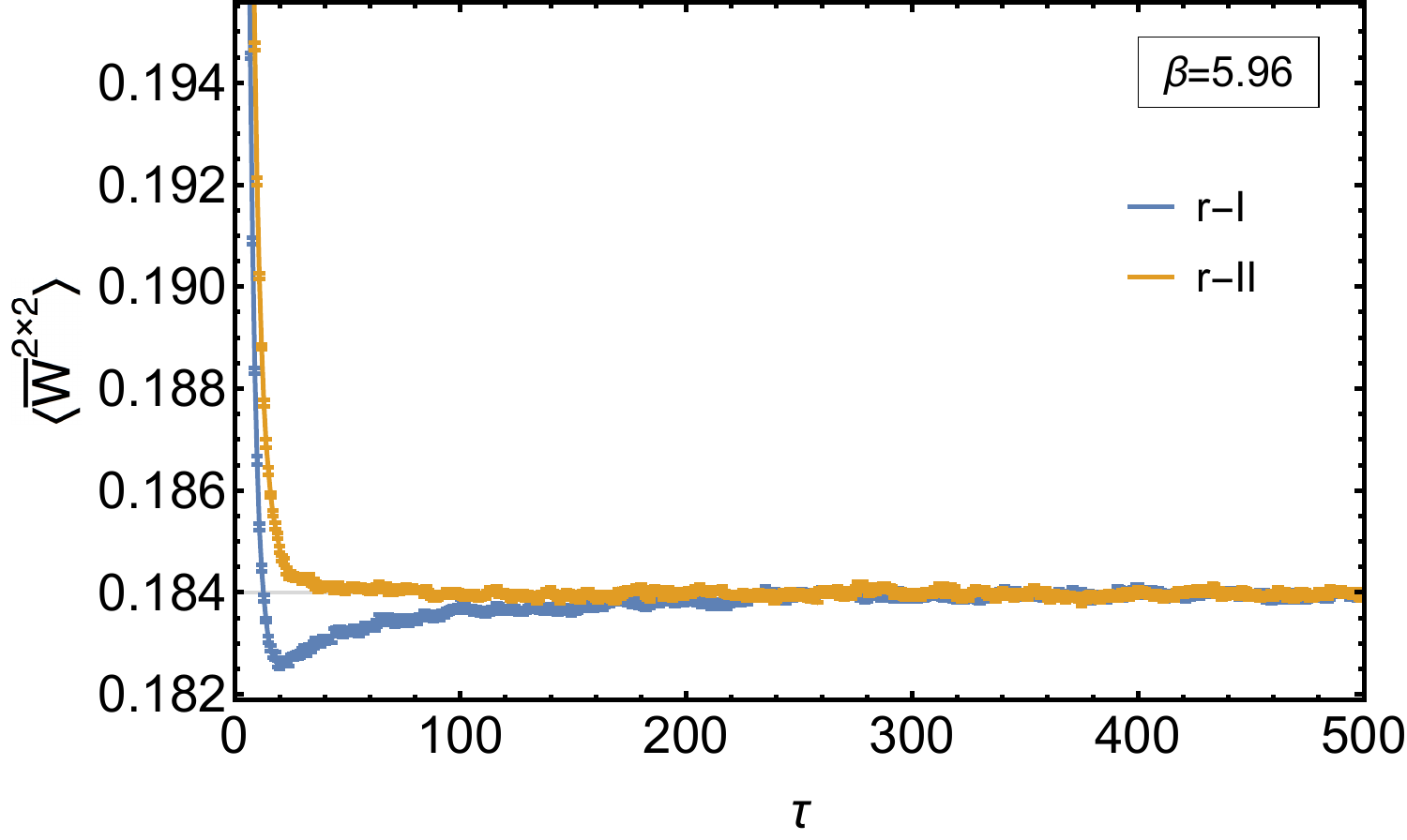}
\hspace{10pt}
\includegraphics[width=\figWidthHalf]{\figdir 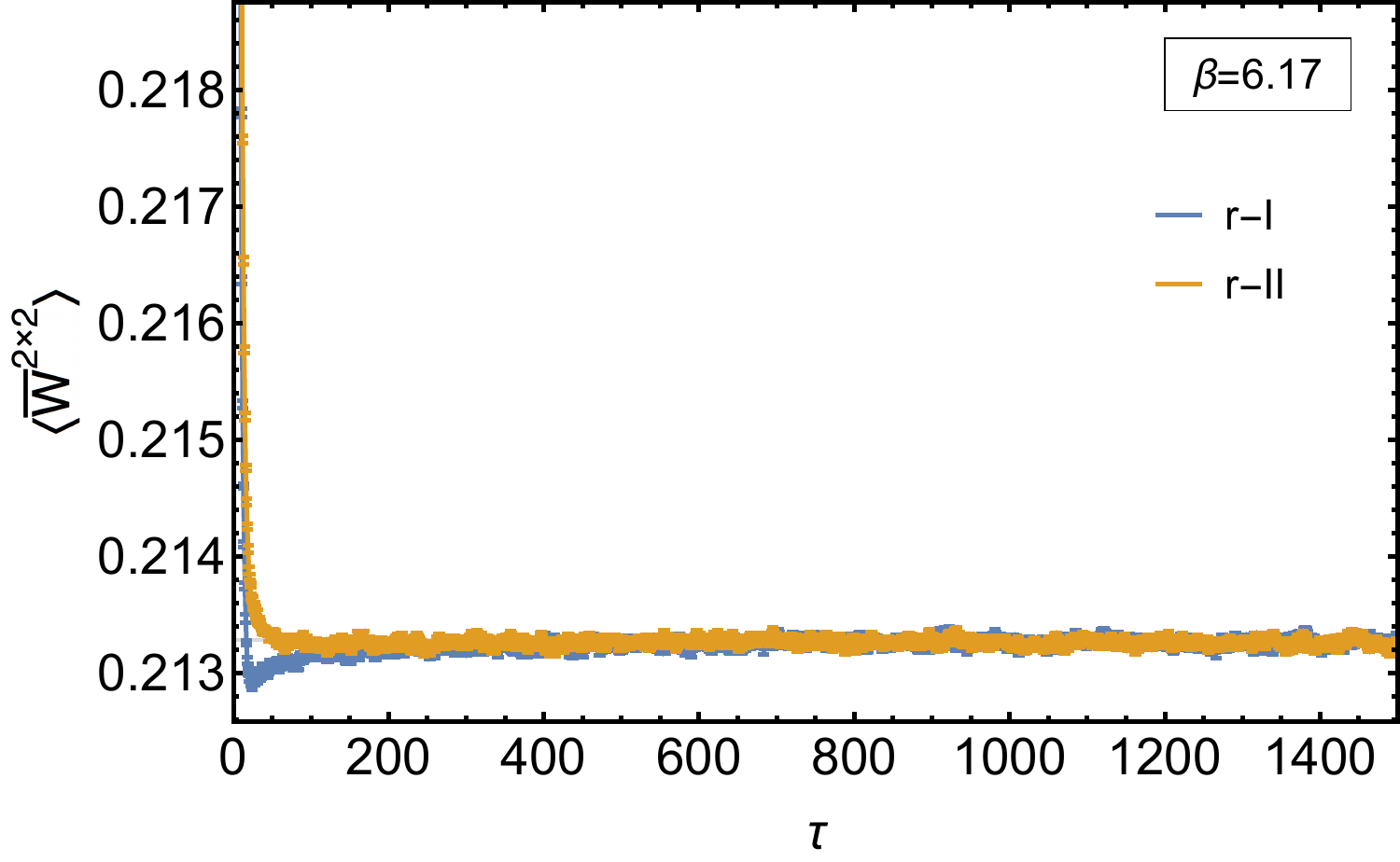}
\caption{\label{fig:hmc_retherm_disp}%
Top: HMC rethermalization of average displaced $2\times 2$ Wilson loops (r-I).
Center: HMC rethermalization of average displaced $2\times 2$ Wilson loops (r-II).
Bottom: HMC rethermalization of $2\times 2$ Wilson loops.
}
\end{figure}

We begin by considering the (re)thermalization properties of the average displaced $2\times 2$ Wilson loops, defined in \Eq{displaced_2x2}.
Recall from \Fig{partial_displaced_plaq} (right) that initially the displaced Wilson loops vary widely with the degree of displacement, $d$.
For example, at $\beta=6.17$, the average value ranges from approximately $0.2$ at $d=0$ to $0.8$ at $d=4$.
As illustrated in \Fig{hb_retherm_disp} (top, center) for HB and \Fig{hmc_retherm_disp} (top, center) for HMC, despite this wide initial variation, the displaced plaquettes converge to a single value for all $d$ after several Monte Carlo updates.
This holds true for both r-I (top) and r-II (center) ensembles, and indicates that the translational symmetry of the fine ensemble is restored rapidly as a function of the rethermalization time.
In the same figures, we show the total Wilson loop (bottom) obtained by an appropriately weighted average over the five displaced loops.
Here we see that the r-I ensemble ``overshoots'' the thermalized average, whereas the r-II ensemble converges more rapidly, and without overshooting.
Although further investigation is needed to better understand these differences, it is encouraging to see that the case r-II converges so well, given that that is the case where coarse ensembles had been generated (as they would be in practical applications) rather than produced artificially by restriction.

\begin{figure} 
\includegraphics[width=\figWidthHalf]{\figdir 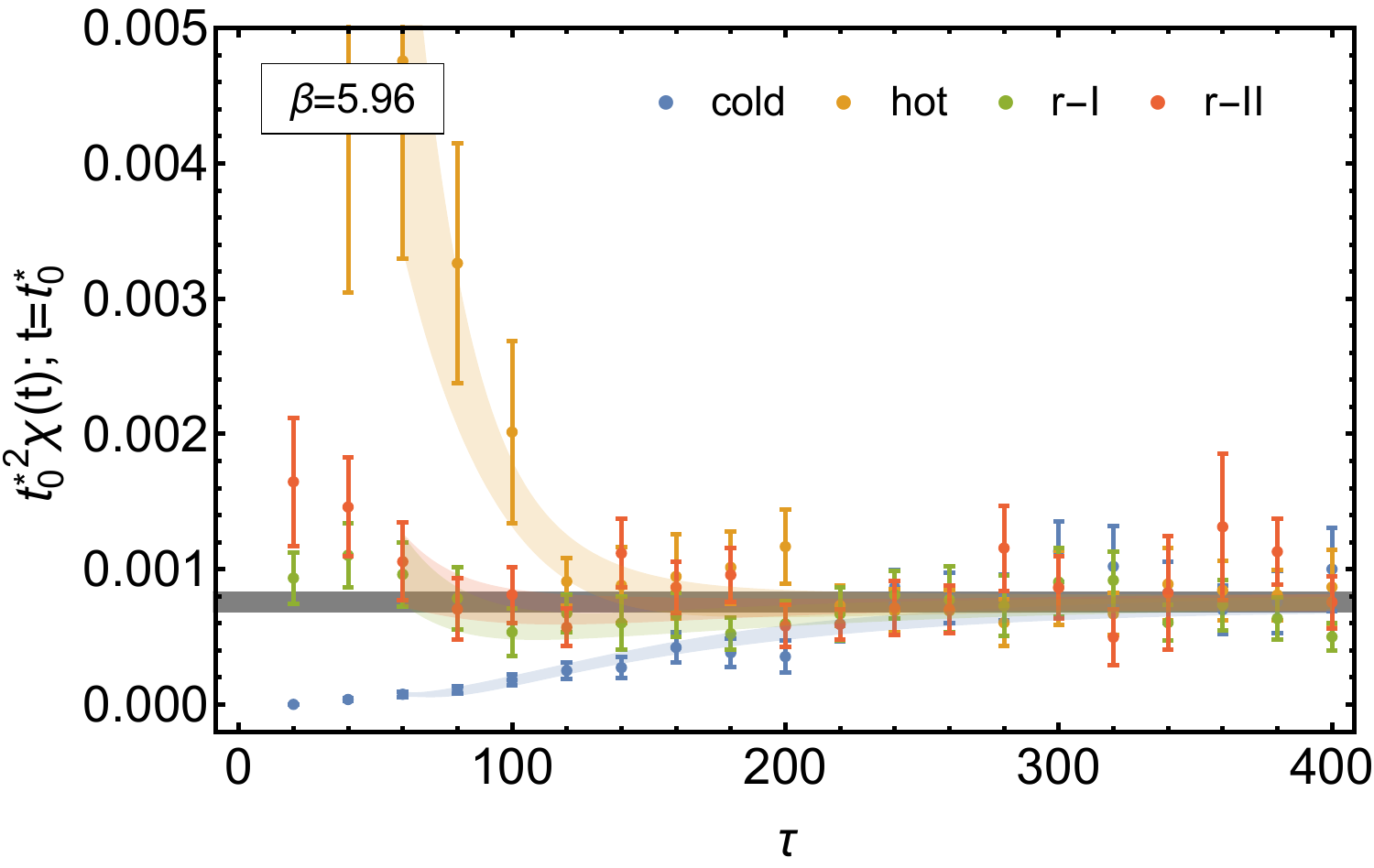}
\hspace{10pt}
\includegraphics[width=\figWidthHalf]{\figdir 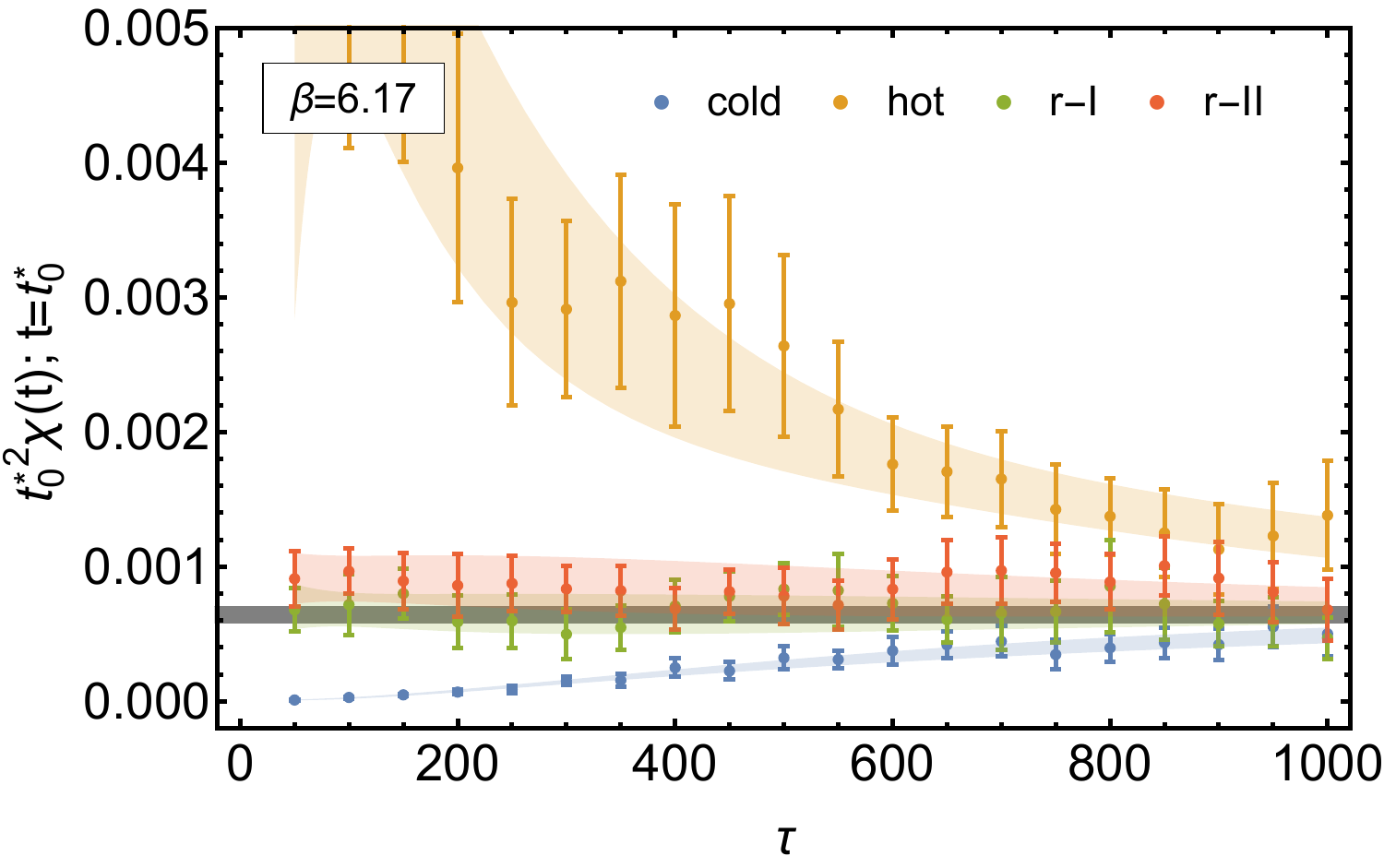} \\
\vspace{10pt}
\includegraphics[width=\figWidthHalf]{\figdir 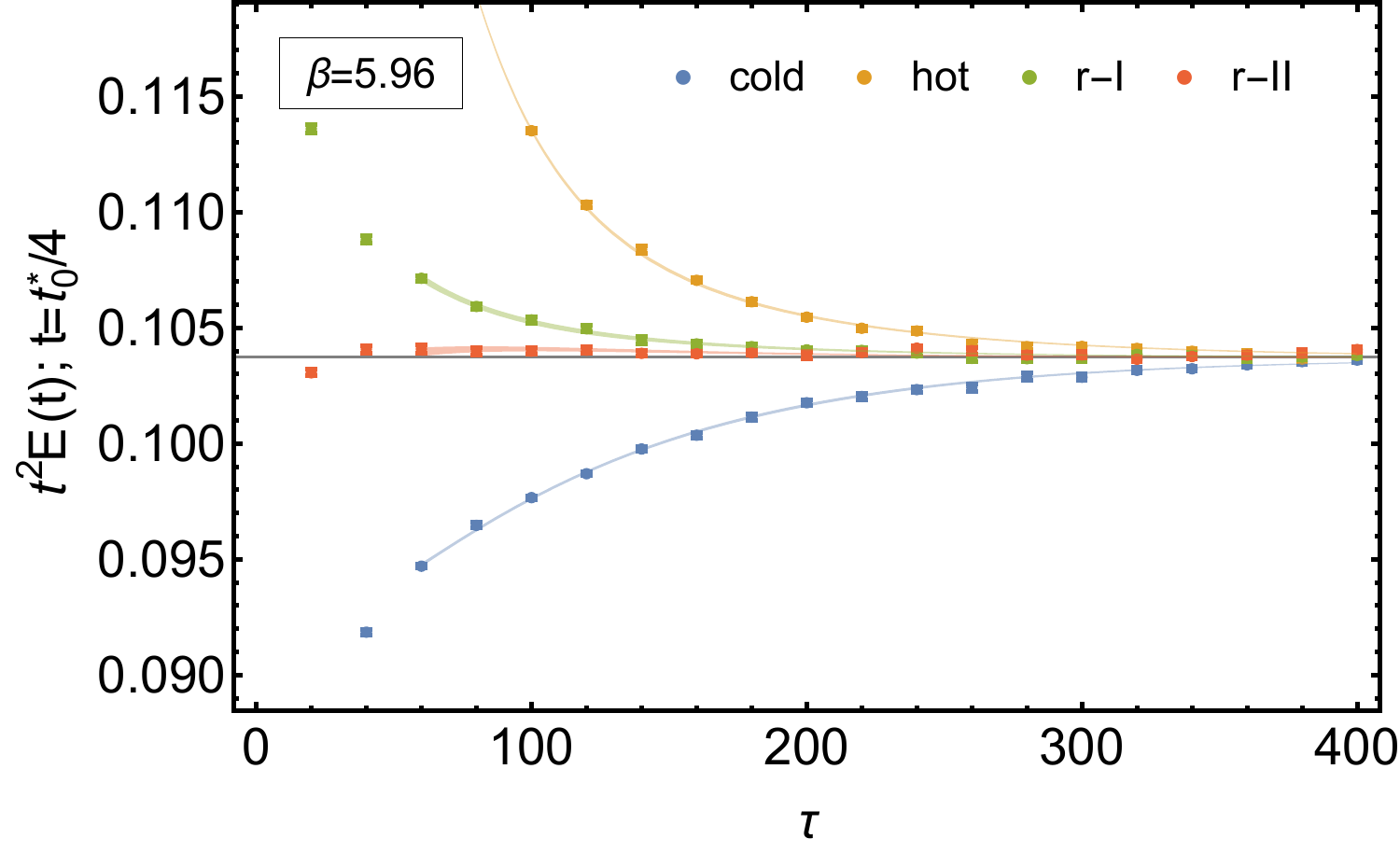}
\hspace{10pt}
\includegraphics[width=\figWidthHalf]{\figdir 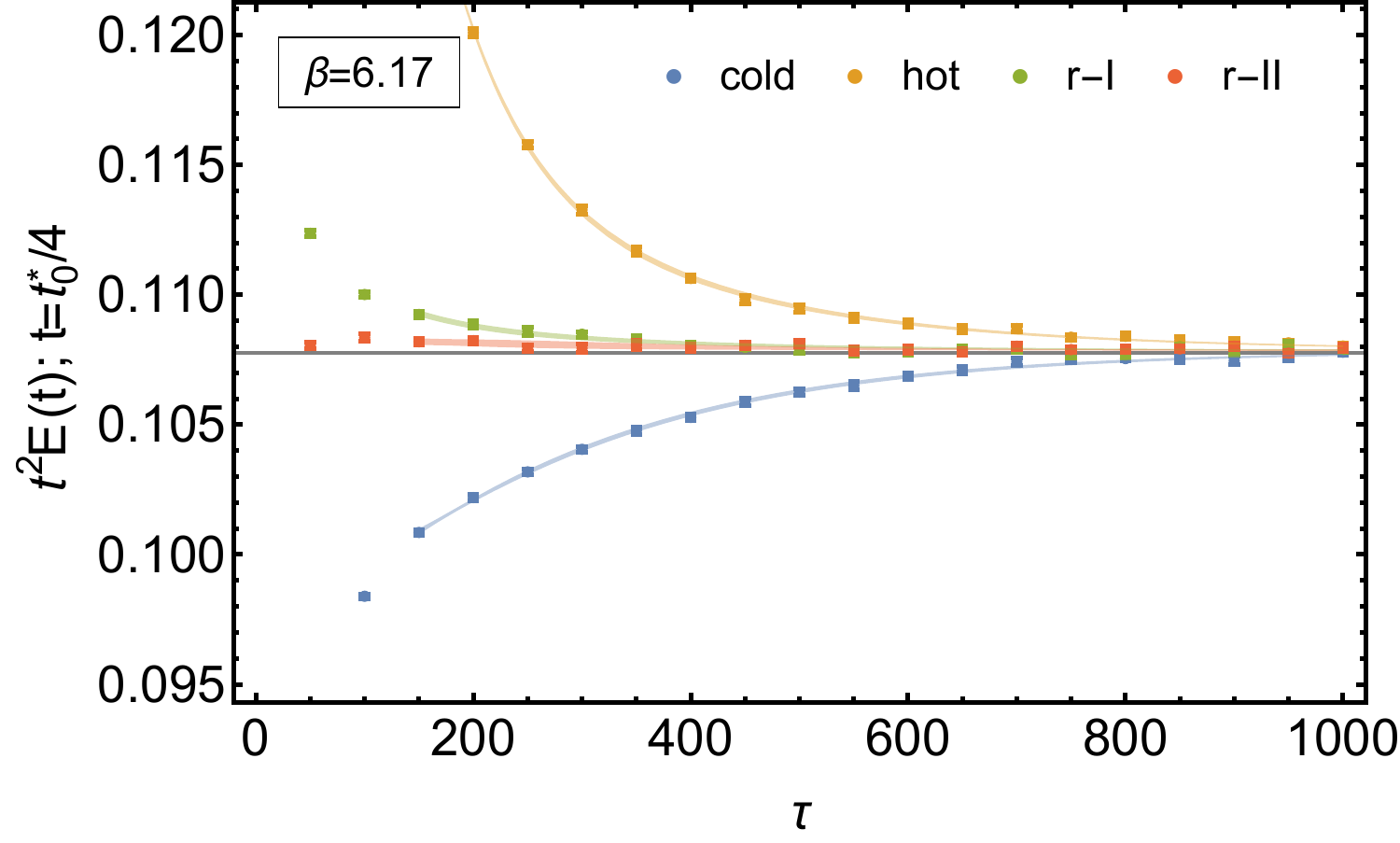} \\
\vspace{10pt}
\includegraphics[width=\figWidthHalf]{\figdir 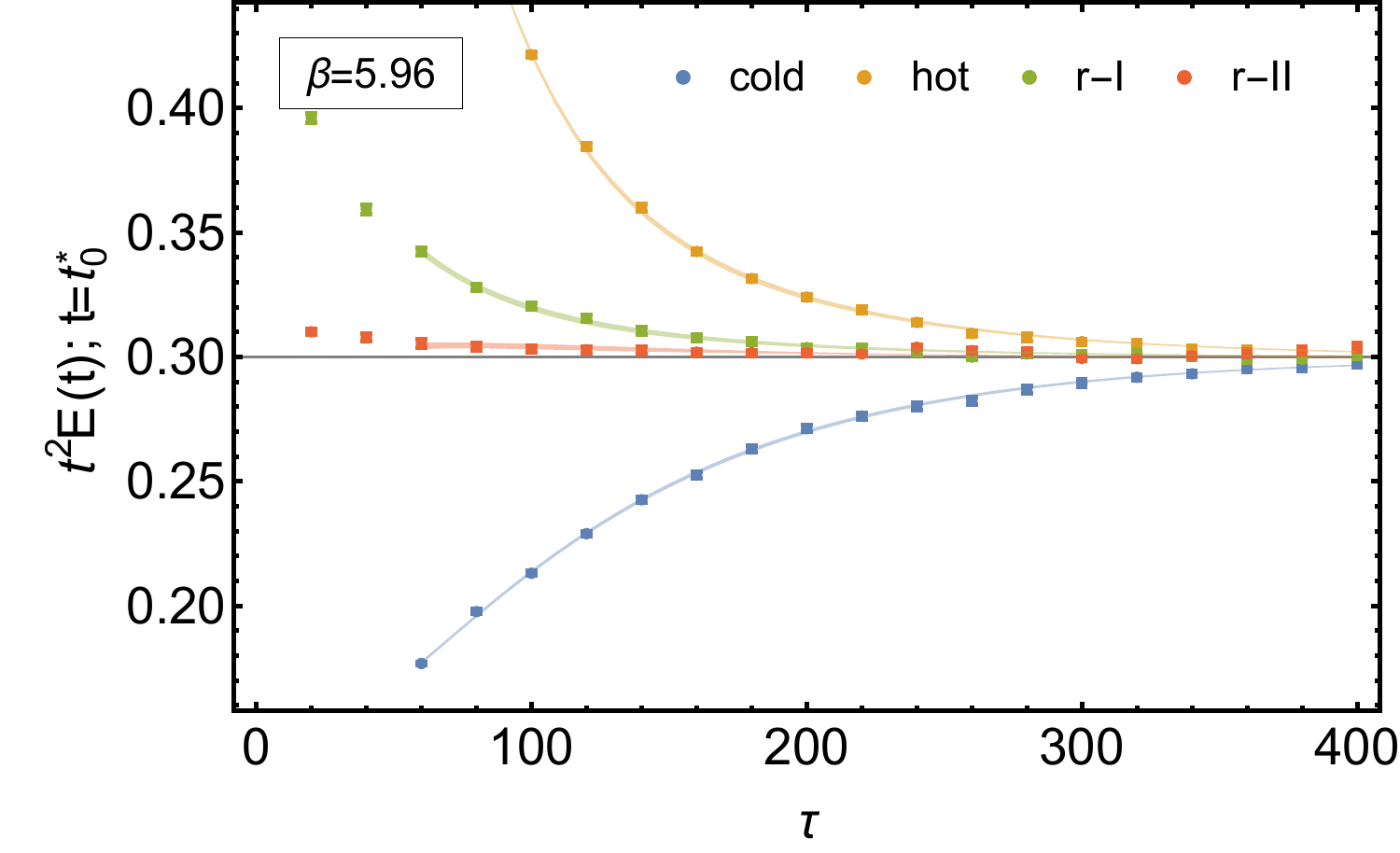}
\hspace{10pt}
\includegraphics[width=\figWidthHalf]{\figdir 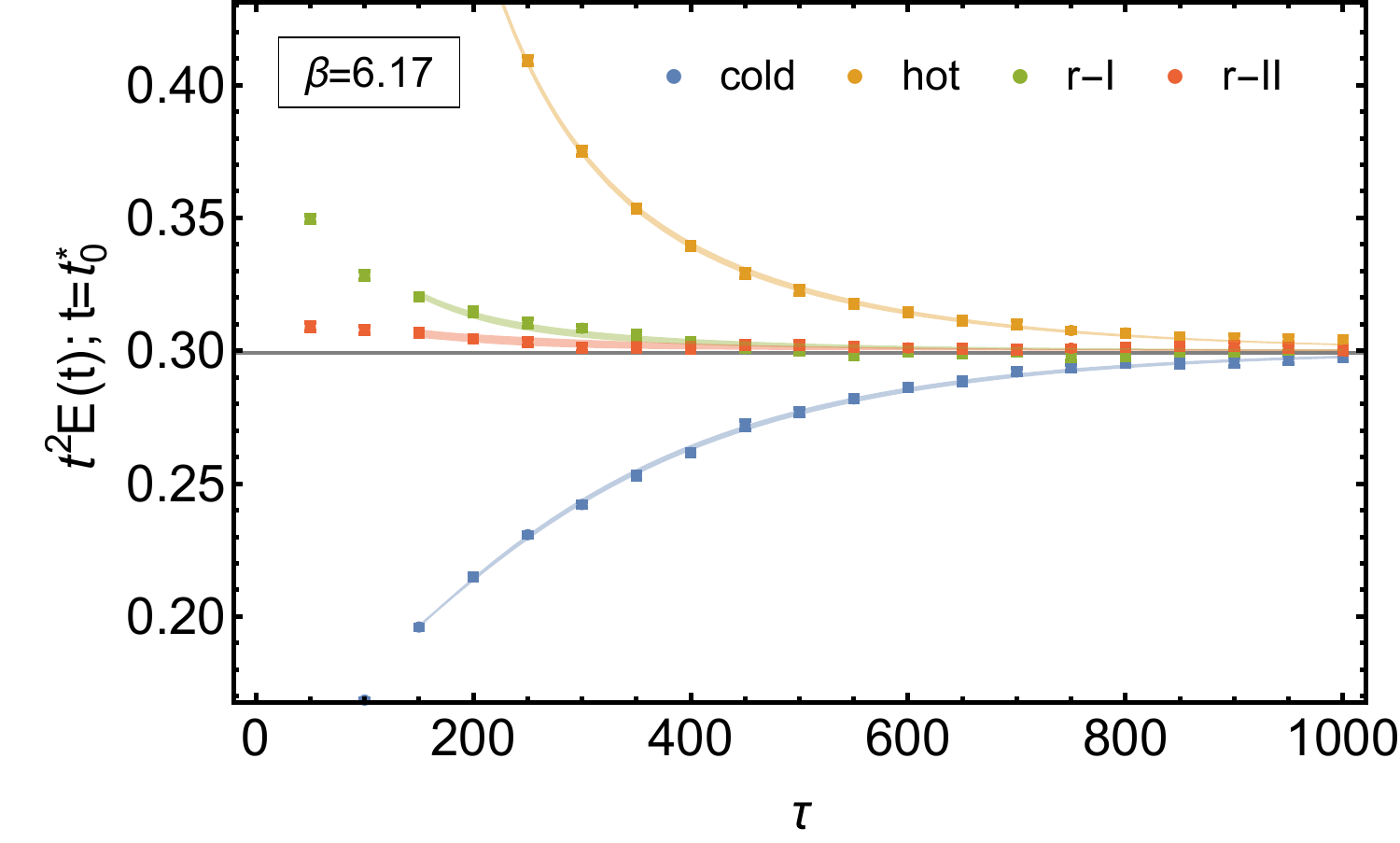} \\
\vspace{10pt}
\includegraphics[width=\figWidthHalf]{\figdir 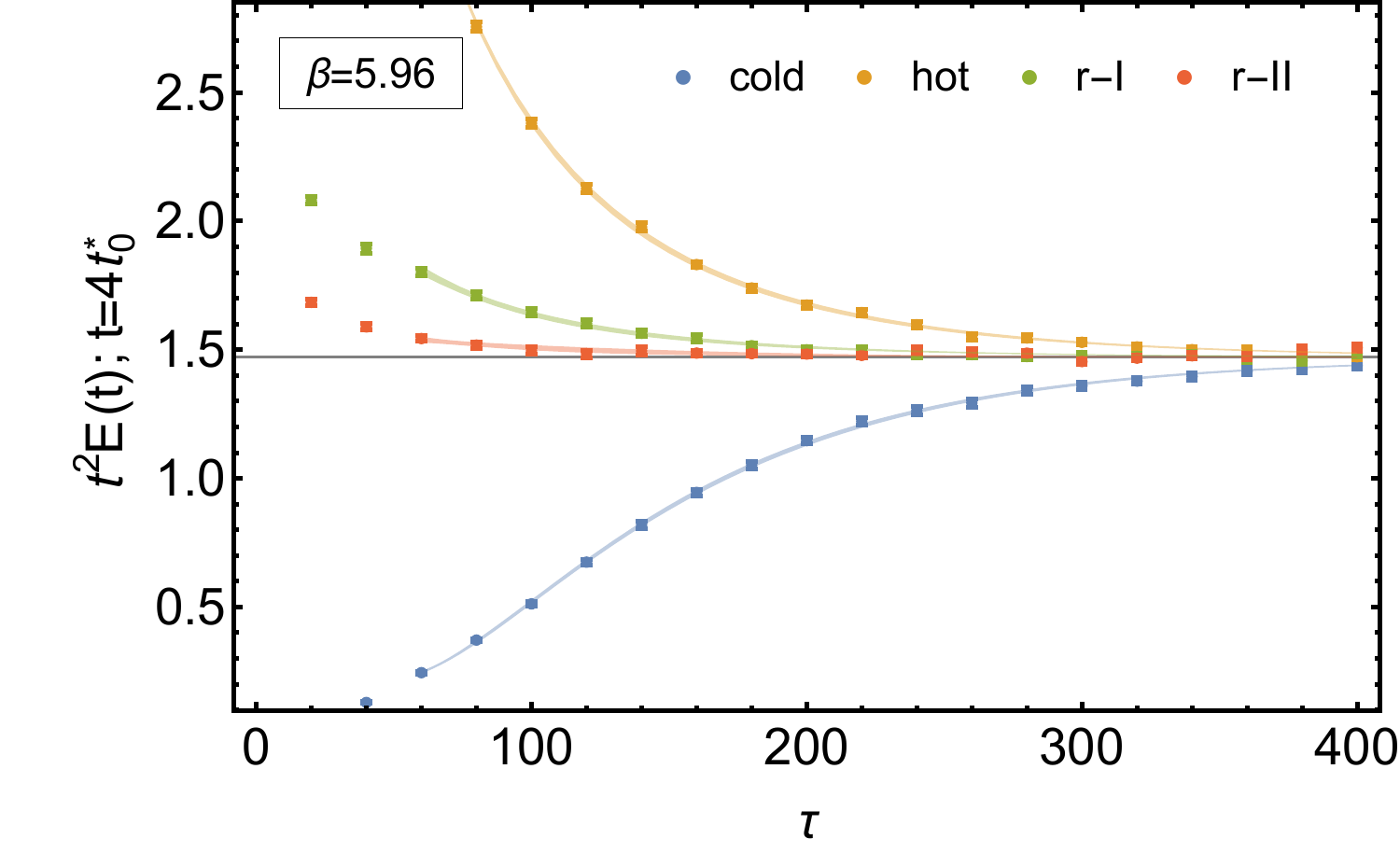}
\hspace{10pt}
\includegraphics[width=\figWidthHalf]{\figdir 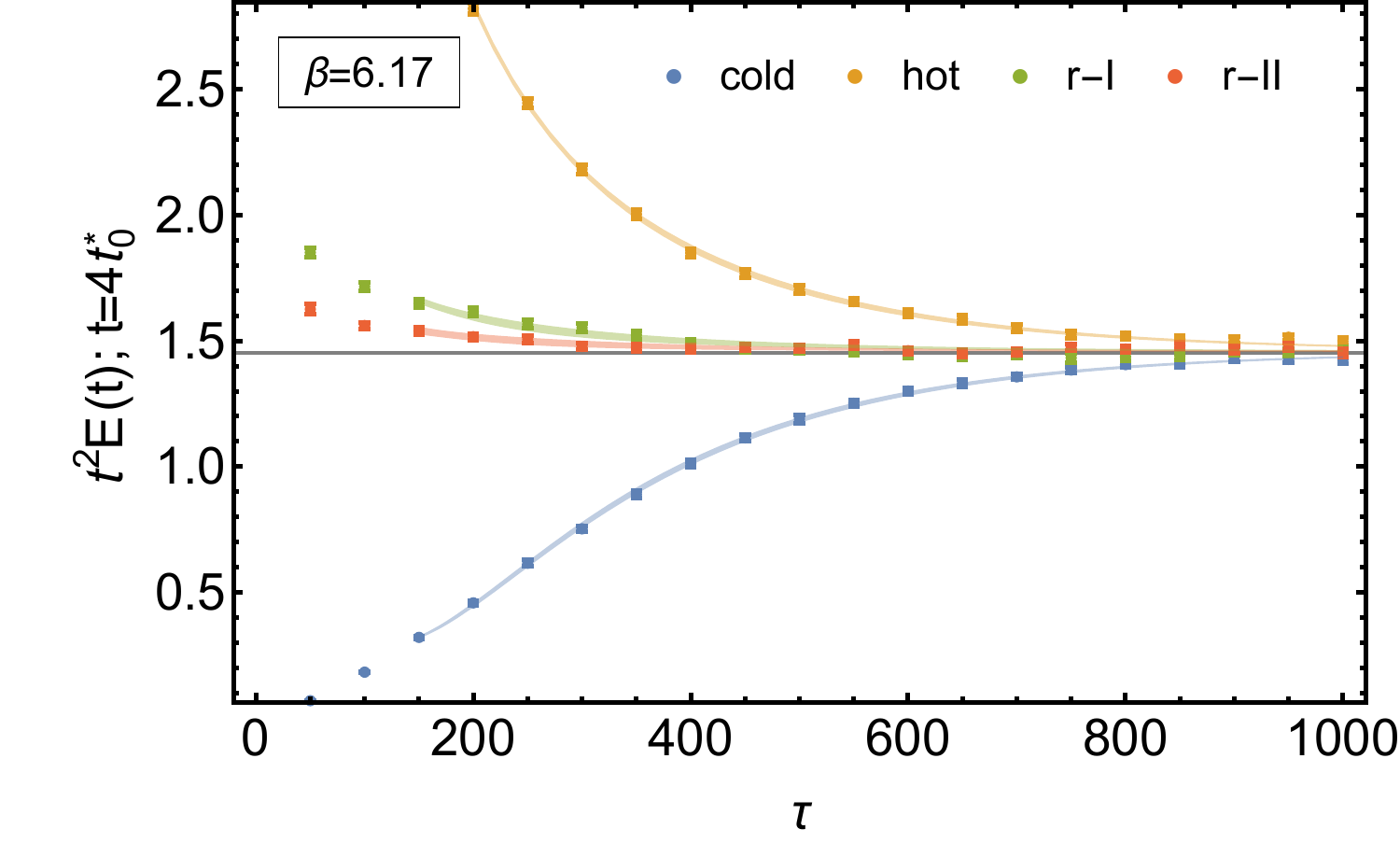}
\caption{\label{fig:purgaug_retherm}%
HB (re)thermalization of topological susceptibility and $E(t)$ as a function of the number of sweeps, for various values of $t$.
}
\end{figure}

\begin{figure} 
\includegraphics[width=\figWidthHalf]{\figdir 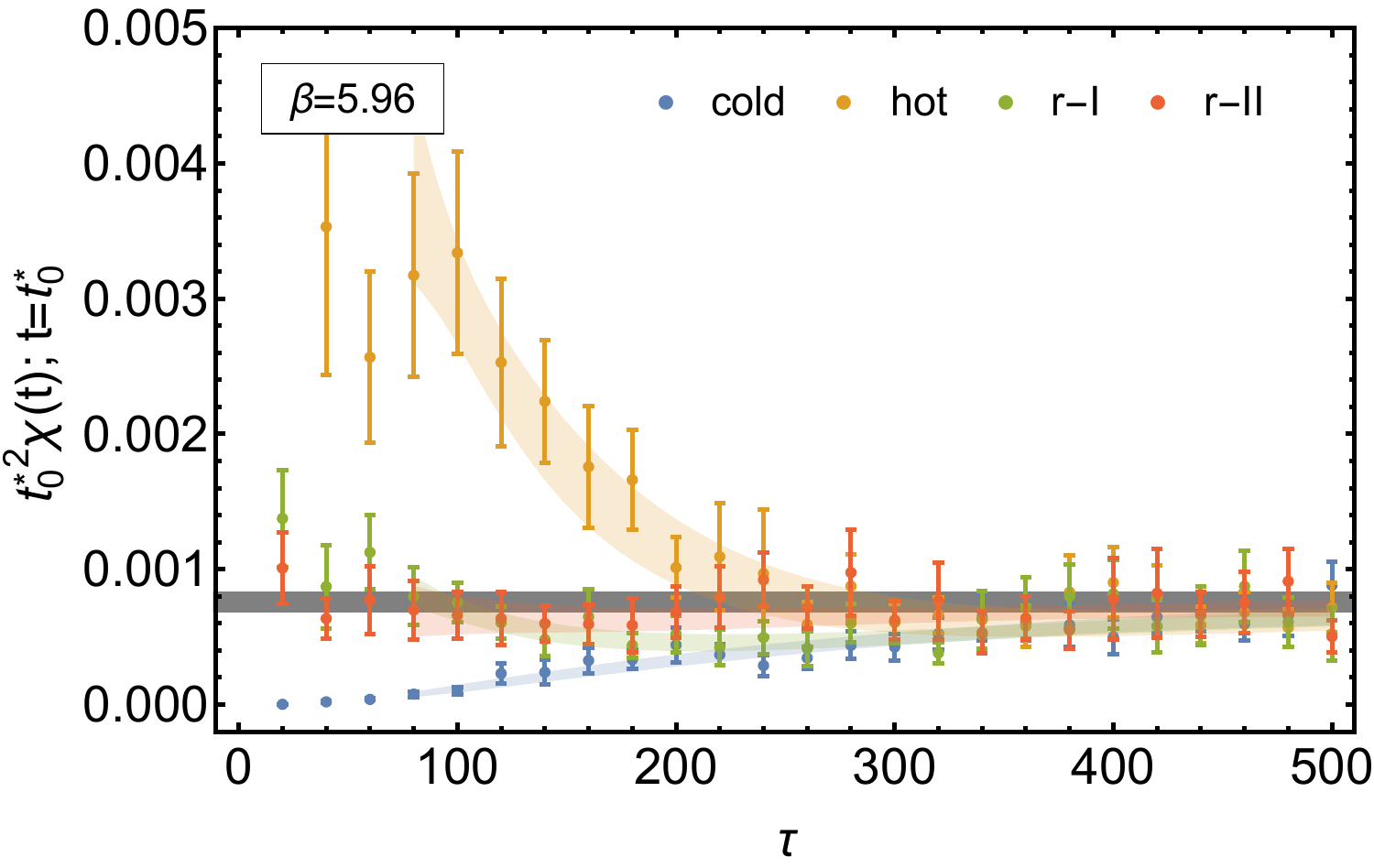}
\hspace{10pt}
\includegraphics[width=\figWidthHalf]{\figdir 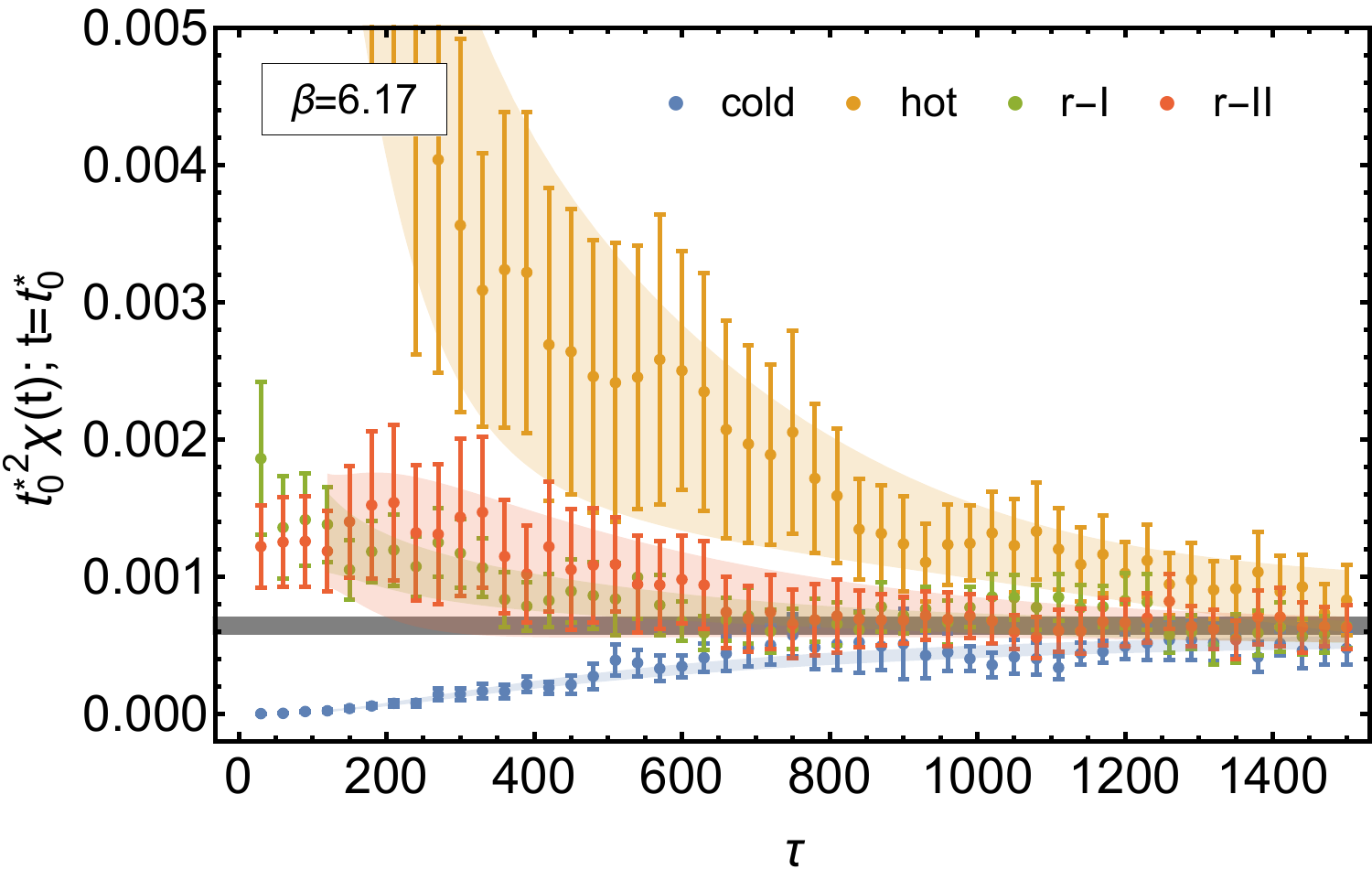} \\
\vspace{10pt}
\includegraphics[width=\figWidthHalf]{\figdir 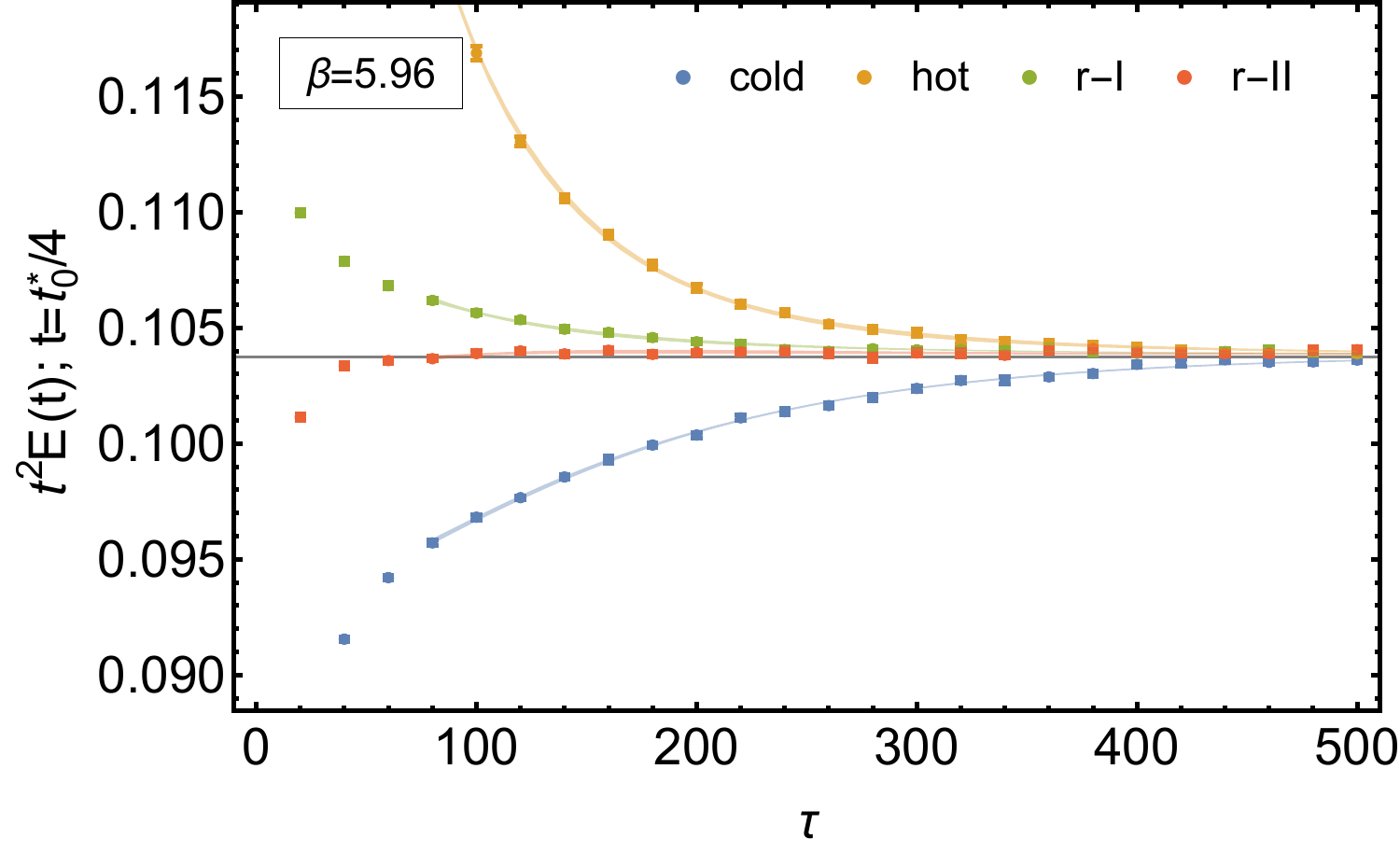}
\hspace{10pt}
\includegraphics[width=\figWidthHalf]{\figdir 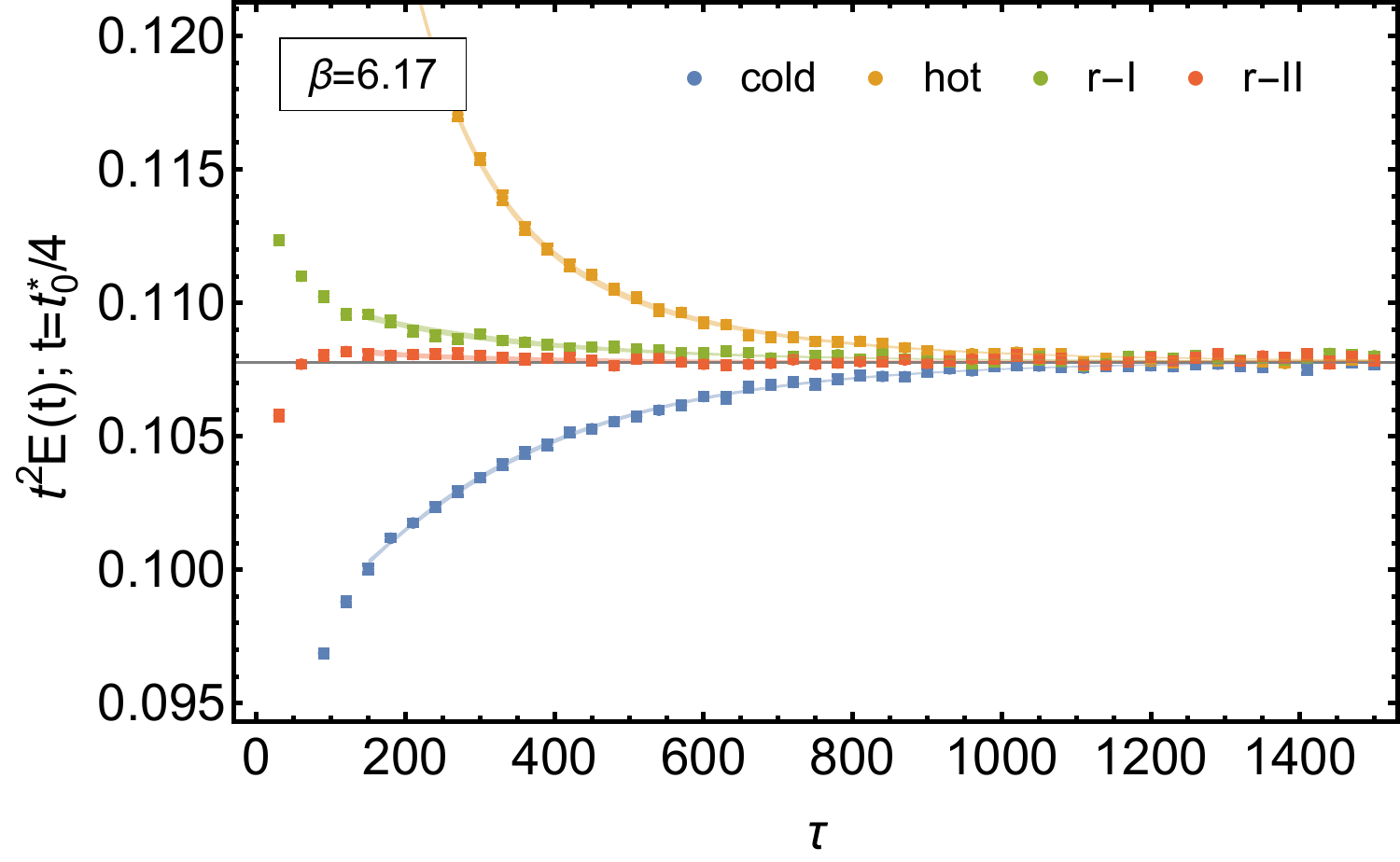} \\
\vspace{10pt}
\includegraphics[width=\figWidthHalf]{\figdir 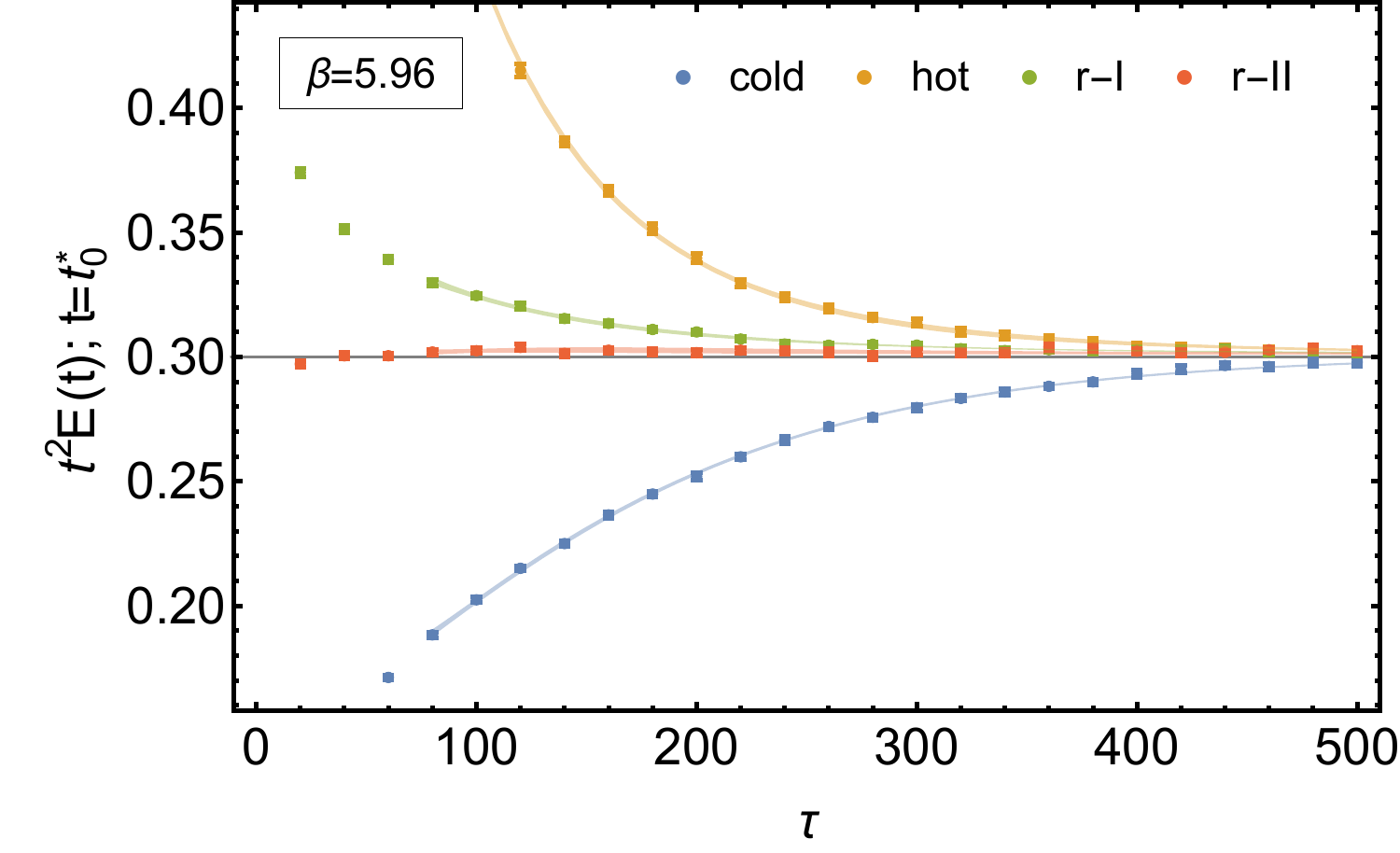}
\hspace{10pt}
\includegraphics[width=\figWidthHalf]{\figdir 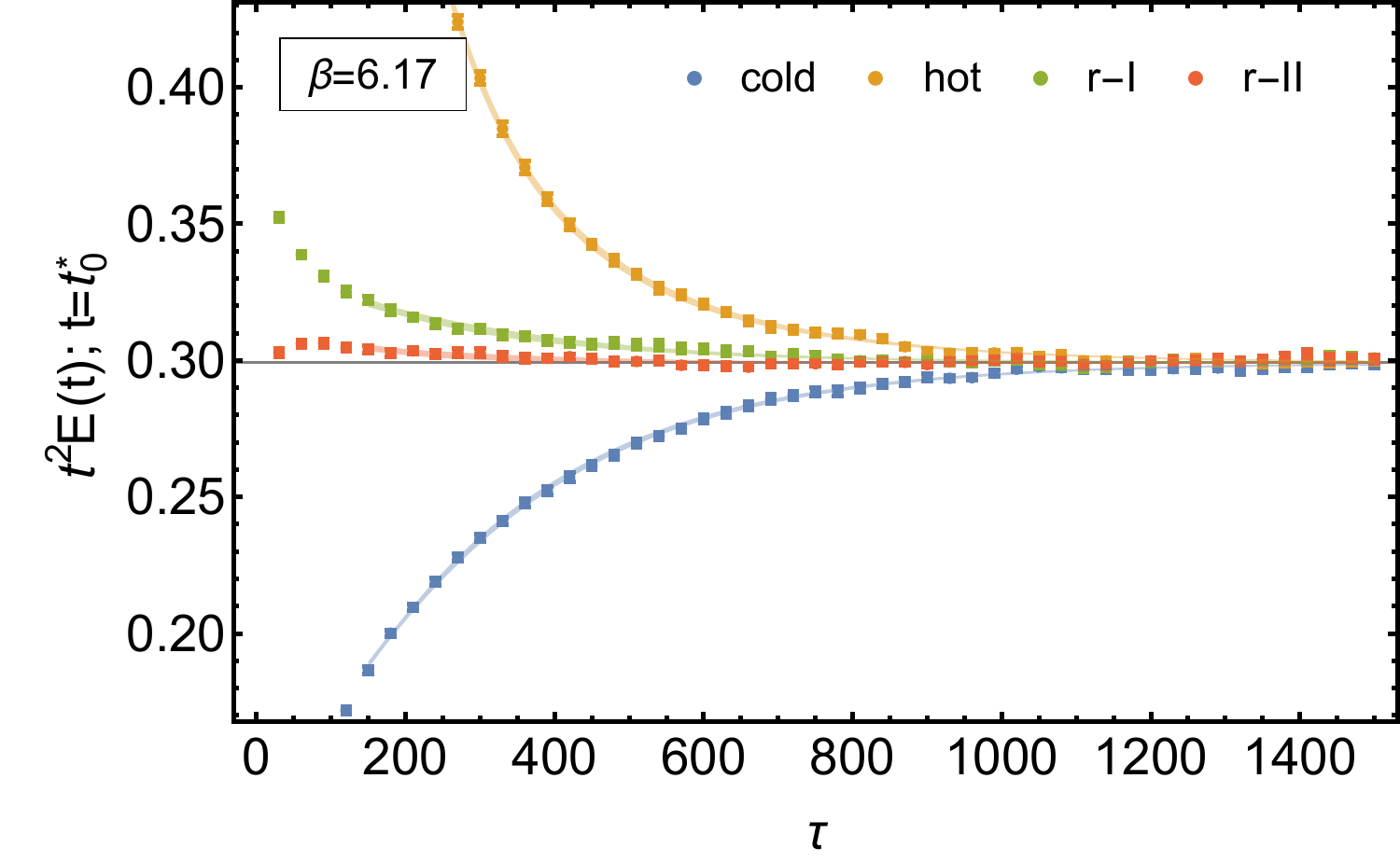} \\
\vspace{10pt}
\includegraphics[width=\figWidthHalf]{\figdir 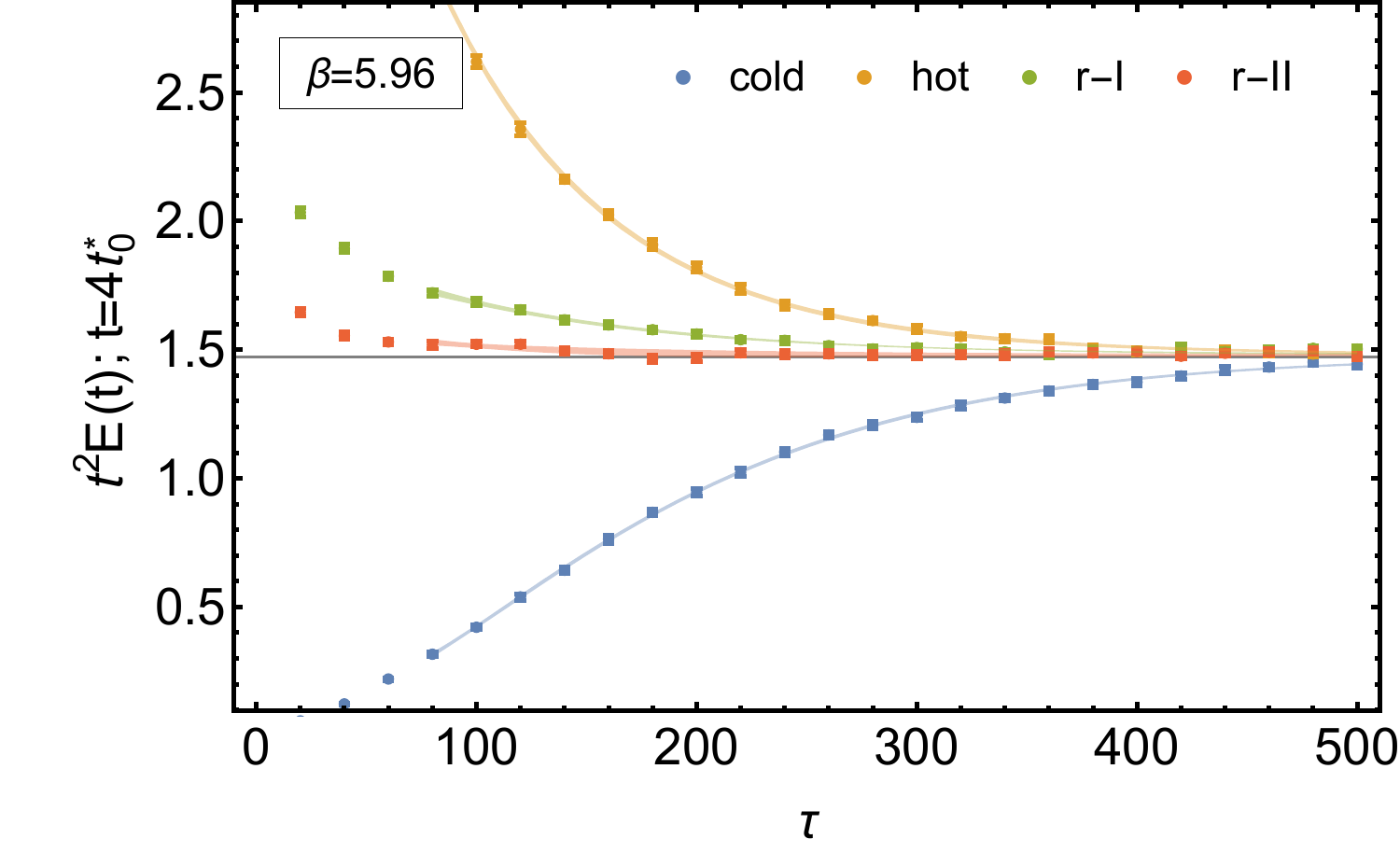}
\hspace{10pt}
\includegraphics[width=\figWidthHalf]{\figdir 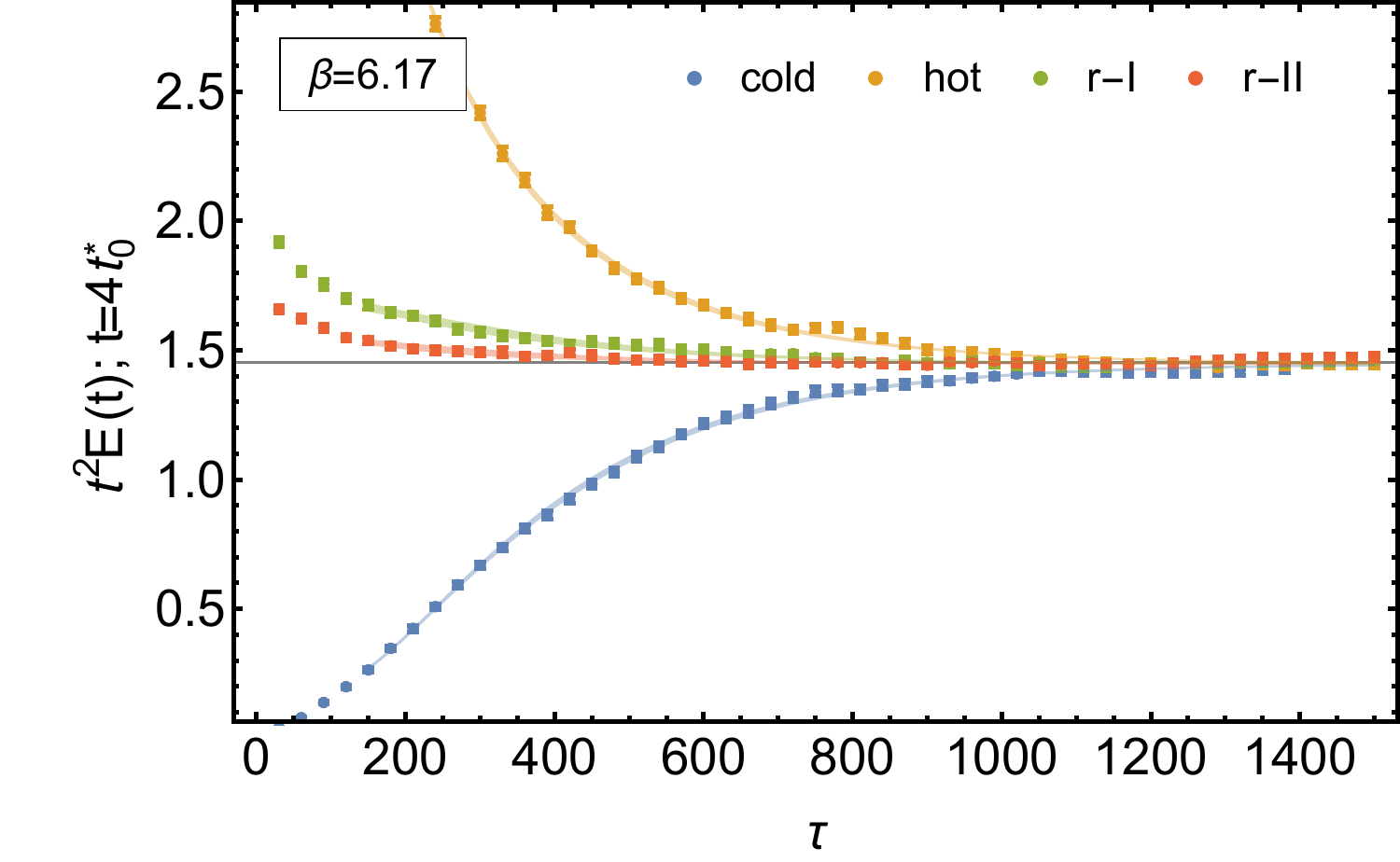}
\caption{\label{fig:hmc_retherm}%
HMC (re)thermalization of topological susceptibility and $E(t)$ as a function of the number of trajectories, for various values of $t$.
}
\end{figure}

In \Fig{purgaug_retherm} and \Fig{hmc_retherm}, we show (re)thermalization curves for the topological susceptibility, $(t_0^\star)^2\chi(t^\star_0)$, and Wilson flow quantity $t^2 E(t)$, for the flow times $t = t_0^\star/4, t_0^\star$ and $4 t_0^\star$.
Recall from the previous section that the topological charge distributions for these ensembles are well preserved upon prolongation.
Consequently, observables derived from topology, such as the susceptibility, are by construction thermalized up to lattice artifacts.
From studies of the topological susceptibility, this indeed appears to be the case.
Furthermore, for nontopological quantities measured on ensembles obtained by prolongation, we find that rethermalization times appear significantly shorter than the thermalization times of either hot and cold starts.

A quantitative comparison of the (re)thermalization times for each observable requires determination of both the exponents and the overlap factors for each (re)thermalization curve.
For each observable, we therefore perform a combined multiexponential fit to all four (hot, cold, r-I, and r-II) (re)thermalization curves, as a function of (re)thermalization time.
We include in these fits estimates of the observable obtained from the much larger thermalized target ensemble (therm), effectively corresponding to $\tau=\infty$.
We considered fits of the functional form
\begin{eqnarray}
f^\alpha(\tau) = \sum_{n=0}^{2} z^\alpha_n e^{-E_n \tau}\ ,
\end{eqnarray}
where $\alpha$ labels the ensemble, and impose the constraints $E_0=0$, $z^\alpha_0 = z_0$ for all $\alpha$ and $z^\textrm{therm}_n =0$ for $2\ge n>0$.
The least squares fits were performed using the variable projection method~\cite{0266-5611-19-2-201}.
A brief description of how the constraints were imposed in this approach is provided in \Appendix{varpro}.
Reliable correlated multiexponential fits to the data were difficult to achieve due to the small ensemble sizes; consequently uncorrelated fits to data were performed, with errors estimated via a bootstrap analysis.
Our aim is to obtain estimates of the relevant evolution time scales rather than precise values, so this simplified analysis is sufficient.
With larger ensembles and more frequent measurements, coupled fits to multiple observables could be performed, potentially constraining higher states in the evolution.
The leading exponents $\tau_n = 1/E_n$ ($n=1,2$) determined from each fitted observable are provided in \Tab{retherm_fit_results} at flow time $t_0$.
Note that generally $\tau_1 \neq \tau_\textrm{exp}$, since the observable under study may not couple strongly enough to the slowest mode, and furthermore, the statistics may be insufficient to resolve the effects of that mode.
Nonetheless, we expect the bound $\tau_\textrm{int} \le \tau_1$ to hold for each observable that is considered.
A comparison of these time scales can be made from the data provided in \Tab{autocorr_ensembles} and \Tab{retherm_fit_results}, and it suggests that this is indeed the case.

\begin{figure} 
\includegraphics[width=\figWidthHalf]{\figdir 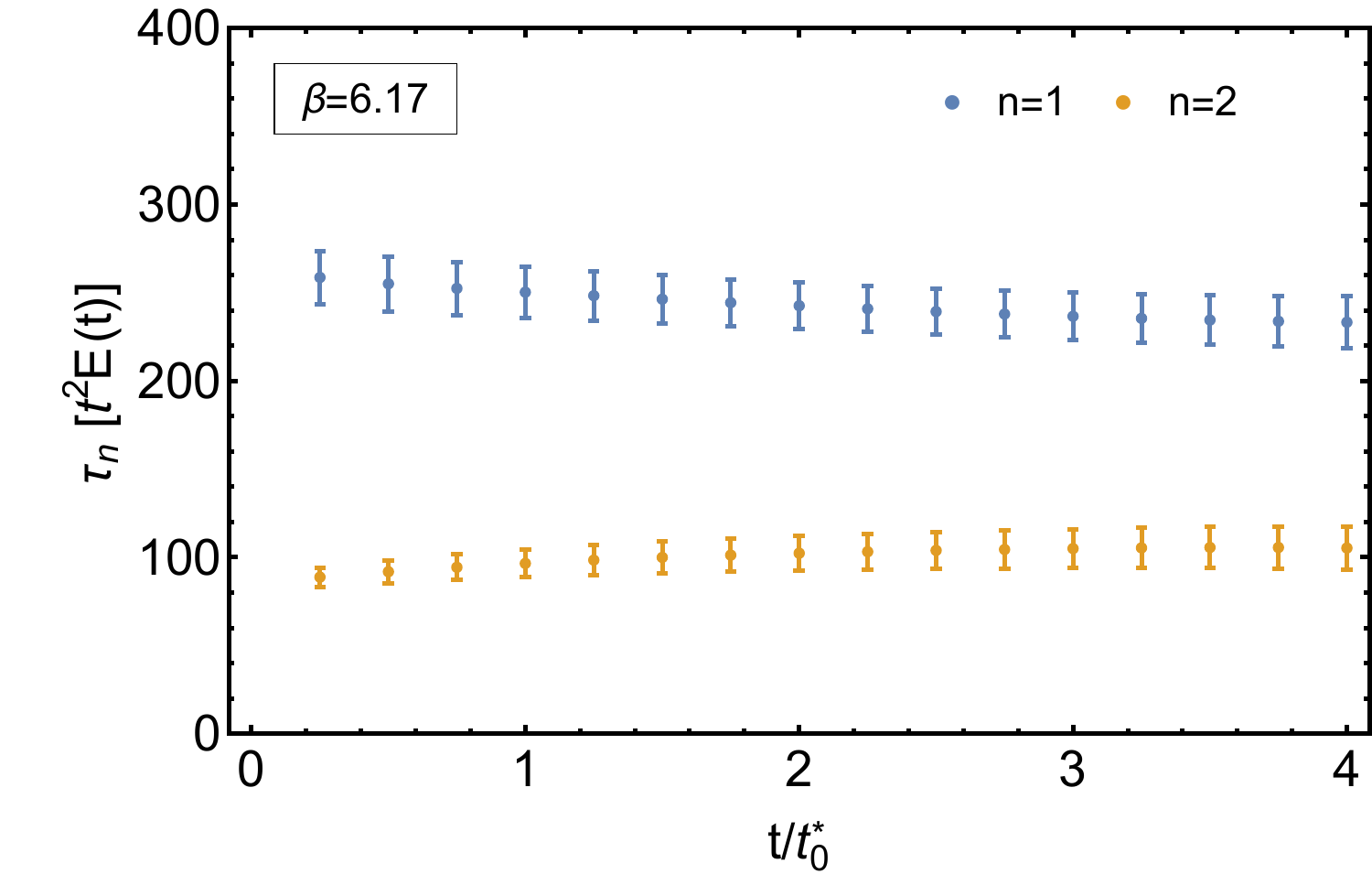}
\hspace{10pt}
\includegraphics[width=\figWidthHalf]{\figdir 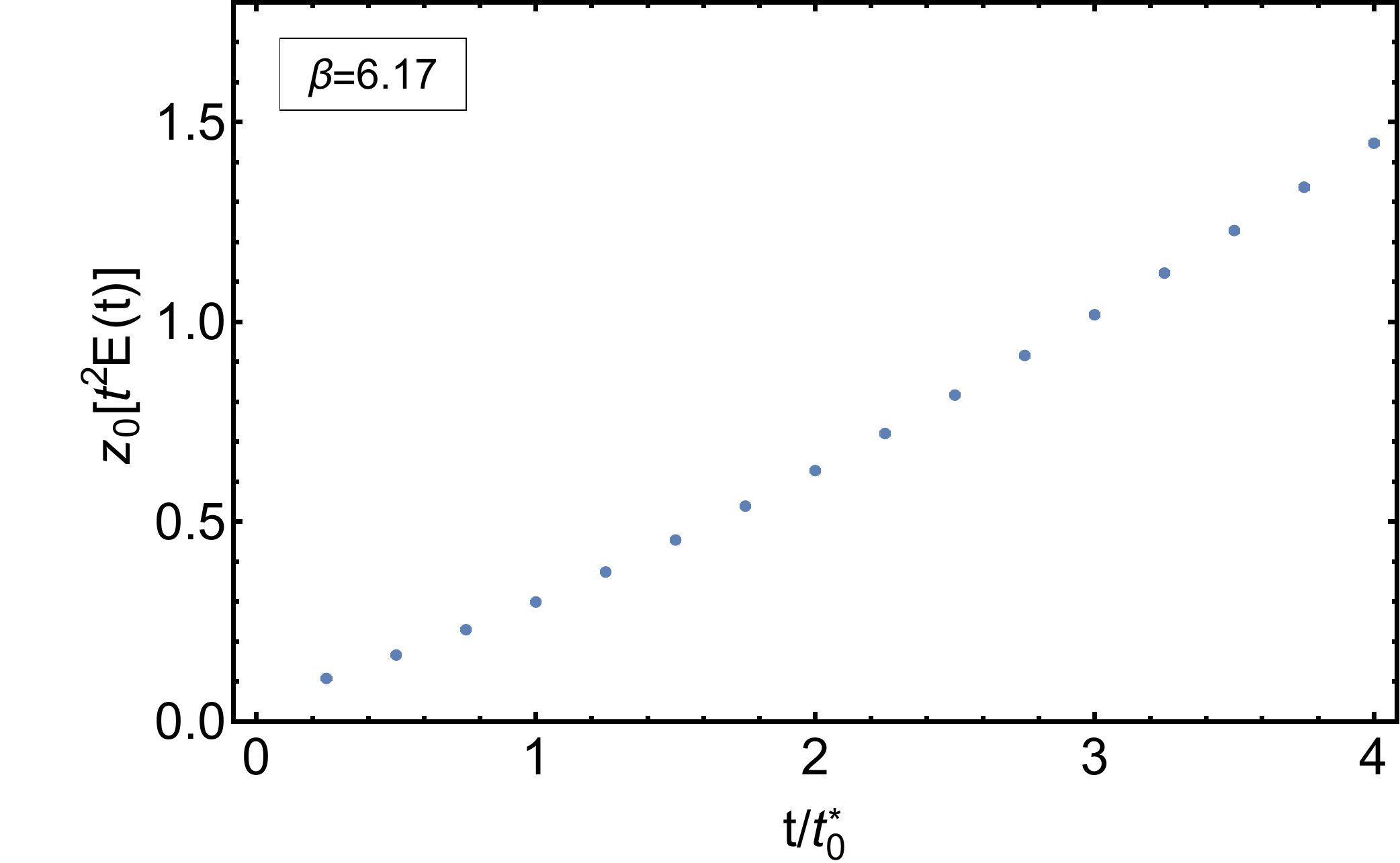}
\vspace{10pt}
\includegraphics[width=\figWidthHalf]{\figdir 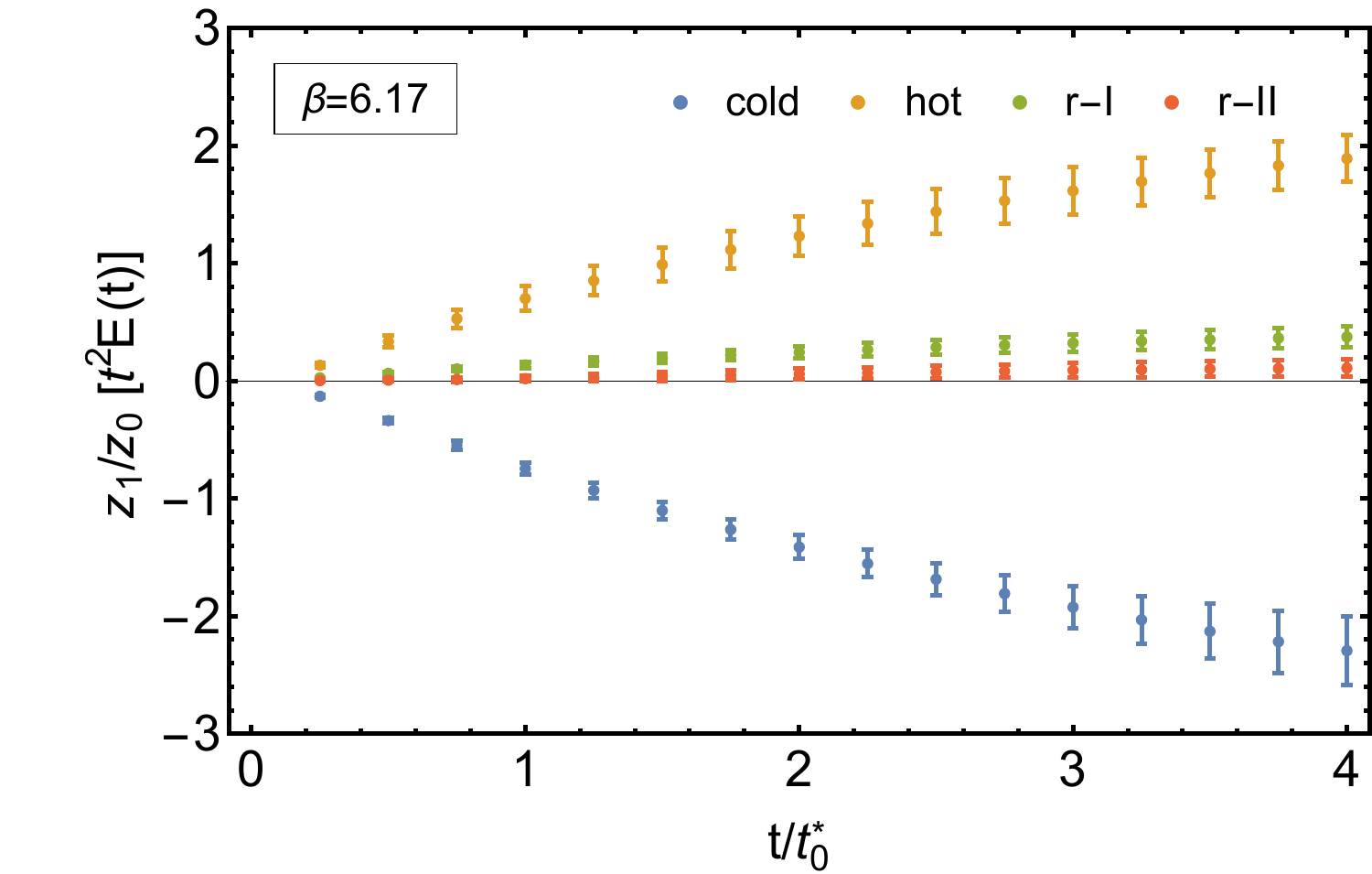}
\hspace{10pt}
\includegraphics[width=\figWidthHalf]{\figdir 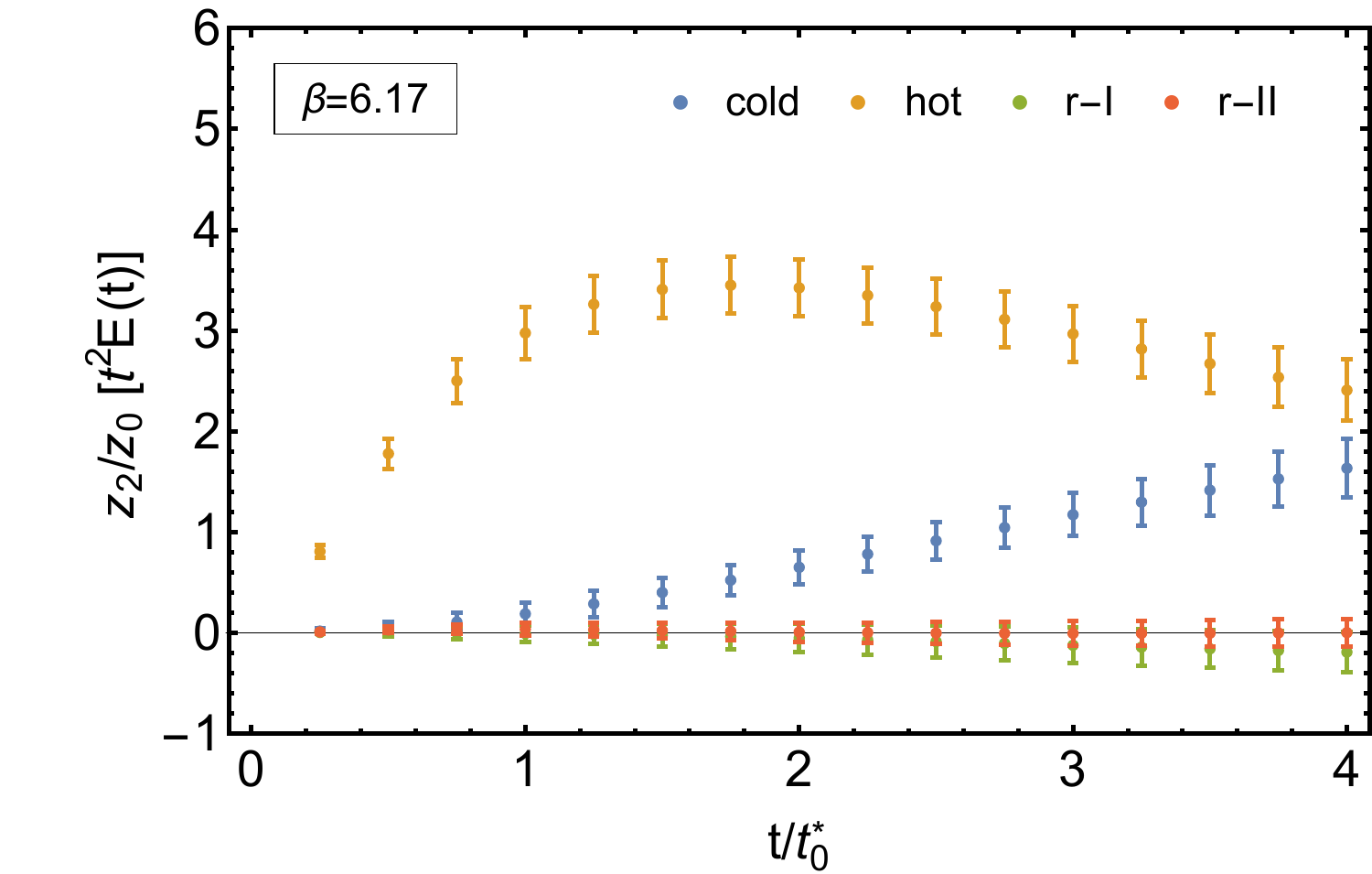}
\caption{\label{fig:hmc_retherm_e_fits}%
Multiexponential fit results (decay constants and overlap factors) for HMC (re)thermalization of $E(t)$ as a function of the number of trajectories, for various values of $t$.
}
\end{figure}

Finally, in \Fig{hmc_retherm_e_fits} we plot extracted values for $\tau_n$ and $z_n^\alpha$ as a function of the Wilson flow time for the observable $t^2 E(t)$, for $32^3\times 72$ ensembles (re)thermalized via HMC.
For this illustrative case, the fitted exponents are insensitive to the flow time, with $\tau_1/\tau_2 \sim 2.5$.
This stability suggests that the fits are picking out the true exponents governing the evolution dynamics.
The overlap factors, on the other hand, need not be independent of the flow time.
For flow times $t/t_0^\star>1$ we find $z_1/z_0 \gtrsim \calO(1)$ and $z_2/z_0 \gtrsim \calO(1)$ for hot and cold ensembles.
The ensemble r-I has significantly reduced overlaps by comparison, with $z_2/z_0$ consistent with zero over the full range of flow times.
The ensemble r-II exhibits the most impressive behavior, with both $z_1/z_0$ and $z_2/z_0$ consistent with zero over the full flow time range.
The result suggests a lower bound on the rethermalization time scale, given by $\tau_3$.
It would be particularly interesting in this example to determine with higher precision the number of low lying states that have been eliminated, thus further constraining this bound.

From analysis of the autocorrelation times in the preceding section, and the (re)thermalization time scales determined here, we may draw several conclusions.
First, the rethermalization times for prolongated ensembles are significantly shorter than the thermalization times for hot and cold starts.
This result implies that the simulation strategy advocated in \Fig{mcDiagram} (b) is more efficient than that of \Fig{mcDiagram} (a).
Second, the rethermalization times for nontopological long-distance observables are significantly shorter than the decorrelation time scale for fine evolution, which is bounded from below by twice the integrated autocorrelation time for topological charge.
An explicit comparison of these time scales can be made from \Tab{autocorr_ensembles} and \Tab{retherm_fit_results}.
For example, in the case r-II (HMC, $a\sim0.07$ fm), where $\tau_3$ (which was undetermined from fits) appears to be the dominant rethermalization time scale, rethermalization to quarter-percent levels can be achieved in the time $\tau_\textrm{retherm} \sim  6 \tau_3 < 6 \tau_2 \sim 600$.
The decorrelation time for the topological charge provides a lower bound $2\hat \tau_\textrm{int} > 2 \tau(Q(t^*_0)) \sim 2600$, and therefore the efficiency of the algorithm, as described by \Fig{mcDiagram} (c), is conservatively estimated to be greater than $2\hat \tau_\textrm{int}/\tau_\textrm{retherm} \sim 4$.\footnote{This accounting does not include any additional reduction in computational cost attributed to reduced communication overhead from having multiple streams.}
The result provides compelling evidence that an ensemble generation strategy along the lines of \Fig{mcDiagram} (c) is not only viable but also a superior alternative to approaches presently available.
Note that this assessment becomes particularly conclusive in the regime of ultrafine lattice spacings, where proper sampling of the topological charge is presently impractical by conventional means due to topological freezing.
As mentioned in \Sec{introduction}, the two strategies described in \Fig{mcDiagram} (b) and \Fig{mcDiagram} (c) represent only the extremes in a range of algorithms, defined by different choices of $N_s$ and $N_e$.
In general, the optimal choice for these parameters depends on the time scales observed for a particular target action and the computational facilities that are available.

\begin{table}
\caption{%
\label{tab:retherm_fit_results}%
Fit results for (re)thermalization curves.
}
\begin{ruledtabular}
\begin{tabular}{cccccccc}
Algorithm & Lattice & $\tau$ range & $\tau_1\left( E(t^\star_0) \right)$ & $\tau_2\left( E(t^\star_0) \right)$ & $\tau$ range & $\tau_1\left( \chi(t^\star_0) \right)$ & $\tau_2\left( \chi(t^\star_0) \right)$ \\
\hline
HB  &  $24^3\times 48$ &  60-400  & 89.1(2.3)   & 32.5(1.2) &  60-400  & 95.9(49.9)   & 24.9(8.9)   \\
    &  $32^3\times 72$ & 150-1000 & 219.3(7.6)  & 72.2(4.5) &  50-1000 & 623.6(84.9)  & 140.4(149.6)   \\
HMC &  $24^3\times 48$ &  80-500  & 115.6(4.0)  & 53.2(5.4) &  80-500  & 187.5(117.3) & 86.9(33.0) \\
    &  $32^3\times 72$ & 150-1500 & 250.3(10.8) & 96.5(6.2) & 120-1500 & 511.2(227.9) & 83.8(52.1)  \\
\end{tabular}
\end{ruledtabular}
\end{table}

\section{Conclusion}

We have proposed a multiscale equilibration strategy for Yang-Mills gauge theories, which can be used to rapidly initialize large numbers of Monte Carlo streams, thereby increasing the efficiency of simulations.
This algorithm shares many features with multigrid solvers~\cite{Babich:2010qb,Babich:2009pc,Frommer:2013fsa,Brannick:2014vda} which have been used to dramatically decrease the computational cost of matrix inversion, a large component of lattice QCD calculations.
The effectiveness of our multilevel strategy for equilibration was demonstrated for the case of pure $SU(3)$ gauge theory using both heat bath and hybrid Monte Carlo updating procedures.
In both cases, the time scales governing the rethermalization of prolongated ensembles were measured to be considerably shorter than the decorrelation times for conventional evolution, as estimated from the autocorrelations in the topological charge.
Furthermore, the ratio of these quantities decreases parametrically as the continuum limit is approached.
As a consequence, prolongation of a RG-matched coarse ensemble followed by rethermalization provides a new way to reduce critical slowing down in lattice gauge theory simulations.
In particular, the poor sampling of topology at fine lattice spacing is ameliorated by evolving multiple fine-action streams derived from a coarse-action ensemble with well-sampled topology.
Although unexplored in this study, multiple levels of refinement would offer additional speed-ups in thermalization.

The successful application of our strategy requires a nonperturbative real space renormalization group procedure to match the physical scales at the coarse and fine levels of evolution.
Generally the matching need not be precise to realize improvements since the subsequent evolution (rethermalization) eliminates any effects of the mistuning of the coarse action.
However, the precision with which the tuning is carried out will influence the rethermalization times of the prolongated ensembles.
A RG transformation of the fine lattice action induces many operators in the coarse action, and these should, in principle, be included in the coarse evolution.
In this study, we have ignored all but the local plaquette coupling in the coarse action, yet still attain impressively short rethermalization times.
Numerical methods have been developed for nonperturbatively determining the induced couplings along a RG flow for simple systems (see, e.g.,~\cite{Hasenbusch:1994ne}), and their use in tuning would likely result in a further reduction in rethermalization times.

From a practical standpoint, the utility of our approach can be realized in several ways.
The method can be directly applied toward generation of very large physical volume ensembles in cases where the matching is already known (e.g., from previous small volume studies).
In this way one can avoid the long initialization time for large lattice streams.
Our expectation is that for a given target lattice spacing, rethermalization times for prolongated ensembles will be insensitive to volume, and therefore the efficiency of the algorithm will be unaffected by the volume scaling.
Alternatively the method can be used to start random ensembles on coarse lattices to initialize a large ensemble of fully independent streams.
These streams will start with different topological charges so that together they will sample the topological charge distribution dictated by the coarse lattice.
In principle, the topological charge distribution could be reweighted if the continuum distribution is determined by some other means.

A potential weakness of our multilevel approach is that its effective use requires prior knowledge of the RG matching of actions, which in turn requires simulations at the fine level.
Nevertheless, there are several ways in which to proceed.
One can exploit the fact that the tuning need not be exact at the coarse level and perform ``sloppy matching'' studies using small volume ensembles with poorly sampled topology; in this case matching would proceed by considering long distance observables which are relatively insensitive to topology.
Another possibility is to use a finite volume scheme such as the Schr{\"o}dinger functional~\cite{Luscher:1996ug}  to define the gauge coupling.
By requiring a fixed coupling constant for a coarse and fine ensemble, one can then obtain a matching of the bare gauge couplings that result from calculations at two lattice spacings with the desired ratio.
Available results that could be used to obtain matched coarse and fine ensembles, including dynamical fermions, can be found in a recent review~\cite{Sommer:2015kza}.
Finally, in the ultrafine limit, one can make contact with perturbative calculations in order to carry out the matching.

The multilevel methods described here naturally extend to simulations of full QCD and, more generally, gauge theories with matter fields.
The presence of fermions has no impact on the details of the prolongation.
The only additional ingredient is that the RG matching must be performed for more than one physical scale (e.g., $\Lambda_\textrm{QCD}$, the pion mass and the kaon mass for $2+1$ flavor QCD), and therefore requires tuning of multiple parameters in the gauge and Dirac actions.
In cases such as this where multiple physical scales are present, rapid thermalization with multiple levels of refinement may be particularly advantageous.
Perfect action constructions \cite{Hasenfratz:1993sp} may be useful in this regard.

Although the multiscale algorithm presented here provides an efficient means for thermalization, it is important to draw a distinction between it and a more ambitious multigrid Monte Carlo dynamics.
The latter implements a fully recursive evolution, including multiple scale evolution, while maintaining exact detailed balance on the finest level.
While there has been some success in constructing such multiscale methods for simpler field theories~\cite{Goodman:1986pv,Edwards:1990hu,Edwards:1991eg,Grabenstein:1993nh,Janke:1993et,Grabenstein:1994ze}, it is an open challenge to construct such an algorithm, particularly for QCD, due to the presence of a nonlocal fermion determinant in the measure.

\begin{acknowledgments}

The authors would like to acknowledge C.-J. David Lin and Evan Weinberg for helpful advice regarding gradient flow and topological charge, and Simon Catterall for information about computing renormalized coupling constants numerically.
All simulations were performed using a modified version of the Chroma Software System for lattice QCD~\cite{Edwards2005832}.
Computations for this study were carried out in part on facilities of the USQCD Collaboration, which are funded by the Office of Science of the U.S. Department of Energy. 
M. G. E. was supported by the U. S. Department of Energy Early Career Research Award No. DE-SC0010495, and moneys from the Dean of Science Office at MIT.
R. C. B. was supported by the U. S. Department of Energy under Grant No. DE-SC0010025.
W. D. was supported in part by the U. S. Department of Energy Early Career Research Award No. DE-SC0010495.
K. O. was supported by the U.S. Department of Energy under Grant No. DE-FG02-04ER41302 and through Contract No. DE-AC05-06OR23177 under which JSA operates the Thomas Jefferson National Accelerator Facility. 
A. V. P. was supported in part by the U.S. Department of Energy Office of Nuclear Physics under Grant No. DE-FC02-06ER41444.

\end{acknowledgments}

\bibliography{quenched}

\appendix

\section{HMC tuning}
\label{appendix:hmc_tuning}

\begin{figure} 
\includegraphics[width=\figWidthHalf]{\figdir 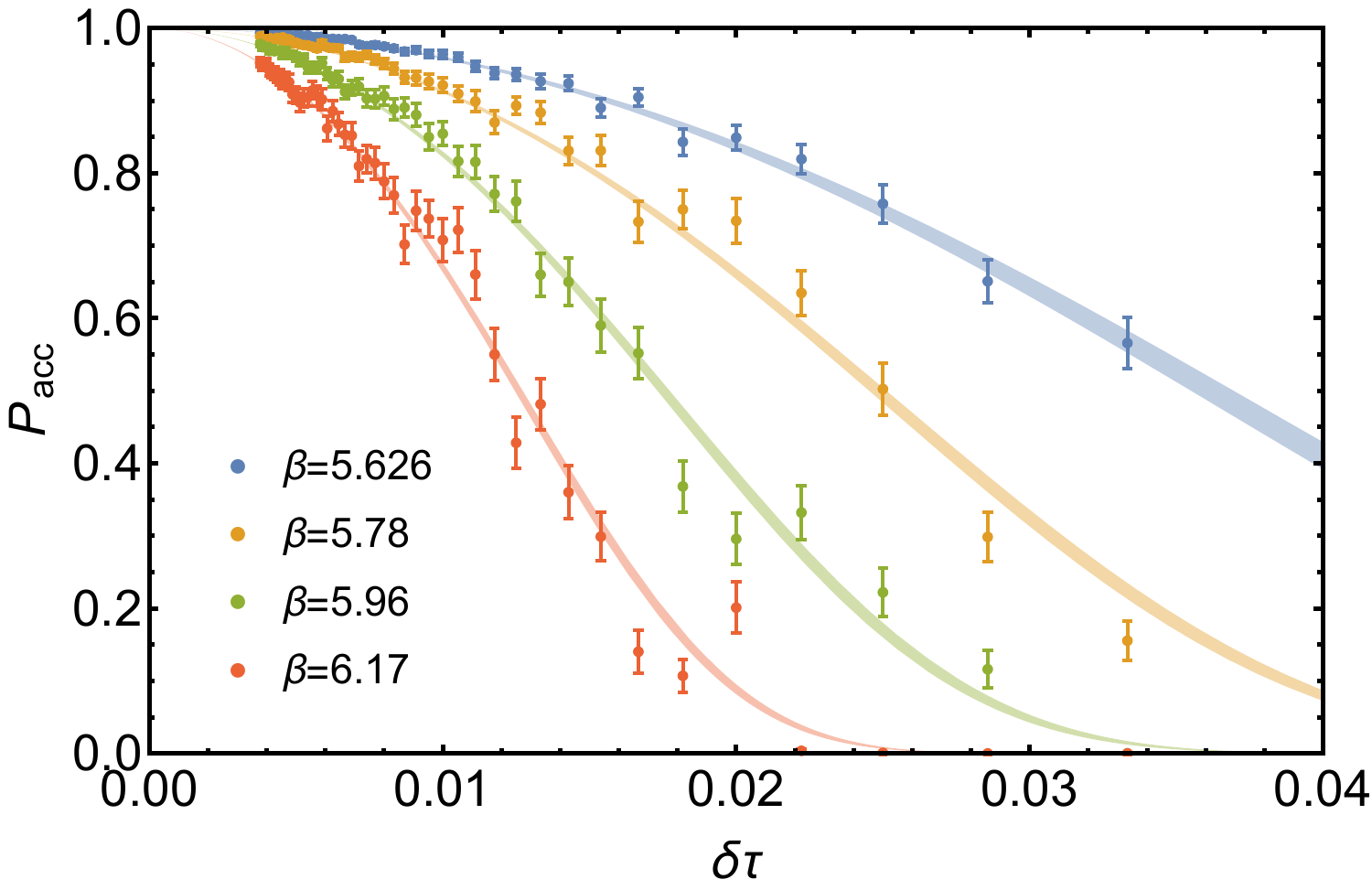}
\hspace{10pt}
\includegraphics[width=\figWidthHalf]{\figdir 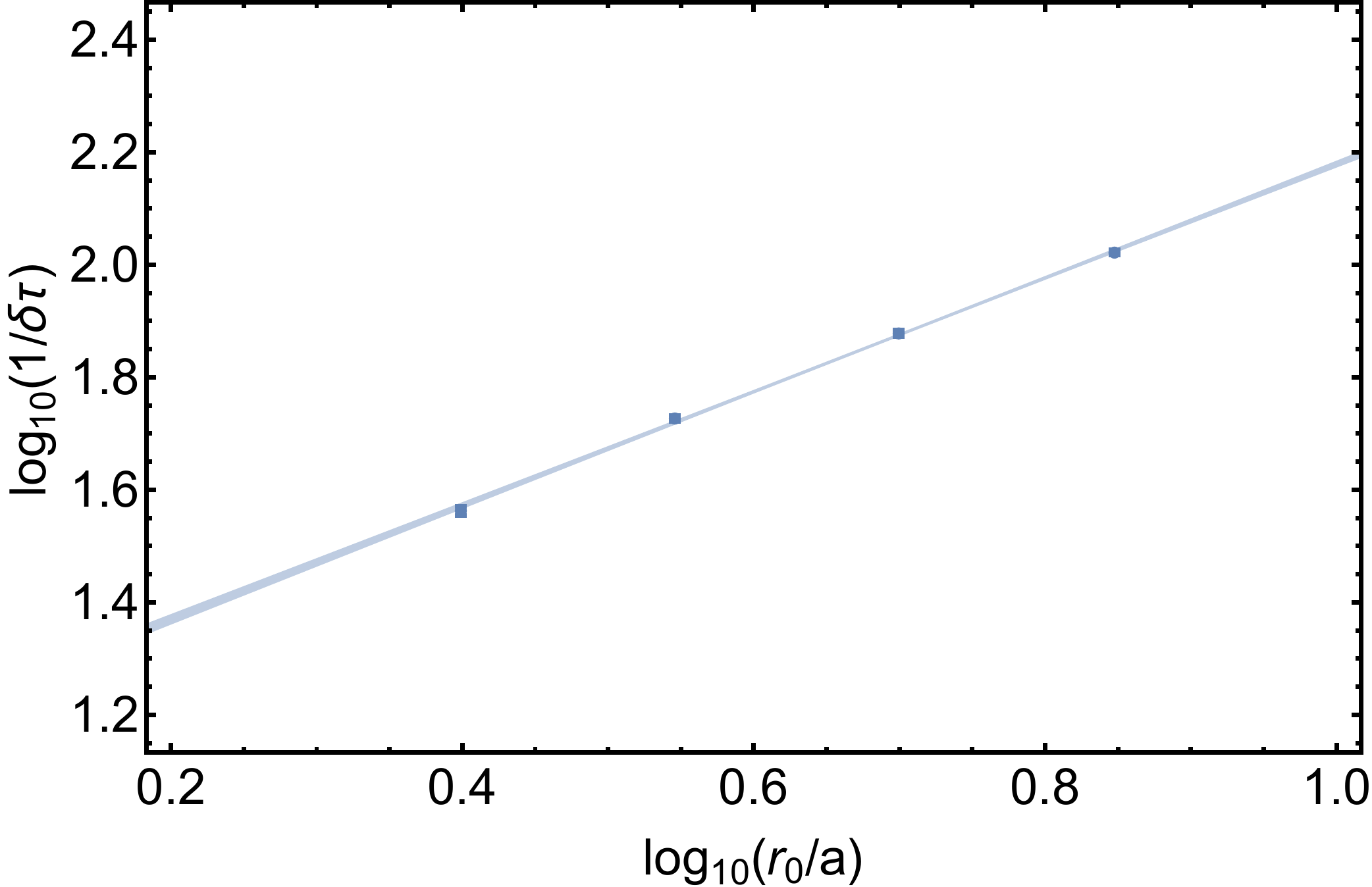}
\caption{\label{fig:tuning}%
Left: HMC acceptance probabilities as a function of step size, $\delta\tau$, for a fixed trajectory of unit length.
Right: Number of steps per trajectory ($1/\delta\tau$) as a function of lattice spacing, for a fixed acceptance probability of 70\%.
}
\end{figure}

Acceptance probabilities for each HMC step size $\delta\tau$ and $\beta$ were estimated from 100 independent trajectories, starting from thermalized configurations.
Estimated acceptance rates are displayed in \Fig{tuning} (left).
The estimated acceptance rates were then fit to the expected functional form
\begin{eqnarray}
P_\textrm{acc} = \textrm{erfc}\left( \frac{1}{2} \Omega^{1/2} \delta\tau^2 \right)\ ;
\end{eqnarray}
the extracted fit values for $\Omega^{1/2}$ are provided in \Tab{tuning_tab} in units of the scale $t_0$.
Fitting the extracted values of $\Omega^{1/2}$ as a function of scale parameter $t_0$, we find $\Omega^{1/2}(t_0) = 1096(14) t_0$.
For each of the HMC studies in this work, the trajectory length is fixed to unity, and the acceptance rate is chosen to be approximately $70\%$.
The nominal number of steps per trajectory, $\delta\tau^\star$, used in these studies, is provided in \Tab{tuning_tab}. 
As a function of the lattice spacing, the requisite number of steps per trajectory length to achieve such an acceptance rate exhibits a power-law behavior, as demonstrated in \Fig{tuning} (right).
Fitting the data, we find $1/\delta\tau \propto \left(r_0/a \right)^{1.012(17)}$.
In light of the fact that the HMC algorithm is not renormalizable~\cite{Luscher:2011qa}, the scaling behavior observed in this study may be regarded as empirical in nature.

\begin{table}
\caption{%
\label{tab:tuning_tab}%
Fit parameters for $\Omega^{1/2}$ in units of $t_0$, estimated HMC step size $\delta\tau$, and  nominal HMC step size $\delta\tau^\star$ required to achieve a 70\% acceptance rate, for a trajectory of unit length.
}
\begin{ruledtabular}
\begin{tabular}{ccccc}
Lattice & $\beta$ & $\Omega^{1/2}/t_0$ & $1/\delta\tau$  & $1/\delta\tau^\star$ \\
\hline
$12^3\times 24$   & 5.626           & 1044(39) & 36.5(0.7) & 37   \\
$16^3\times 36$   & 5.78\phantom{0} & 1138(28) & 53.3(0.5) & 54   \\
$24^3\times 48$   & 5.96\phantom{0} & 1084(27) & 75.5(0.9) & 75   \\
$32^3\times 72$   & 6.17\phantom{0} & 1098(23) & 105.1(1.1) & 105  \\
\end{tabular}
\end{ruledtabular}
\end{table}

\section{Autocorrelations}
\label{appendix:autocorrelations}

Consider a set of data, $X$, comprising arbitrarily spaced measurements $x_\tau$, labeled by $\tau\in[1,N]$.
The autocorrelation function at lag-time $\Delta$ is defined by
\begin{eqnarray}
\rho_\Delta(X) = \frac{\Gamma_\Delta(X) }{\Gamma_0(X) }\ ,
\end{eqnarray}
where $\Gamma_\Delta(X)$ is the corresponding autocovariance function for $X$.
To estimate $\Gamma_\Delta(X)$, first consider the set $\delta X$ comprising the elements $\delta x_\tau  = x_\tau - \bar x$ for all $\tau$, where
\begin{eqnarray}
\bar x = \frac{1}{N} \sum_{\tau=1}^N x_\tau\ ,
\end{eqnarray}
and the function
\begin{eqnarray}
\hat\Gamma_\Delta(\delta X) = \frac{1}{N} \sum_{\tau=1}^{N-\Delta} \delta x_\tau \delta x_{\tau+\Delta}\ ,
\end{eqnarray}
defined on this set.
If $X$ consists of uniformly sampled data, then the autocovariance function is simply given by
\begin{eqnarray}
\Gamma_\Delta(X) = \frac{N}{N-\Delta} \hat \Gamma_\Delta(\delta X)\ .
\end{eqnarray}
Note that the computational cost of evaluating $\hat\Gamma_\Delta(\delta X)$ is $\calO(N\log N)$ using standard discrete Fourier transform methods, whereas the computational cost for evaluating $\Gamma_\Delta(X)$ for arbitrarily spaced samples is generally $\calO(N^2)$.

Assuming $X$ consists of uniformly spaced data, the errors on $\Gamma_\Delta(X)$ can be computed efficiently via the following jackknife procedure.
First partition the measurements $x_\tau$ into $N/N_{J}$ consecutive blocks of size $N_J$, where it is assumed that $N$ mod $N_J$ equals zero.
Labeling the partitions by the integers $j\in[1,N/N_J]$, define the jackknife ensemble $X^j$ comprising the $N-N_J$ elements of $X$, with elements $x_\tau$ on the interval $\tau \in ( (j-1) N_J , j N_J] $ omitted.
Furthermore, define the set $\delta X^j$ which comprises elements
\begin{eqnarray}
\delta x_\tau^j = \left\{
\begin{array}{ll}
0   & (j-1) N_J < \tau \le j N_J \\
x_\tau -\bar x^j & \textrm{otherwise}
\end{array}
\right.\ .
\end{eqnarray}
Note that although $X^j$ comprises $N-N_J$ elements, $\delta X^j$ comprises $N$ elements, of which $N_J$ vanish.
The autocorrelation function on the $j$th jackknife ensemble is given by
\begin{eqnarray}
\Gamma_\Delta(X^j) = \frac{N}{g_\Delta(X^j)}\hat\Gamma_\Delta(\delta X^j)\ ,
\end{eqnarray}
where the piecewise function $g_\Delta(X^j)$ quantifies the degeneracy of distances $\Delta$ for the ensemble $X^j$.
If the set of integers $m_\alpha$ for $\alpha \in [1,5]$ label the quantities $\{ N_J, (j-1)N_J, j N_J, N-j N_J, N-(j-1)N_J \}$ in ascending order, then
\begin{eqnarray}
g_\Delta(X^j) = \left\{
\begin{array}{ll}
N-N_J-2\Delta      & \qquad \,\,\, 0 \le \Delta < m_1 \\
N-N_J -m_1 - \Delta    & \qquad m_1 \le \Delta < m_2 \\
N-N_J -m_1-m_2             & \qquad m_2 \le \Delta < m_3 \\
N-N_J -m_1-m_2+m_3 -\Delta       & \qquad m_3 \le \Delta < m_4 \\
N-N_J -m_1-m_2+m_3-m_4      & \qquad m_4 \le \Delta < m_5 \\
N - \Delta         & \qquad m_5 \le \Delta < N
\end{array}
\right. \ .
\end{eqnarray}

Once the jackknife estimates $\rho_\Delta(X^j)$ are obtained, the standard error is determined by
\begin{eqnarray}
\delta \rho^2_\Delta(X) =  \frac{N}{N-N_J} \sum_{j=1}^{N/N_J} \left[\rho_\Delta(X^j) - \rho_\Delta(X) \right]^2 \ .
\end{eqnarray}
The integrated correlation time is given by
\begin{eqnarray}
\tau_\textrm{int}(X) = \frac{1}{2} + \sum_{\Delta=1}^{\Delta_\textrm{max}} \rho_\Delta(X)\ ,
\end{eqnarray}
where the cutoff lag-time $\Delta_\textrm{max}$ is defined as the minimum time at which
\begin{eqnarray}
\rho_\Delta(X) > \sqrt{ \delta \rho^2_\Delta(X) }
\end{eqnarray}
following~\cite{Luscher:2005rx}.
Once $\Delta_\textrm{max}$ is selected, the errors on $\tau_\textrm{int}$ may be determined from the jackknife estimates $\tau_\textrm{int}(X^j)$ via
\begin{eqnarray}
\delta\tau^2_\textrm{int}(X) = \frac{N}{N-N_J} \sum_{j=1}^{N/N_J} \left[\tau_\textrm{int}(X^j) - \tau_\textrm{int}(X) \right]^2\ .
\end{eqnarray}

\section{Variable projection with constraints}
\label{appendix:varpro}

Assume we have a set of measurements $y^\alpha(\tau)$ and covariance matrix $\Gamma^{\alpha\beta}(\sigma,\tau)$.
Given the fit function
\begin{eqnarray}
f^\alpha(\tau) = \sum_n z^\alpha_n e^{-E_n \tau}\ ,
\end{eqnarray}
and the constraint $\phi z=0$, we may construct
\begin{eqnarray}
\chi^2(z,E) = \sum_{\alpha\beta}\sum_{\sigma\tau} \left[y^\alpha(\sigma)-f^\alpha(\sigma) \right] V^{\alpha\beta}(\sigma,\tau) \left[y^\alpha(\tau)-f^\alpha(\tau) \right] + \sum_{\alpha\beta}\sum_{mn} \xi^\alpha_m \phi^{\alpha\beta}_{mn} z^\beta_n \ ,
\label{eq:chi_sq}
\end{eqnarray}
where $V=\Gamma^{-1}$, and $\xi^\alpha_n$ are Lagrange multipliers for each constraint.
Note that $\phi$ is generally a rectangular matrix.
Next, we may express \Eq{chi_sq} as
\begin{eqnarray}
\chi^2(z,E) = \chi^2(0,0) -2 \sum_\alpha\sum_n z^\alpha_n u^\alpha_n(E) + \sum_{\alpha\beta} \sum_{mn} z^\alpha_m W^{\alpha\beta}_{mn}(E) z^\beta_n + \sum_{\alpha\beta}\sum_{mn} \xi^\alpha_m \phi^{\alpha\beta}_{mn} z^\beta_n \ ,
\end{eqnarray}
where
\begin{eqnarray}
u^\alpha_n(E) = \sum_\beta \sum_{\sigma\tau} e^{-E_n \sigma} V^{\alpha\beta}(\sigma\tau) y^\beta(\tau)\ ,
\end{eqnarray}
and
\begin{eqnarray}
W^{\alpha\beta}_{mn}(E) = \sum_{\sigma\tau} e^{-E_m \sigma} V^{\alpha\beta}(\sigma,\tau) e^{-E_n \tau}\ .
\end{eqnarray}
Minimizing this function with respect to $z$ yields
\begin{eqnarray}
z(E) = W^{-1}(E) u(E) - W^{-1}(E) \phi^\Transpose \left[ \phi W^{-1}(E) \phi^\Transpose  \right]^{-1} \phi W^{-1}(E) u(E)\ ,
\label{eq:overlaps}
\end{eqnarray}
where indices $\alpha$ and $n$ have been suppressed.
One can then construct a reduced $\chi^2$, which is only a function of $E$ and given by
\begin{eqnarray}
\chi^2_r(E) = \chi^2(z(E),E)\ .
\end{eqnarray}
Numerical minimization of $\chi^2_r(E)$ proceeds with standard methods, yielding $E_\textrm{min}$; the corresponding  overlaps $z(E_\textrm{min})$ can be reconstructed from \Eq{overlaps}.

\section{Thermalization and rethermalization of Wilson loops}
\label{appendix:wilson_loops}

\begin{figure} 
\includegraphics[width=\figWidthHalf]{\figdir 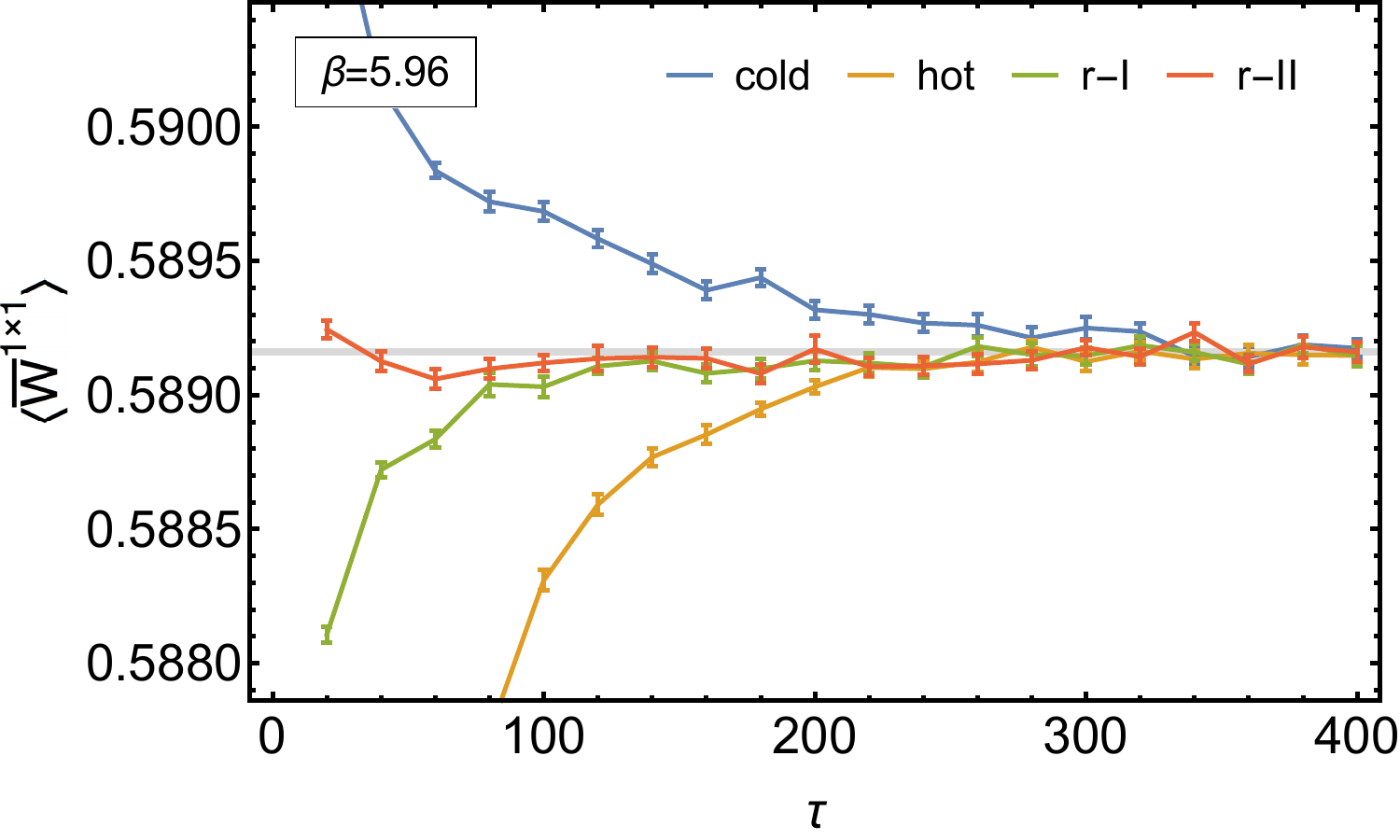}
\hspace{10pt}
\includegraphics[width=\figWidthHalf]{\figdir 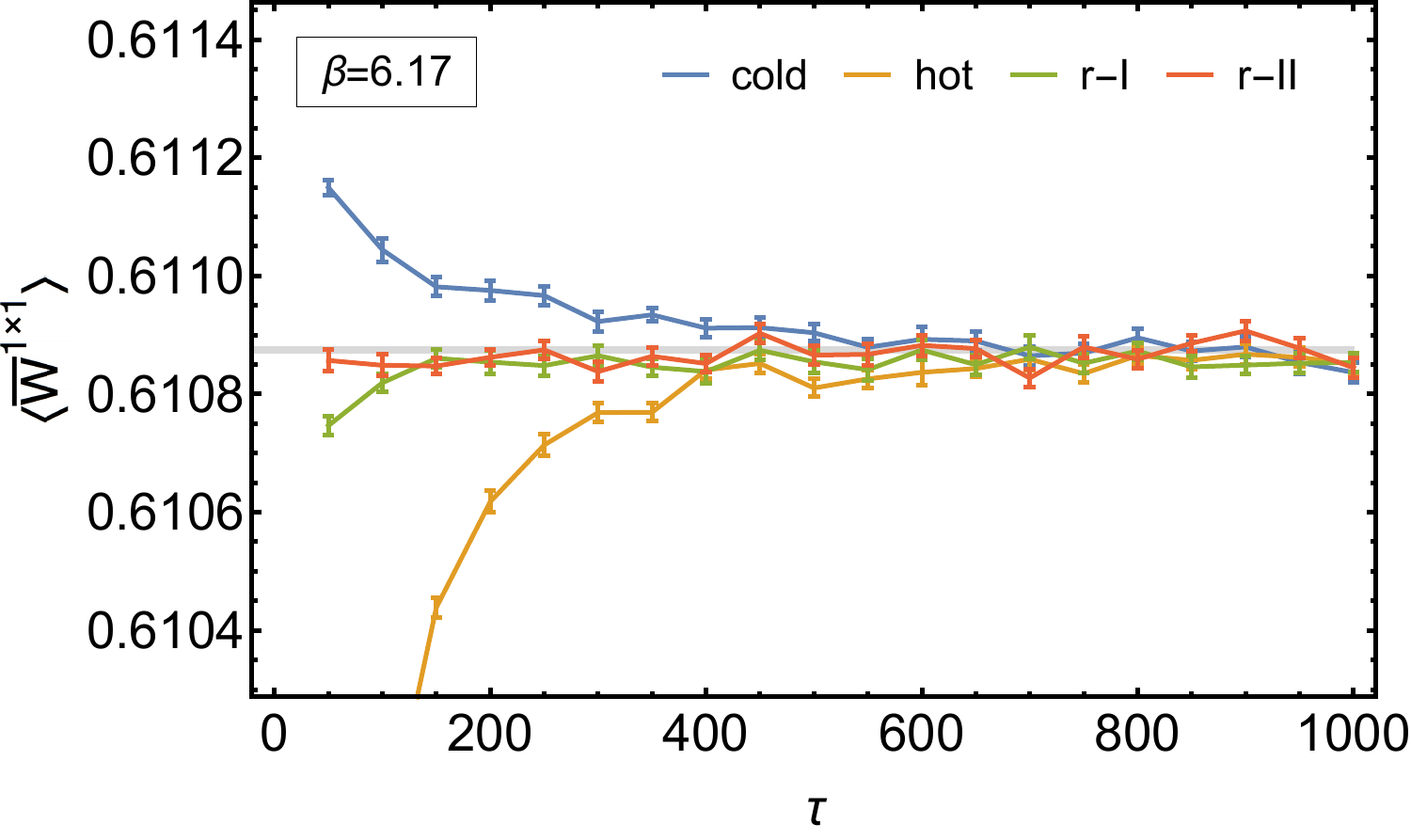}
\vspace{10pt}
\includegraphics[width=\figWidthHalf]{\figdir 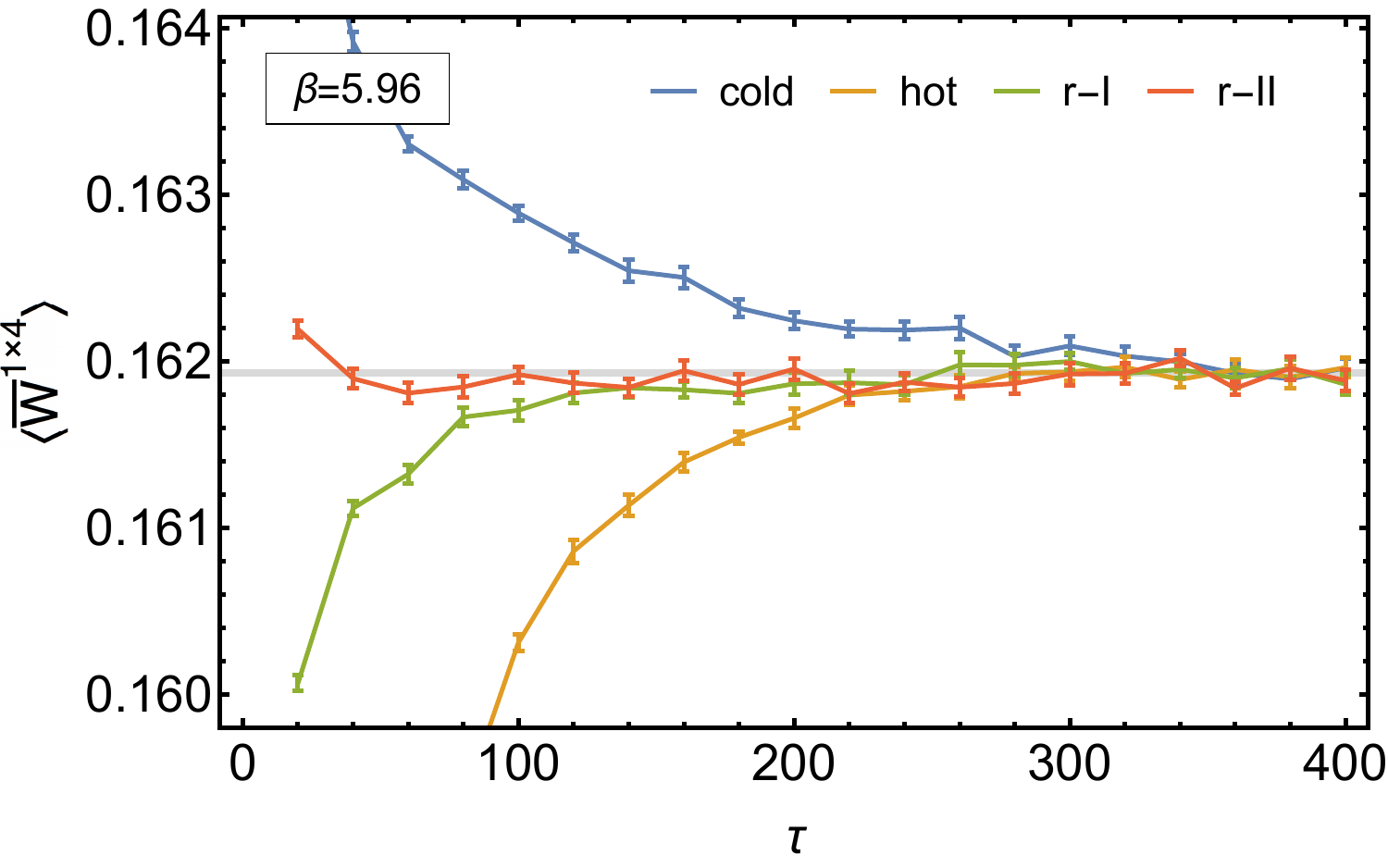}
\hspace{10pt}
\includegraphics[width=\figWidthHalf]{\figdir 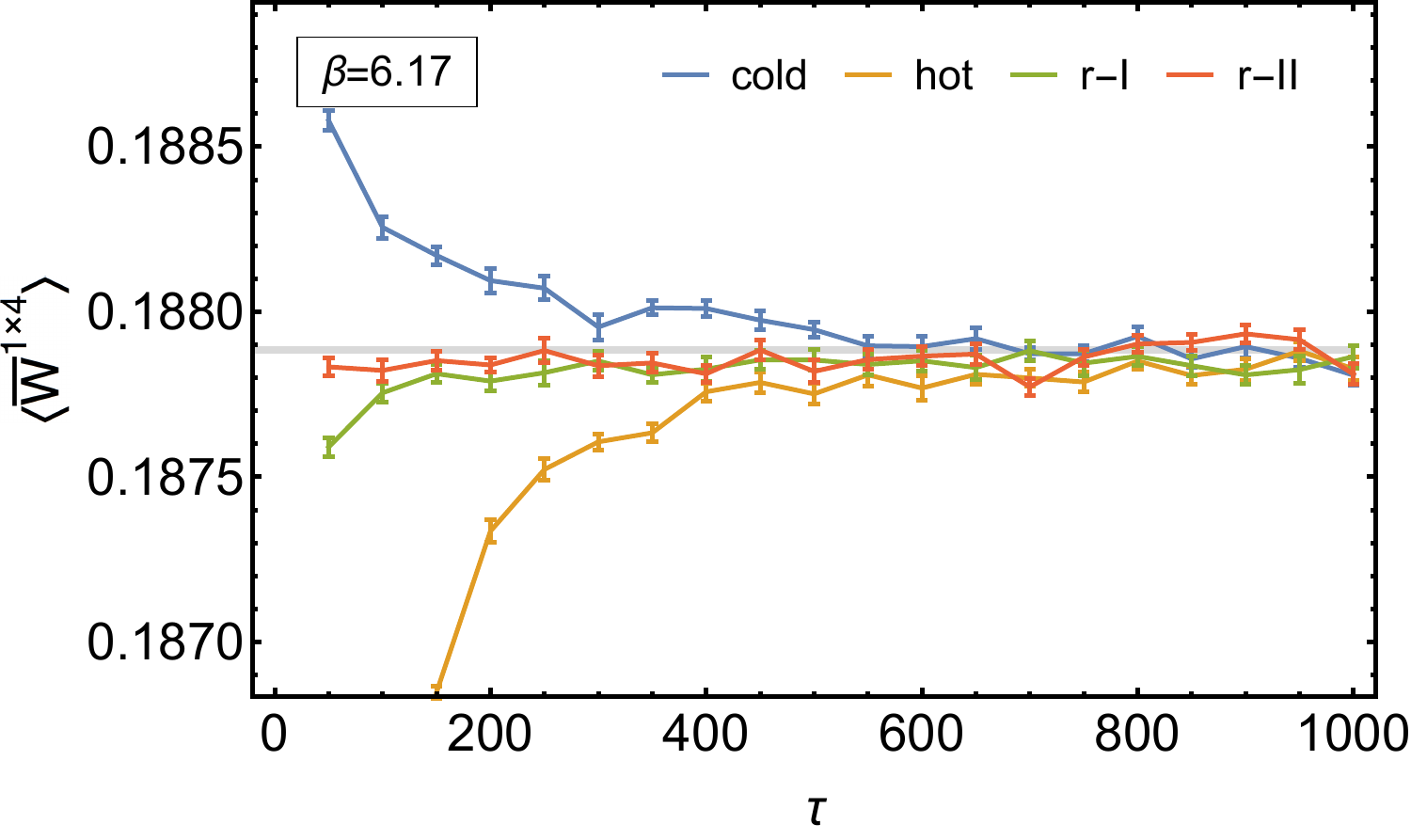}
\vspace{10pt}
\includegraphics[width=\figWidthHalf]{\figdir 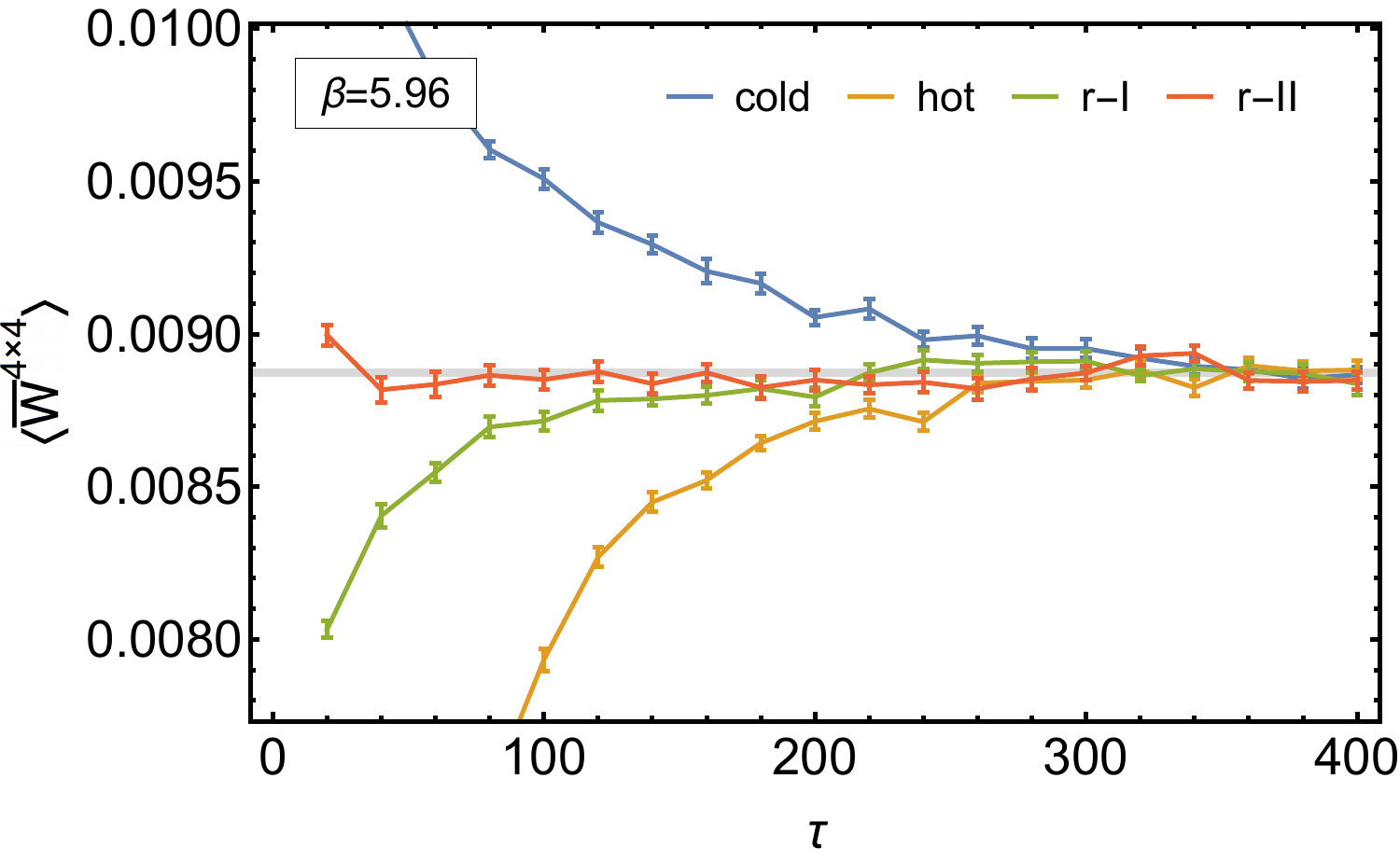}
\hspace{10pt}
\includegraphics[width=\figWidthHalf]{\figdir 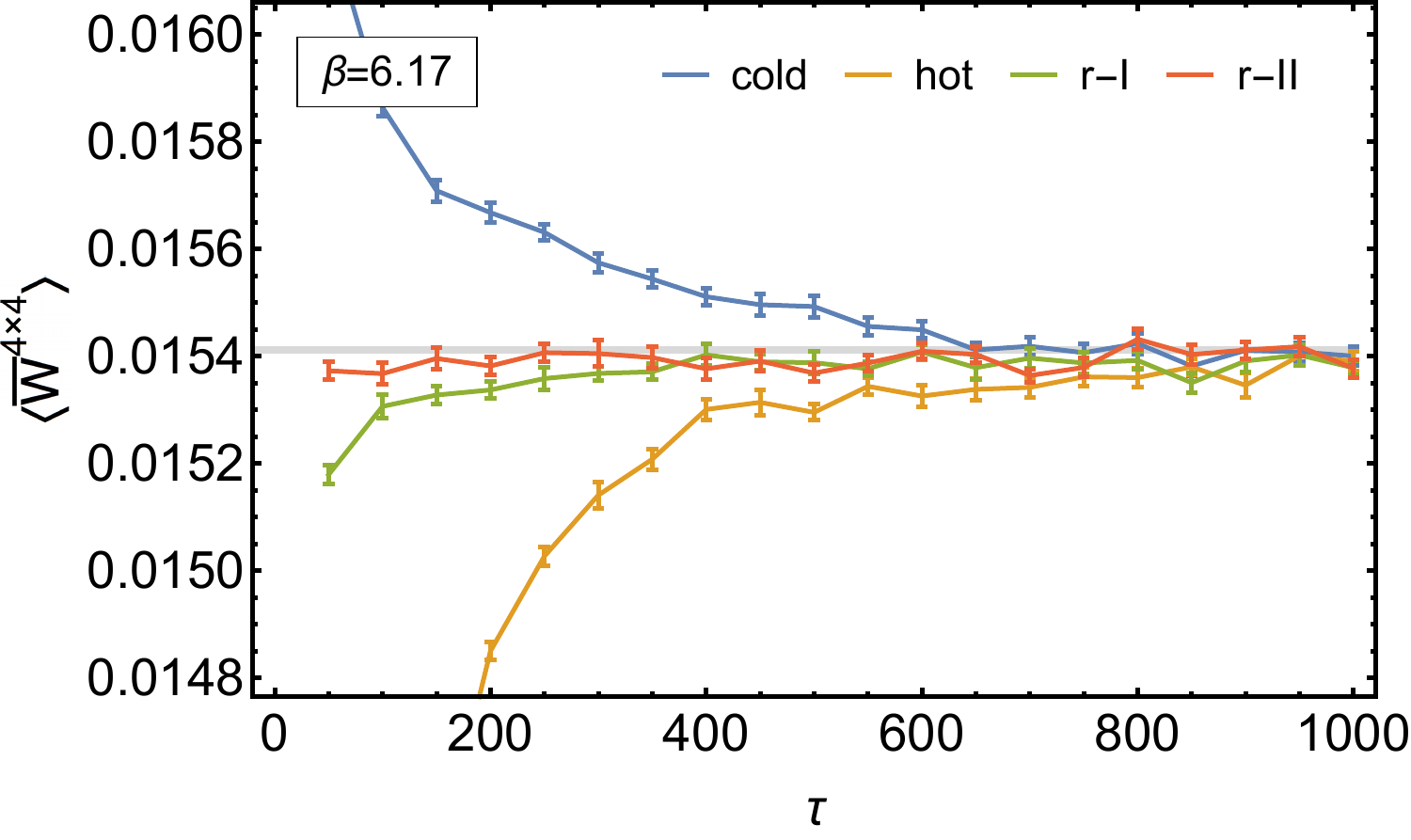}
\caption{\label{fig:purgaug_retherm_w}%
HB (re)thermalization of space-time Wilson loops $W$ as a function of the number of sweeps.
}
\end{figure}

\begin{figure} 
\includegraphics[width=\figWidthHalf]{\figdir 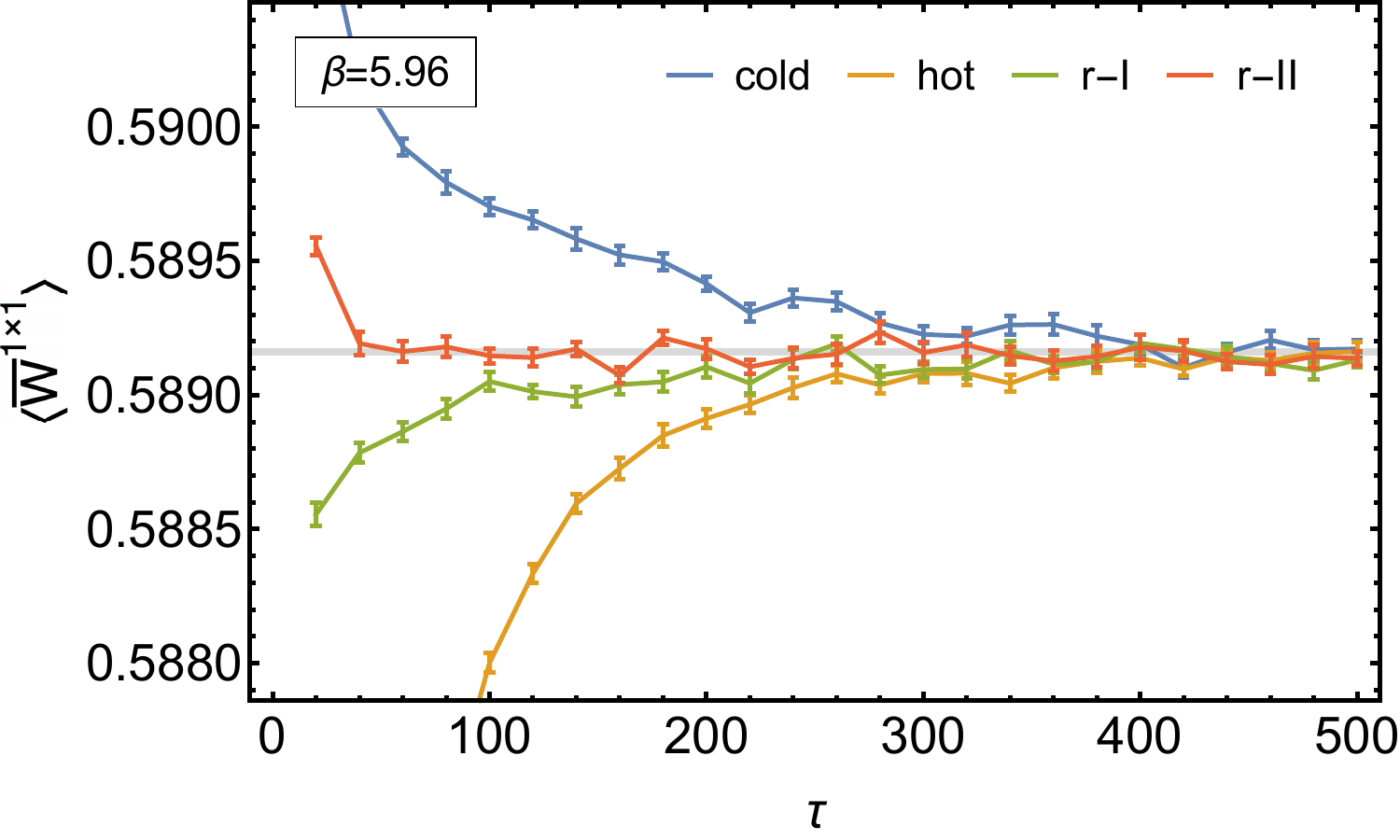}
\hspace{10pt}
\includegraphics[width=\figWidthHalf]{\figdir 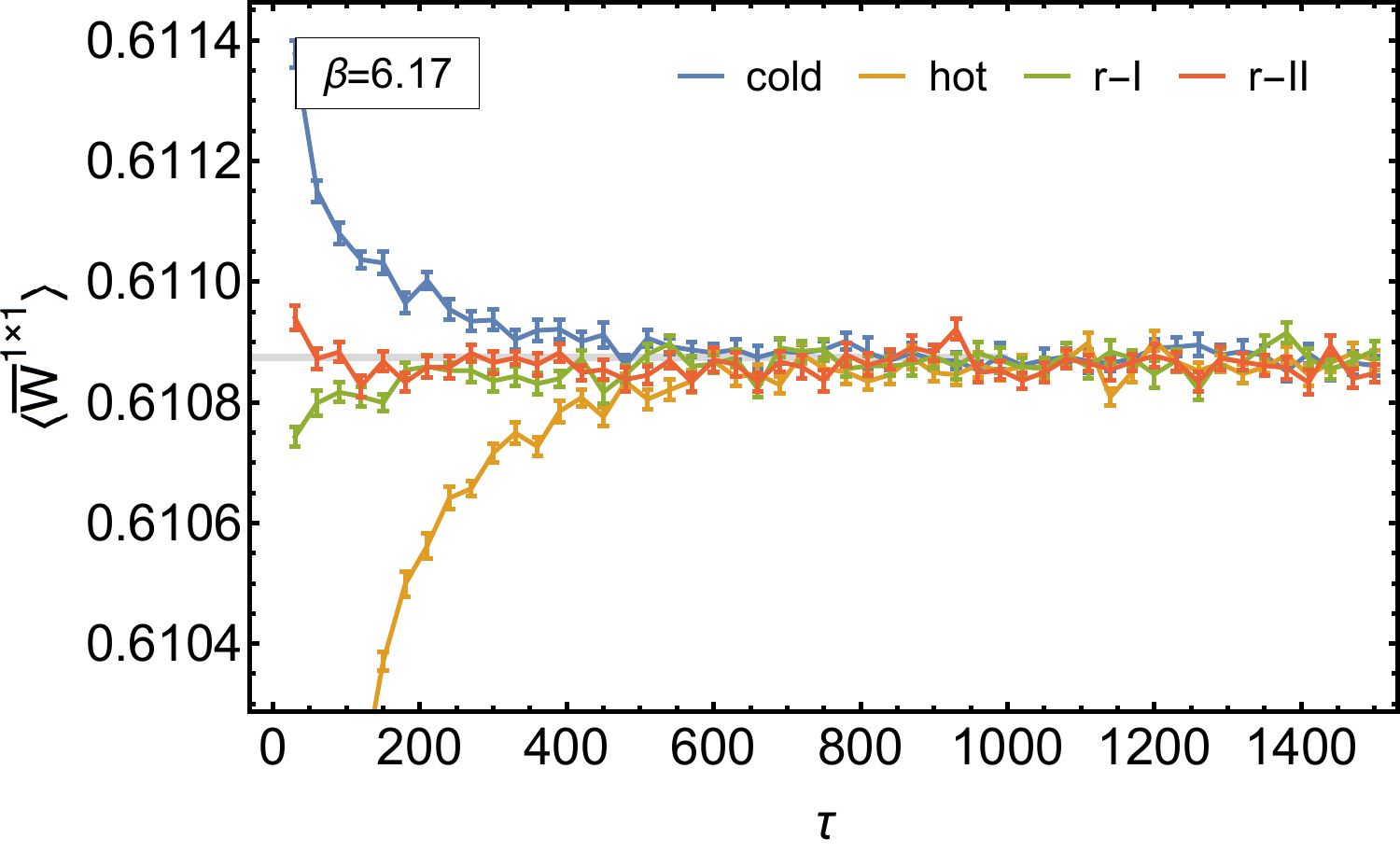}
\vspace{10pt}
\includegraphics[width=\figWidthHalf]{\figdir 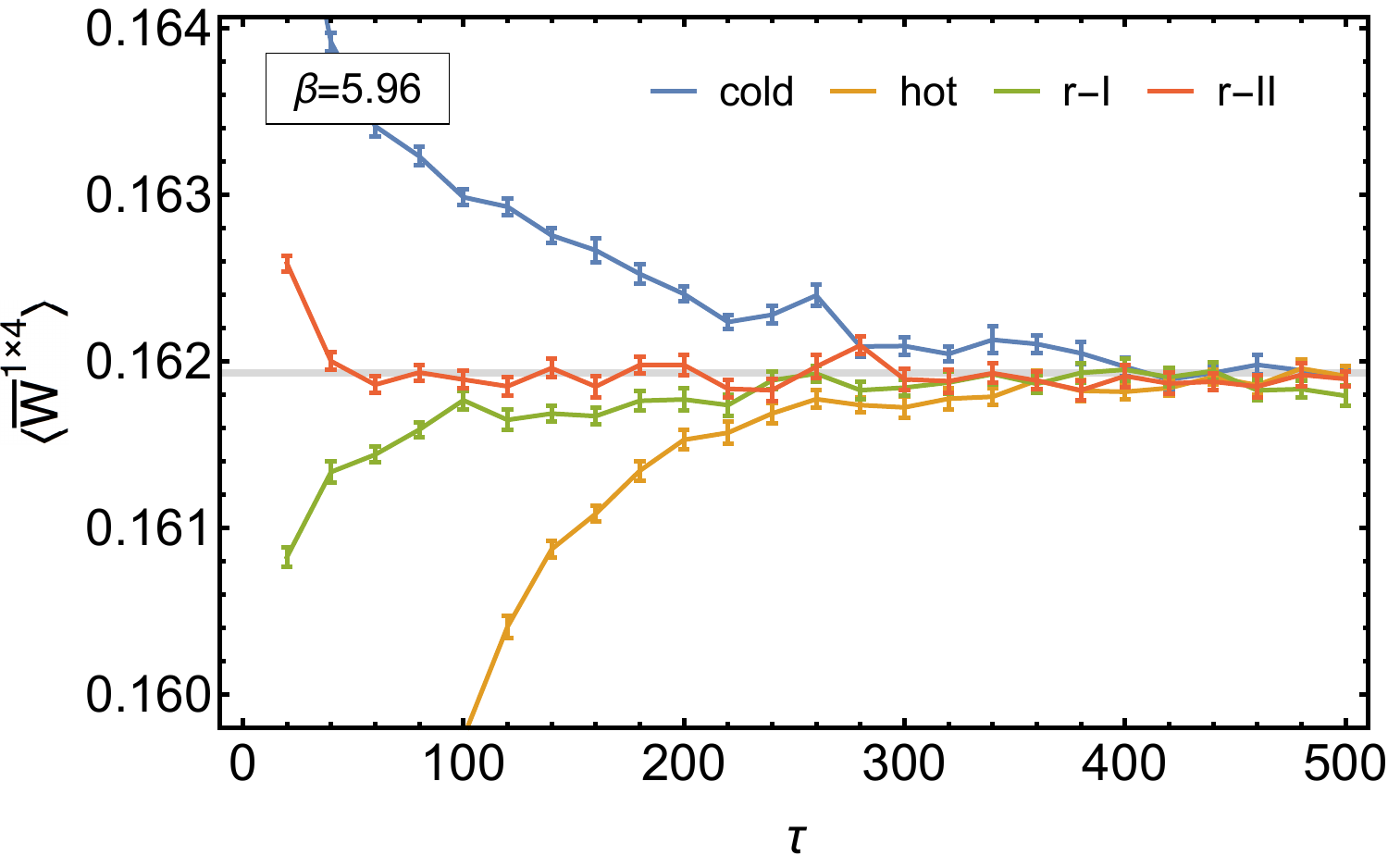}
\hspace{10pt}
\includegraphics[width=\figWidthHalf]{\figdir 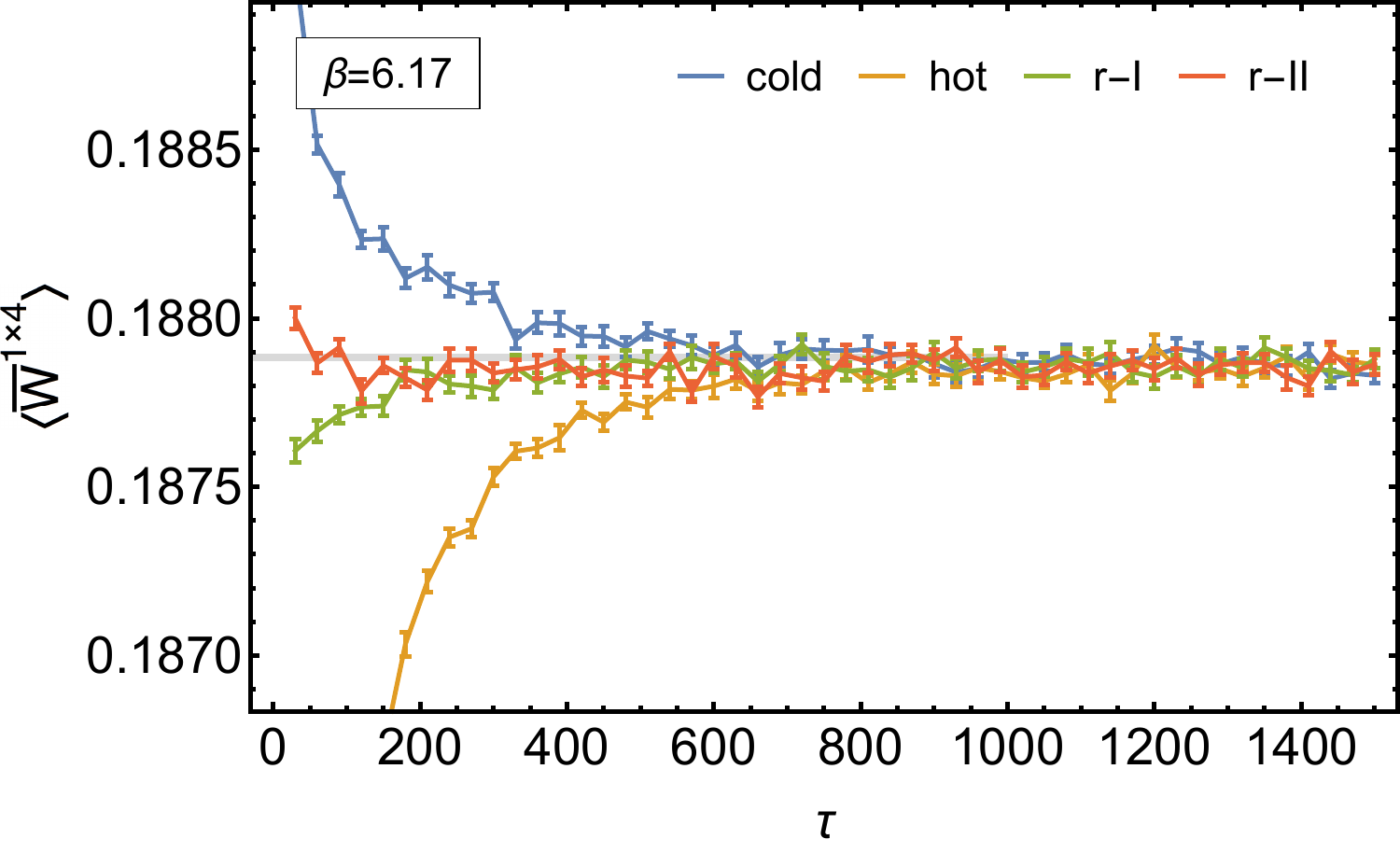}
\vspace{10pt}
\includegraphics[width=\figWidthHalf]{\figdir 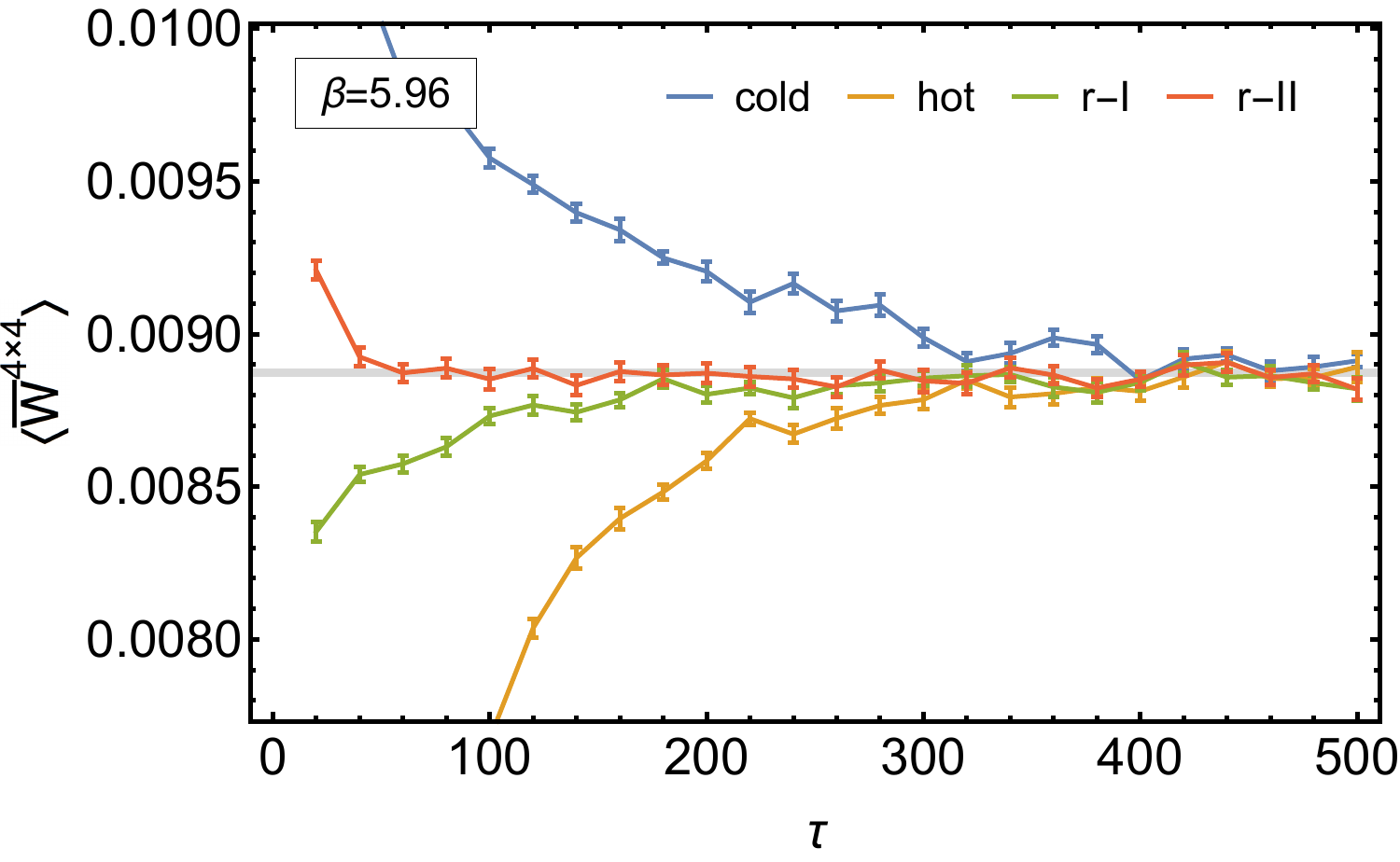}
\hspace{10pt}
\includegraphics[width=\figWidthHalf]{\figdir 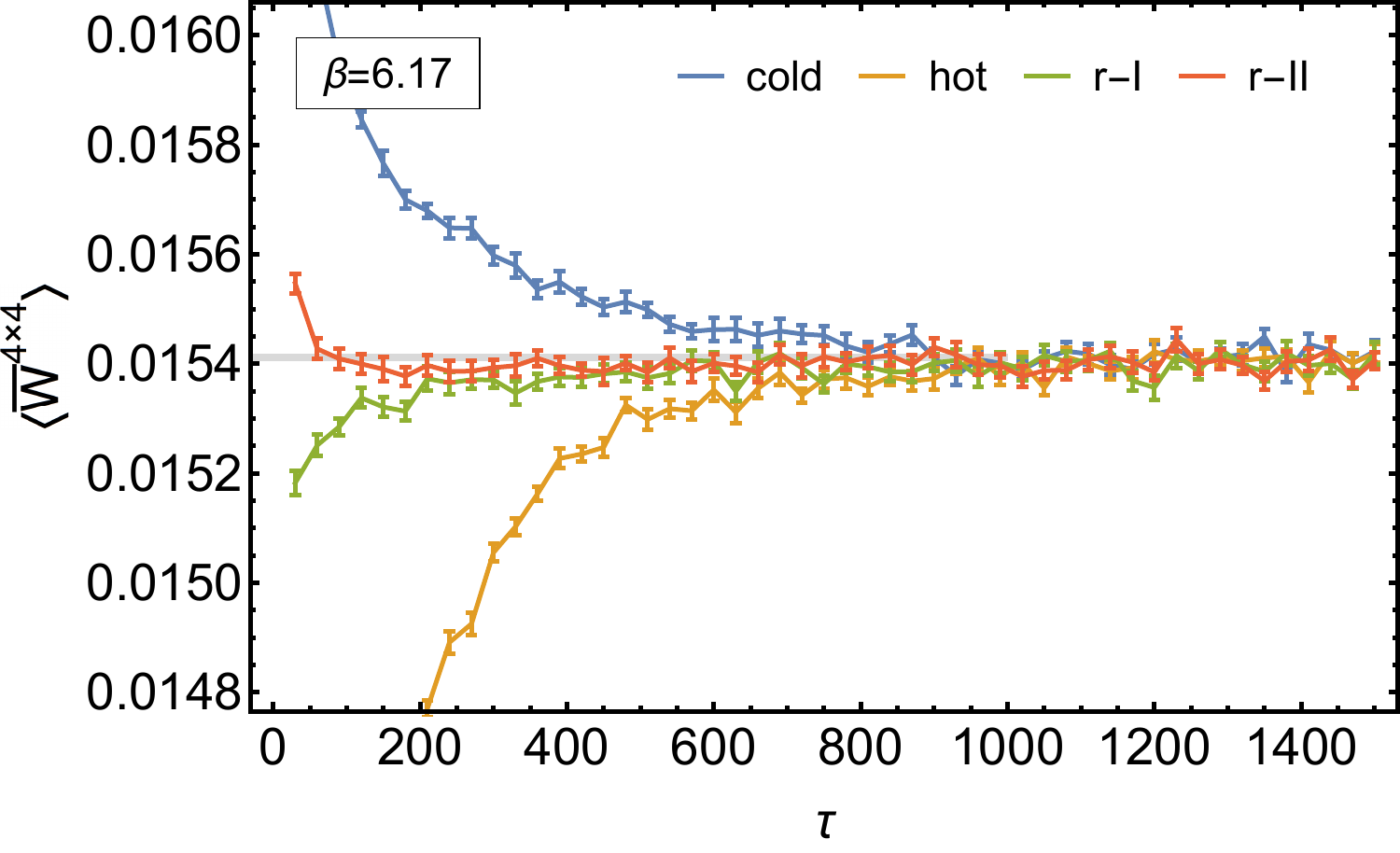}
\caption{\label{fig:hmc_retherm_w}%
HMC (re)thermalization of space-time Wilson loops $W$ as a function of the number of trajectories.
}
\end{figure}

(Re)thermalization curves for Wilson loops of various shapes are provided in \Fig{purgaug_retherm_w} for HB and \Fig{hmc_retherm_w} for HMC.
As with the Wilson flow quantity $t^2 E(t)$, we find that the rethermalization time for the r-II ensemble is significantly shorter than that of r-I.
Furthermore, both rethermalization times are shorter than their hot and cold counterparts.
The (re)thermalization times for Wilson loops generally appear to be significantly shorter than that of $t^2 E(t)$, even for physically large Wilson loops of size $0.5$ fm.
This is likely attributed to the fact that our prolongator preserves all even size Wilson loops originating at sites $\bfn$ satisfying $\chi(\bfn) =0$.

\end{document}